\documentclass[12pt,a4paper]{article}
\pdfoutput=1
\usepackage{jheppub}

\usepackage{bm}
\usepackage{ifsym}
\usepackage{amsthm}
\usepackage{amsfonts}
\usepackage{ccaption}
\usepackage{mathrsfs}
\usepackage{booktabs}
\usepackage{enumerate}

\makeatletter
\@addtoreset{equation}{section}

\makeatletter
\renewcommand\section{\@startsection {section}{1}{\z@}%
                                   {-3.5ex \@plus -1ex \@minus -.2ex}
                                   {2.3ex \@plus.2ex}%
                                   {\normalfont\large\bfseries}}
\renewcommand\subsection{\@startsection{subsection}{2}{\z@}%
                                     {-3.25ex\@plus -1ex \@minus -.2ex}%
                                     {1.5ex \@plus .2ex}%
                                     {\normalfont\bfseries}}

  \captionnamefont{\bfseries}
  \captiontitlefont{\small\sffamily}
  \captiondelim{: }
  \hangcaption

\def\sec#1{\S\ref{#1}}
\def\fig#1{Fig.\,\ref{#1}}
\def\req#1{(\ref{#1})}
\def\App#1{Appendix \ref{#1}}

\def\thuss{\qquad \Longrightarrow \qquad}

\definecolor{rust}{rgb}{0.8,0.2,0.2}
\definecolor{green}{rgb}{0.1,0.8,0.2}
\definecolor{hue55}{rgb}{0.0,0.5,0.67}
\definecolor{hue75}{rgb}{0.65,0.15,0.6}

\def\AdS#1{AdS$_{#1}$}
\def\SAdS#1{Schwarzschild-AdS$_{#1}$}

\def\SAdS{Schwarzschild-AdS}
\def\io{\alpha}  
\def\rh{r_h}   
\def\tA{t_{{\cal A}}}   
\def\phA{\varphi_{{\cal A}}}  
\def\rE{r_{{\mathfrak E}}}   
\def\rX{r_{\Xi}}  
\def\rxi{r_{\xi}}  
\def\Xicurve{{\rm x}}
\def\rs{r_s}   
\def\phs{\varphi_s}  
\def\ts{t_s}  
\def\og{{\scriptscriptstyle \nearrow}}
\def\ig{{\scriptscriptstyle \nwarrow}}

\def\CA{{\cal A}}

\def\ph{\varphi}

\def\ddtipp{{q^\vee}}  
\def\ddtipf{{q^\wedge}}  
\def\cwedge{\blacklozenge_{{\cal A}}} 
\def\domd{\Diamond_{\cal A}}  

\def\ellk{k}
\def\rbdy{{\bar r}}  
\def\zmax{{z_{\rm max}}}  
\def\zcutoff{{z_{\rm co}}}  

\def\bcwedgef{\partial_+(\blacklozenge_{{\cal A}})}
\def\bcwedgep{\partial_-(\blacklozenge_{{\cal A}})}

\def\extr#1{{\mathfrak E}_{#1}}   
\def\csf#1{{\Xi}_{#1}}  

\title{Thermalization of Causal Holographic Information}
\author{Veronika E. Hubeny$^a$}
\author{  Mukund Rangamani$^a$}
\author{\& Erik Tonni$^b$}

\affiliation[a]{ Centre for Particle Theory \& Department of Mathematical Sciences,\\
Science Laboratories, South Road, Durham DH1 3LE, UK.}
\affiliation[b]{SISSA and INFN, via Bonomea 265,  34136, Trieste, Italy}
\emailAdd{veronika.hubeny@durham.ac.uk}
\emailAdd{mukund.rangamani@durham.ac.uk}
\emailAdd{erik.tonni@sissa.it}

\abstract{We study causal wedges associated with a given sub-region in the boundary  of asymptotically AdS spacetimes. Part of our motivation is to better understand the recently proposed holographic observable, causal holographic information, $\chi$, which is given by the area of a bulk co-dimension two surface lying on the boundary of the causal wedge. It has been suggested that $\chi$ captures the basic amount of information contained in the reduced density matrix about the bulk geometry. To explore its properties further we examine its behaviour in time-dependent situations. As a simple model we focus on null dust collapse in an asymptotically AdS spacetime, modeled by the Vaidya-AdS geometry. We argue that while $\chi$ is generically quasi-telelogical in time-dependent backgrounds, for suitable choice of sub-regions in conformal field theories, the temporal evolution of $\chi$ is entirely causal. We comment on the implications of this observation and more generally on features of causal constructions and contrast our results with the behaviour of holographic entanglement entropy. Along the way we also derive the rate of early time growth and late time saturation (to the thermal value) of both $\chi$ and entanglement entropy in these backgrounds. } 


\begin{document}
\begin{flushright} \small{DCPT-13/01} \end{flushright}

\maketitle

\flushbottom



\section{Introduction}
\label{s:intro}
One of the important questions in holography is to understand the precise dictionary between the bulk spacetime and its avatar in the dual boundary quantum field theory.
Over the years we have learnt to encode geometry in terms of field theory observables. While there has been considerable success in identifying key geometrical features in terms of the field theory data, it is  nevertheless clear that the translation between the two descriptions is far from complete. We are still trying to ascertain the sharpest statement about geometry. The present work, which is exploratory in spirit,
examines the features of observables generated from purely causal constructs of the bulk spacetime.

One class of questions which probe the CFT encoding of the bulk geometry starts by restricting the boundary region on which one has access to CFT data.  For example, supposing we know the full reduced density matrix $\rho_\CA$ on some spatial region $\CA$ on the boundary, how much information does $\rho_\CA$ contain about the bulk geometry?
This question was examined recently in \cite{Bousso:2012sj, Czech:2012bh, Hubeny:2012wa,Bousso:2012mh}, though no consensus on the final answer was reached. Instead of confronting this question head-on, \cite{Hubeny:2012wa} took an indirect approach of asking: what is the most natural (i.e., simplest, nontrivial) bulk region associated to $\CA$?
Given such a natural and therefore important bulk construct, we expect that there should exist a correspondingly important dual quantity in the field theory, perhaps waiting to be found.  If we succeed in identifying such an object within the field theory, we will obtain a more direct handle on the gauge/gravity mapping of the geometry and consequently on bulk locality.  

The most immediate geometrical construct associated with $\CA$ that probably springs into the reader's mind is an extremal co-dimension two bulk surface which is anchored at the AdS boundary on $\partial \CA$. Indeed, this is a well-known and important construct: In  \cite{Ryu:2006bv,Ryu:2006ef}, Ryu and Takayanagi conjectured that for an equilibrium state, the entanglement entropy of $\CA$  is given by the area of precisely such a bulk surface: a co-dimension two minimal area surface at constant time which is homologous to $\CA$ and anchored on $\partial \CA$ (entangling surface). More generally, for states that evolve non-trivially in time, one should use extremal surfaces as argued in \cite{Hubeny:2007xt}. Though we have no proof to date, there is mounting evidence that entanglement entropy is indeed given by the extremal surface area; see \cite{Headrick:2007km, Hayden:2011ag,Allais:2011ys,Callan:2012ip,Wall:2012uf} for arguments to show that the holographic constructions satisfy entropy inequalities and \cite{Casini:2011kv} for a derivation of the holographic entanglement entropy in some special circumstances.

But is this the most {\it natural} bulk construct associated with $\CA$?
Finding an extremal surface in the bulk requires the knowledge of the bulk geometry.  Although this is what we are ultimately after, there is a more primal and perhaps more fundamental concept, namely the bulk causal structure (the knowledge of which requires a proper subset of the information contained in the bulk geometry). 
In \cite{Hubeny:2012wa} we argued that the simplest and most natural  construct is in fact the {\it causal wedge} (which we will denote $\cwedge$) corresponding to the boundary region $\CA$, and associated quantities.  We will review the definition of $\cwedge$ in detail below; but to orient the reader, a succinct description is as follows: Take the boundary domain of dependence $\domd$ of $\CA$; this is the boundary-spacetime region where the physics is fully determined by the initial conditions at $\CA$.  The bulk causal wedge is the intersection of the causal past and future of $\domd$.  Hence any causal curve through the bulk which starts and ends on $\domd$ must be contained inside the causal wedge $\cwedge$, and conversely we may think of $\cwedge$ as consisting of the set of all such curves.\footnote{Note that \cite{Bousso:2012sj} shows that the causal wedge $\cwedge$ is equivalently defined in terms of the intersection of future and past going light-sheets emanating from $\domd$.  They further argue using the covariant holographic entropy bounds \cite{Bousso:1999xy} that this implies that the causal wedge $\cwedge$ must be the maximal region of the bulk that can be described by observables restricted to $\domd$. Since the extremal surfaces computing entanglement entropy necessarily lie outside the causal wedge \cite{Hubeny:2012wa,Wall:2012uf} it however seems more natural that the boundary theory restricted to $\domd$ is cognizant of a larger part of the bulk as argued in \cite{Czech:2012bh}.}

\begin{figure}
\begin{center}
\includegraphics[width=5in]{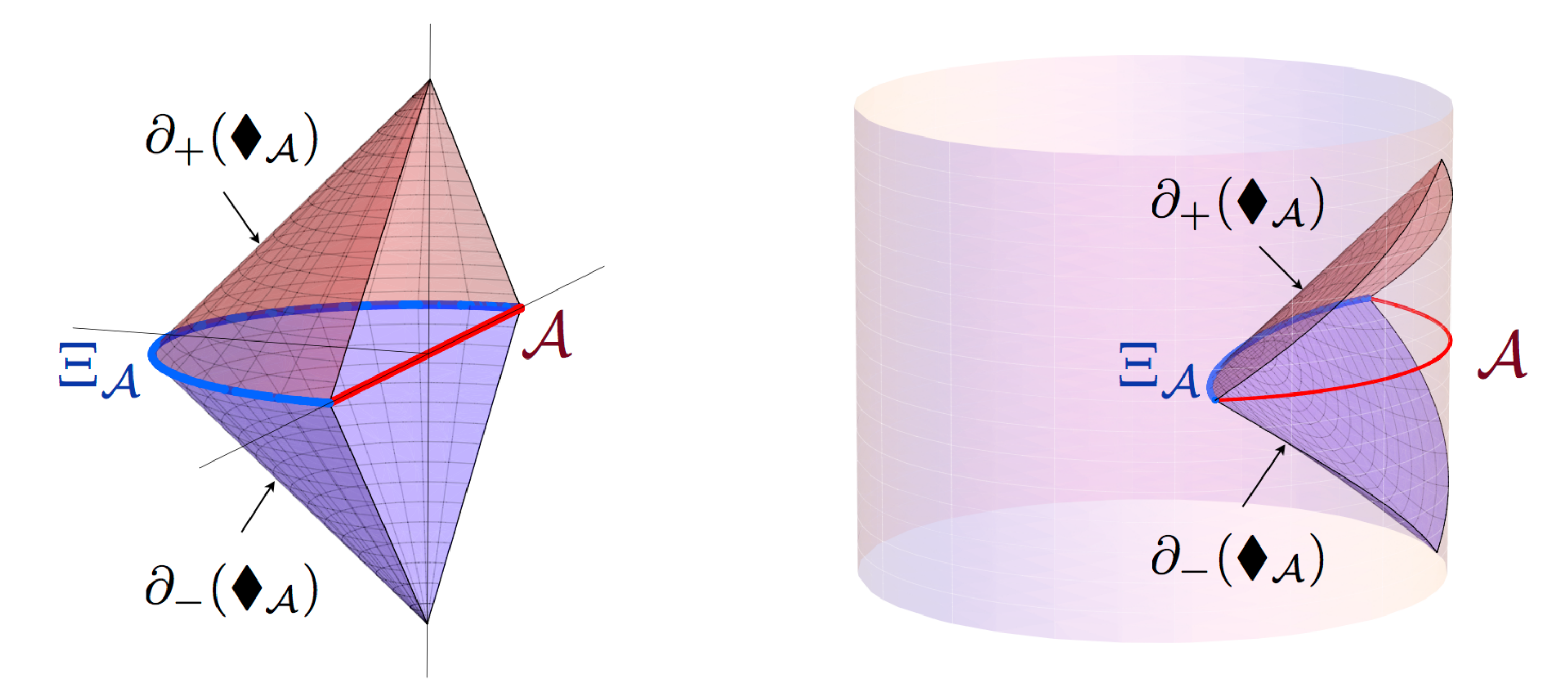} 
\caption{
A sketch of the causal wedge $\cwedge$ and associated quantities in planar AdS (left) and global AdS (right) in 3 dimensions: in each panel, the region $\CA$ is represented by the red curve on right, and the corresponding surface $\Xi_\CA$ by blue curve on left; the causal wedge $\cwedge$ lies between the AdS boundary and the null surfaces $\bcwedgef$ (red surface) and $\bcwedgep$ (blue surface).}
\label{f:CWsketch}
\end{center}
\end{figure}

The causal wedge is a (co-dimension zero) spacetime region; but we can immediately identify  associated lower-dimensional quantities constructed from it, namely bulk co-dimension
one null surfaces, forming the  `future part' $\bcwedgef$ and `past part' $\bcwedgep$ of the boundary of the causal wedge, as well as a bulk co-dimension  two spacelike surface $\Xi_\CA$ lying at their intersection.
For orientation, these constructs are illustrated in \fig{f:CWsketch}, for planar AdS (left) and global AdS (right).
Hence, $\Xi_\CA$,
dubbed the {\it causal information surface} in  \cite{Hubeny:2012wa},
 is a spacelike surface lying within the boundary of the causal wedge which penetrates deepest into the bulk and is anchored on $\partial \CA$.
  In \cite{Hubeny:2007xt,Hubeny:2012wa} we demonstrated that while $\Xi_\CA$ must in fact be a minimal surface within $\partial (\cwedge)$ that is anchored on  $\partial \CA$, it is in general {\it not} an extremal surface in the full spacetime.  There however are certain situations where the causal information surface $\Xi_\CA$ actually coincides with the extremal surface $\extr{\CA}$ as noted in \cite{Hubeny:2012wa}. It was conjectured there  that the corresponding density matrix $\rho_{\cal A}$ was maximally entangled with the rest of the field theory degrees of freedom. Below, we will consider these special situations further and provide additional evidence for this suggestion.
  
So far we have utilized solely the causal structure of the bulk to construct our natural bulk region $\cwedge$ and associated surface $\Xi_\CA$; but now we finally recourse to geometry.  In particular, in analogy with entanglement entropy $S_\CA$ related to the proper area of $\extr{\CA}$, we identify a quantity $\chi_\CA$ related to the proper area of $\Xi_\CA$,
\begin{equation}
\chi_{\cal A} \equiv \frac{\text{Area}(\csf{\cal A})}{4\,G_N} \ .
\label{}
\end{equation}	
  In \cite{Hubeny:2012wa} we called this quantity the {\it causal holographic information}, and studied its properties in equilibrium.  In particular, we conjectured that $\chi_\CA$ provides the lower bound on the holographic information contained in the boundary region $\CA$; however to make this more precise or meaningful, we need to understand better {\it what sort of quantity  $\chi$ is from the field theory standpoint.}
To that end, we will continue the exploration of the properties and behavior of $\chi$ under various circumstances.  

Since \cite{Hubeny:2012wa} constructed $\chi$ and causal wedge in equilibrium configurations, we will explore the properties of $\chi$ in more general out-of-equilibrium situations.  Rather than examining the qualitative features of our constructs in arbitrary spacetimes, it will be instructive to obtain more detailed quantitative results. 
To that end, it is useful to pick  a specific class of examples,  which are not only tractable, but also far out of equilibrium.  The more extreme the time variation, the more easily we can sample the  `dynamics' of $\cwedge$, $\Xi_\CA$, and $\chi_\CA$, e.g.\ when the spatial position of $\CA$ is fixed but we study it at different times.
We focus on a particularly simple time-dependent bulk geometry, describing a collapse of a thin null spherical shell to a black hole in AdS, namely the Vaidya-AdS spacetime.  Since a (sufficiently large) AdS black hole corresponds to a thermal state in the field theory, this geometry has been much-used   to study thermalization in the field theory via black hole formation in the bulk 
(see 
\cite{Danielsson:1999fa,Hubeny:2006yu,AbajoArrastia:2010yt,Albash:2010mv,Balasubramanian:2010ce, Balasubramanian:2011ur,Keranen:2012vn,Galante:2012ys,Caceres:2012em, Balasubramanian:2011at,Balasubramanian:2012zr}
for a sampling of references). Moreover, since the shell is null, the collapse to a black hole (and hence the corresponding boundary thermalization) happens maximally quickly. Also, since the shell is thin (and so starts out from the boundary at a single instant in time), the change in the boundary corresponding to the introduction of the shell is sudden: we deform the boundary Hamiltonian and then let the system equilibrate -- in other words, such process in an example of a quantum quench.\footnote{We  should nevertheless emphasize that the modeling of quantum quench using the Vaidya-AdS spacetime is somewhat contrived. 
Indeed, in the boundary theory the final state is thermal and known by construction, while in a typical global quench protocol one changes a parameter of the Hamiltonian at some time without knowing the final state, which is not guaranteed to be thermal. For instance, for simple integrable models like the Ising chain it is known that the final state is given by a Generalized Gibbs Ensemble (GGE) \cite{Calabrese:2012kx} where in principle all the (infinite) integrals of motion  occurs.
Furthermore, it is more realistic to introduce localized sources to deform the theory, in contrast to the homogeneous disturbance (injected in the UV) used in the null shell collapse. The primary advantage of the models we describe below  is their tractability. This caveat should be borne in mind before drawing general conclusions from our analysis.}  
We refer the reader to \cite{Calabrese:2005uq,Calabrese:2006ve,Calabrese:2007kx,Calabrese:2007fk,Sotiriadis:2008dq} for further discussions of quantum quenches (including computation of observables) in conformal field theories and to \cite{Kovchegov:2007pq, Chesler:2008hg, Bhattacharyya:2009uu, Chesler:2009cy, Das:2010yw, Heller:2011ju, Basu:2011ft,  Heller:2012km, Heller:2012je, Buchel:2012gw,Bhaseen:2012gg,Taliotis:2012sx} for discussion of holographic quenches and thermalization.

To study the behavior of $\cwedge$, $\Xi_\CA$, and $\chi_\CA$ in the Vaidya-AdS class of geometries, we set up the geometrical construction in \sec{s:prelims}.   We discuss the general expectations for the behavior of these quantities in \sec{s:pgenexp} and derive the explicit equations to construct $\cwedge$  in \sec{s:geodvaidya}.  In subsequent sections we examine the detailed `dynamics' of our causal constructs, focusing on the causal holographic information $\chi_\CA$, for Vaidya-AdS$_3$ in \sec{s:adsv3} and for higher-dimensional Vaidya-AdS in \sec{s:adsvd}.
While the thin shell Vaidya-AdS class of geometries studied hitherto illustrates many of the essential features of the  causal wedge and $\chi$, some of these results derive from the large amount of symmetry which has rendered these examples tractable in the first place.  To surmount potential bias towards special situations, and  in order to gain more intuition on  the requirements which any putative CFT duals of these quantities must satisfy in general, we comment on some general properties of the causal construction in \sec{s:discuss}. These will be further explicated in a companion paper exploring more formal aspects \cite{Hubeny:2013fk}.  The appendices collect some of the technical computations for three dimensional Vaidya-AdS spacetimes (in particular \App{s:3dS} contains some new results on entanglement entropy for global collapse and details of early and late time behaviour).

\section{Preliminaries}
\label{s:prelims}

As described in \sec{s:intro}, we would like to understand the behavior of the causal wedge $\cwedge$ and the causal holographic information $\chi_\CA$ introduced in \cite{Hubeny:2012wa}, in situations where the reduced density matrix $\rho_\CA$ associated with the given spatial region $\CA$ is time dependent.  We will concentrate on the 
 process of thermalization following a sudden disturbance which has oft been used as a convenient toy model of a quantum quench. In particular, we consider a field theory on  globally hyperbolic background geometry  ${\cal B}_d \equiv \Sigma_{\cal B}\times {\mathbb R}_t$ in which we introduce a homogeneous disturbance at an instant in time $ t = \ts$. Of specific interest will be the cases where the background is either Minkowski spacetime 
 $\Sigma_{\cal B} = {\mathbb R}^{d-1}$
 or the Einstein Static Universe (ESU), $\Sigma_{\cal B} = {\bf S}^{d-1}$. We can  generate deformations via explicit (relevant) operators introduced into the Lagrangian, and ensure homogeneity by smearing the insertion over the constant time slices $\Sigma_{\cal B}$.
 The resulting configuration will then undergo some non-trivial evolution, whose consequences we wish to examine for a specified boundary region $\CA \subset \Sigma_{\cal B}$. 

In the gravity dual, the said process of  thermalization will be described by a simple spherically symmetric null shell collapse geometry. We model this by the Vaidya-\AdS{d+1} spacetime with the metric:
\begin{equation}
ds^2 = 2 \, dv \, dr -  f(r,v)\, dv^2 + r^2\, d{\bf \Sigma}_{d-1,K}^2  \ , \qquad f(r,v) = r^2\left(1 + \frac{K}{r^2}- \frac{m(v)}{r^d}\right)
\label{vaidya}
\end{equation}	
where $r$ is the bulk radial coordinate such that $r=\infty$ corresponds to the boundary, the null coordinate $v$ coincides with the boundary time (i.e.\ we fix $v=t$ on the boundary of the spacetime), and $d{\bf \Sigma}_{d-1,K}^2$ describes the metric on a plane (sphere) $\Sigma_{\cal B}$ for $K =0 \ (K=1)$, so that $K$ keeps track of the spatial curvature of the boundary spacetime geometry.
The bulk spacetime \req{vaidya} interpolates between vacuum \AdS{d+1} and a  \SAdS$_{d+1}$ black hole if $m(v) \to \{ 0, m_0\}$ for $v \to \mp \infty$  respectively.  While any monotonically increasing interpolating function $m(v)$ will do the trick,  the simplest examples are obtained in the so-called thin shell limit, when the transition is sharp,
\begin{equation}
m(v) = m_0 \, \Theta(v-\ts) \ ,
\label{thinshellm}
\end{equation}	
with $\Theta(x)$ being the Heaviside step-function. 
In this case the shell is localized at constant $v=\ts$, imploding from the boundary $r=\infty$ to the origin $r=0$.
Moreover, we can write the metric in a piecewise-static form,
\begin{equation}
ds_\io^2 = - f_\io(r) \, dt_\io^2 + {dr^2 \over f_\io(r)} + r^2 \, d{\bf \Sigma}_{d-1,K}^2 \ , 
\label{metio}
\end{equation}	
where the subscript $\io$ stand for $i$ inside the shell and $o$ outside, and the shell separating the two spacetime regions is at some radius $r=R_\io(t_\io)$ corresponding to a radial null trajectory.  Although $r$ is continuous across the shell, the time coordinate $t$ is not. 
Though these thin shell geometries will be main focus of our discussion, 
we will start by setting up the construction in the more general spacetimes \req{vaidya}, allowing the deformation to act more smoothly temporally.  This will enable us to easily derive the jump across the thin shell as well as to check our analytical results by numerical computations.

Having specified the bulk geometry, let us now return to the bulk quantities we wish to construct,  $\cwedge$, $\Xi_\CA$, and $\chi_\CA$, for a given boundary region $\CA$.  
It will be useful to start by recalling the general story, to better understand the simplification afforded by \req{vaidya} and the choice of regions we use below. Following  \cite{Hubeny:2012wa}, we define\footnote{
The notation is explained in more detail in \cite{Hubeny:2012wa}. Briefly, by 
$J^\pm$ we mean the causal past and future in the full bulk geometry, whereas $J_{\partial}^\pm$ indicates the causal past and future  restricted to the boundary.
}
 the causal wedge as
\begin{equation}
\cwedge = J^+[\domd] \cap J^-[\domd]
\label{cwedgedef}
\end{equation}	
where the domain of dependence $\domd \in {\cal B}_d$ contains the set of points through which any inextendible causal boundary curve necessarily intersects $\CA$.
Both $\domd$ and $\cwedge$ are defined as causal sets; as such, their boundaries must be null surfaces, generated by null geodesics (within ${\cal B}_d$ and in the $d+1$ dimensional bulk spacetime, respectively), except possibly at a set of measure zero corresponding to the caustics of these generators.  This means that constructing the causal wedge, for any boundary region $\CA$ and in any spacetime, boils down to `merely' finding null geodesics in that spacetime.  The crux of the computation typically lies in delineating where these future and past null congruences intersect.  In practice, though, it is desirable to simplify the problem still further, by considering convenient regions $\CA$ and convenient asymptotically-AdS bulk geometries.

Consider first the construction of $\domd$ within the boundary spacetime ${\cal B}_d$.
Although this background spacetime is simple (e.g.\ spherically symmetric around any point), for generic regions $\CA$, the domain of dependence is as complicated as the shape of $\CA$, as it  terminates at a set of caustic curves, the locus where the generators (namely null geodesics emanating normal to $\partial \CA$) intersect.  However,  for spherically symmetric regions $\CA$, the symmetry of the setup  guarantees that within each (future and past) congruence, all null geodesic generators intersect at a single point.  
  
Hence for any interval in $1+1$ dimensional boundary or for round ball regions in higher-dimensional boundary, the domain of dependence is fully characterized by a pair of boundary points, corresponding to its future tip $\ddtipf$ and past tip $\ddtipp$.
These two points then likewise characterize the full causal wedge $\cwedge$, since definition \req{cwedgedef} merely extends this construction into the bulk,
\begin{equation}
\domd = J_{\partial}^-[\ddtipf] \cap J_{\partial}^+[\ddtipp]
\thuss
\cwedge = J^-[\ddtipf] \cap J^+[\ddtipp] \ .
\label{}
\end{equation}	
Therefore $\bcwedgef \subset \partial J^-[\ddtipf]$ and $\bcwedgep \subset \partial J^+[\ddtipp]$ are generated by null bulk geodesics which terminate at $\ddtipf$ or emanate from $\ddtipp$, respectively.  

In general spacetimes, finding null geodesics amounts to solving a set of coupled 2nd order nonlinear ODEs, so one would typically need to resort to numerics to construct these.
Though the equations simplify substantially for spherically symmetric geometries of the form \req{vaidya} as presented in \sec{s:geodvaidya},  they still retain the form 2nd order coupled nonlinear ODEs.
 However, for piecewise-static and spherically symmetric geometries \req{metio}, there are enough constants of motion to obtain the geodesics by integration; in fact, in the specific cases of interest, we can even obtain analytic expressions for the geodesics.  Since the thin shell renders Vaidya-AdS merely piecewise static, we need to supplement our expressions for the geodesics in each static piece by a `refraction' law for geodesics passing through the shell.  This, however, is easy to derive from the geodesic equations for the global geometry, as we explain in \sec{s:geodvaidya}.
 
The remainder of this section is organized as follows.  
In \sec{s:pgenexp} we focus primarily on spherical regions $\CA$ and thin shell Vaidya-AdS geometries, to motivate our general expectations for the dynamics of the causal wedge and $\chi$.  In \sec{s:geodvaidya} we go on to derive the equations to calculate these constructs explicitly.

\subsection{General expectations for $\chi_{\cal A}$}
\label{s:pgenexp}

Consider a spherical entangling region ${\cal A}$, specified by its radius $a$, located at time $t = t_{\cal A}$.  As indicated above, one may think of $\domd$ as enclosed by inverted light cones over the region ${\cal A}$, so $\domd$ is equivalently specified by its future and past tip, $\ddtipf$ and $\ddtipp$; clearly for Minkowski or ESU boundary geometry, the time at which these tips are located is simply
\begin{equation}
t_\ddtipp = t_{\cal A} - a \ , \qquad t_\ddtipf = t_{\cal A}+ a \ .
\label{}
\end{equation}	
Armed with this data, we are now ready to describe the qualitative behavior of our bulk constructs  $\cwedge$, $\Xi_\CA$, and $\chi_\CA$ in thin shell global Vaidya-AdS geometry.  

\begin{figure}
\begin{center}
\includegraphics[width=6.5in]{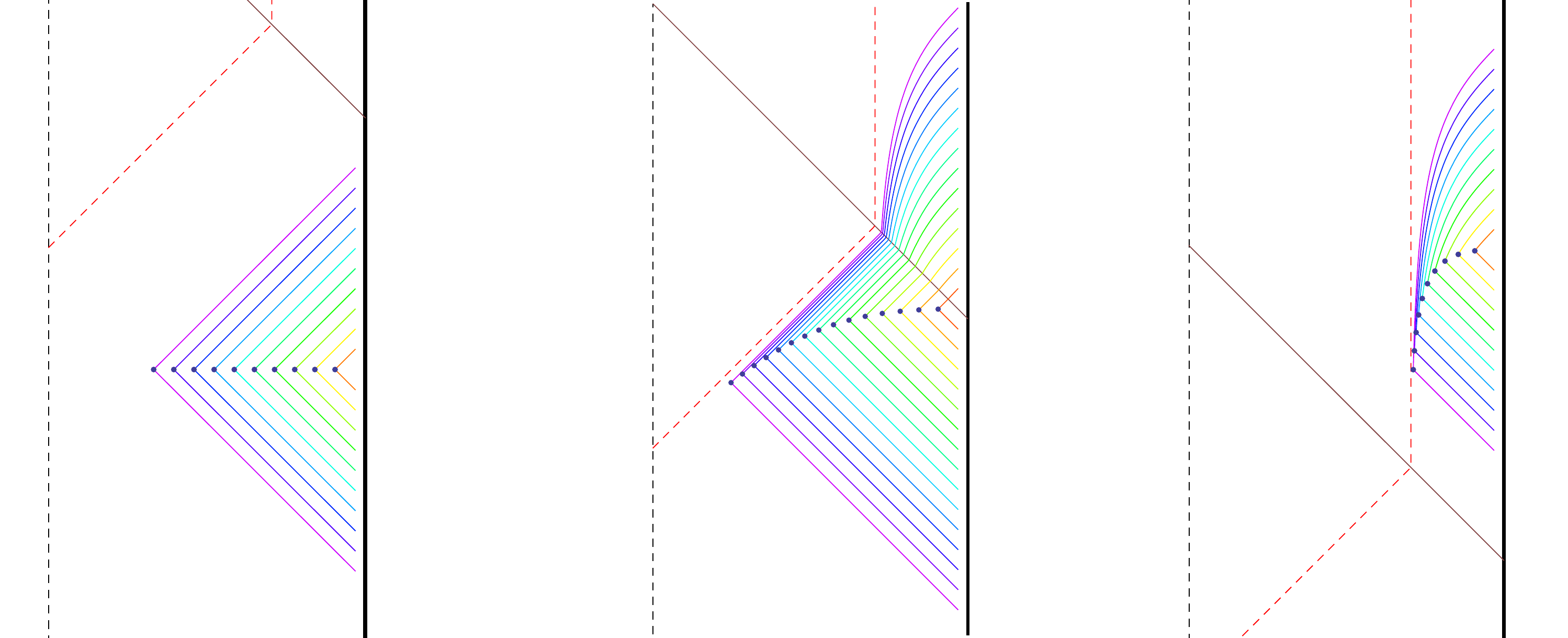}
\caption{
Radial profile of the causal wedge for fixed $\tA = -1.5$ (left), $\tA = 0$ (middle), and $\tA = 1.5$ (right), for  a set of  $\cal A$, color-coded by size $a$.  The thick black curve on right in each panel is the AdS boundary, the dashed black line on left is the origin, the dashed red curve the event horizon (whose final size is $\rh=2$ in AdS units), and the thin brown diagonal line the shell.  The black dots denote the radial position of $\Xi_{\cal A}$ corresponding to the given ${\cal A}$ at time $\tA$ and size $a$.  Our coordinates are such that ingoing radial null geodesics are diagonal everywhere (i.e.\ parallel to the shell).
The plots are made for Vaidya-\AdS{3} spacetime.}
\label{f:CWradprof}
\end{center}
\end{figure}

Let us start by making the following simple observation:  If we choose $\CA$ such that its causal wedge $\cwedge$ lies entirely in the AdS part of the geometry, $\chi_\CA$ will have the same `vacuum' value as in pure AdS.  Similarly, if $\cwedge$ lies entirely in the \SAdS\ (SAdS) part of the spacetime, then $\chi_\CA$ will have the `thermal' value it would have in the corresponding eternal black hole geometry.  The former will be guaranteed if we take $\tA$ to precede $\ts$ by sufficient amount, such that the future tip $\ddtipf$ lies in AdS, namely $t_\ddtipf < \ts$ -- then by causality the rest of $\cwedge$ cannot know about the shell.  
Similarly, for $\cwedge$ to lie entirely outside the shell, in the SAdS part of the spacetime, it suffices that $t_\ddtipp > \ts$, since then the null rays from $\ddtipp$ can never catch up with the null shell and sample the AdS region.  Conversely, if $t_\ddtipp < \ts < t_\ddtipf$, then some part of $\cwedge$ lies in AdS and some part lies in the black hole geometry.  
To illustrate the point, in  \fig{f:CWradprof} we plot the radial profile of the causal wedge for a set of region sizes $a$ for three values of $\tA$:  $\tA<\ts-a$ (left), $\tA=\ts$ (middle), and $\tA>\ts+a$ (right).
%

\begin{table}[tdp]
\begin{center}
\begin{tabular}{|c||c|c|c|c|}
\hline \hline
Regime & time $\tA$ & equivalently & $\bcwedgep$ & $\bcwedgef$ \\
\hline\hline
1 &
$\tA<\ts-a$  \ \ 
	&\ $t_{\ddtipp} < \ts$, \ $t_{\ddtipf} < \ts$ \
	& same as in AdS 
	& same as in AdS 
  \\ \hline
2 &
$\ts-a<\tA<\ts$  
	&\ $t_{\ddtipp} < \ts$, \ $t_{\ddtipf} > \ts$ \
	& same as in AdS 
	& intersects the shell 
  \\ \hline
3 &
\ $\ts<\tA<\ts+a$  
	&\ $t_{\ddtipp} < \ts$, \ $t_{\ddtipf} > \ts$ \
	& intersects the shell  
	& intersects the shell 
  \\ \hline
4 &
\ \ \ \ $\tA>\ts+a$  
	&\ $t_{\ddtipp} > \ts$, \ $t_{\ddtipf} > \ts$ \
	& same as in SAdS 
	& same as in SAdS 
  \\ \hline
\end{tabular}
\end{center}
\caption{Behavior of boundaries of causal wedge $\bcwedgep$ and $\bcwedgef$ depending on $\tA$. (Here \SAdS\ is abbreviated by SAdS.)}
\label{t:tAbreakdown}
\end{table}

To examine this in bit more detail, in Table \ref{t:tAbreakdown} we tabulate how the future and past parts of the causal wedge boundary behave, depending on $\tA$.
If both $\bcwedgep$ and $\bcwedgef$ behave as in AdS, then so does 
$\Xi_\CA$ and $\chi_\CA$.  Similar statement holds for both parts behaving as in SAdS.  However, the intermediate case has a richer behavior:  For $\tA<\ts$, none of the null geodesics starting at $\ddtipp$ can  cross the shell before being intersected by those ending at $\ddtipf$, which means that $\bcwedgep$  still behaves as it would in AdS.  However, despite the fact that $\Xi_\CA$ lies on this surface, it will not be the same curve as in AdS if $t_{\ddtipf}>\ts$, since the other null surface $\bcwedgef$ ends in the SAdS part of the geometry and therefore it no longer behaves as in pure AdS.  So in the regime $\ts-a<\tA<\ts$ indicated in the second line of Table \ref{t:tAbreakdown}, $\Xi_\CA$ lies only within the AdS part of the geometry, but it is nevertheless deformed from the pure AdS behavior.  Since the surface $\Xi_\CA$ is deformed, one would naturally expect that its area $\chi_\CA$ is likewise deformed from the AdS (`vacuum') value.  We will however see later that for the special case of spherical entangling regions $\CA$, this is in fact {\it not} the case; this is one of the surprising revelations of our exploration.

The deformation (from its AdS behavior) of $\Xi_\CA$  will grow as $\tA$ increases from $\ts-a$ to $\ts$, since more and more of $\bcwedgef$ samples the SAdS part of the geometry.  
When $\tA > \ts$,  $\Xi_\CA$ itself can no longer lie entirely within AdS, since its boundary is in SAdS, by virtue of $\partial \Xi_\CA = \partial \CA$.  However, not all of $\Xi_\CA$ can be entirely in SAdS either while $t_{\ddtipp} < \ts$, since the radial null geodesic from $\ddtipp$ must remain to the past of the shell and hence the deepest part of $\Xi_\CA$ remains in AdS until $\tA = \ts +a$, when the thermal regime is entered.  Since the geometry is continuous, we expect $\chi_\CA$ to vary continuously (and in fact monotonically) with $\tA$.

Hence, the expected behavior of $\chi_\CA$, characterized in terms of $\tA$, is:
\begin{align}
&t_{\cal A} < t_s - a && \chi_{\cal A} = \text{vacuum result} \nonumber \\
&t_s - a \leq t_{\cal A} \leq  t_s + a &&  \chi_{\cal A} \ \text{has non-trivial temporal variation} \nonumber \\
&t_{\cal A} > t_s + a && \chi_{\cal A} = \text{thermal result}  
\label{}
\end{align}
This means that by general causality arguments, we expect the following to hold:
\begin{enumerate}
\item The `thermalization' timescale as characterized by $\chi_{\cal A}$ scales linearly with the system size.\footnote{
In other words, here we mean the timescale on which it takes $\chi_\CA$
to achieve its thermal value after the excitation.  This is not necessarily the timescale by which all observables in the field theory would achieve their thermal values; indeed, depending on the diagnostic we use, we may never see true thermalization.}
\item $\chi_{\cal A}$ is mildly teleological; it responds  in advance to the perturbation on a timescale set by light-crossing time of ${\cal A}$.
\end{enumerate}
The fact that $\chi_{\cal A}$ generically responds to the presence of the shell at an earlier time $t_{\cal A}< t_s$ on the boundary follows from the fact our construction involves $\domd$ which samples both the future and past of the boundary region ${\cal A}$.  While ostensibly peculiar, this teleological nature is capped off by the light-crossing time, set by the size of the region.  Hence the teleological nature of $\chi_\CA$ is not as bad as it sounds, since if we imagine measuring any thermodynamic quantity which pertains to the full system, we would need at least this much time anyway.  Of course as the system size goes to infinity (in the planar case), we will see the usual teleological behavior often associated with black hole horizons. 

Both of these timescales (teleology and thermalization) are given simply by $a$, which is not so surprising since it is the only scale characterizing $\CA$.
However, we can in fact also generalize the above statements to any-shaped region $\CA$, with appropriate identification of $a$:  Since the boundary metric is fixed and has a well-defined notion of time $t$, we can define $a$ as the difference between $\tA$ and the earliest time to which $\domd$ reaches (or equivalently, half the timespan of $\domd$).
Such identification provides a natural notion of characteristic size of the region $\CA$, and with this definition, statements 1 and 2 above hold for any simply-connected\footnote{
We will briefly consider non-simply-connected regions in \sec{s:discuss}.
} region $\CA$.
In the following sections we explore these statements in some detail, starting first with the simplest case of $d=2$, where we can carry out most of the constructions analytically.

Before turning to the geodesics which govern our causal constructs, let us make one further remark about the nature of $\chi$.
Above, we have been glibly discussing the `value' of $\chi_\CA$; however, the surface $\csf{\cal A}$ stretches out to the boundary of \AdS{} and hence $\chi_{\cal A}$ is divergent. Moreover, it has already been shown in \cite{Hubeny:2012wa} that the divergence structure of the area of  $\csf{\cal A}$ is generically different from that encountered in the area of the extremal surface $\extr{\cal A}$ relevant for the computation of the holographic entanglement entropy (though in both cases the leading divergence is given by the area law).  Hence it is meaningless to compare $\chi_{\cal A} - S_{\cal A}$ for a given region as a function of time, except in special circumstances (e.g.\ $d=2$). We therefore will most often concentrate on regulated answer obtained by background subtraction, defining
\begin{equation}
\delta \chi_{\cal A} (t) = \chi_{\cal A}(t) -  \chi_{\cal A}^\text{bg} \;; \qquad \chi_{\cal A}^\text{bg} \equiv \lim_{t\to -\infty} \chi_{\cal A}(t)
\label{}
\end{equation}	
and similarly for $\delta S_{\cal A}(t)$.

\subsection{Geodesics in Vaidya-\AdS{d+1} geometry}
\label{s:geodvaidya}

Let us now collect some basic facts about geodesics in the spacetime \eqref{vaidya} that will prove useful in the sequel. 
Since the full $d+1$ dimensional spacetime has spherical (for $K=1$) or planar (for $K=0$) symmetry, one can effectively reduce the problem of finding geodesics to 3 dimensional problem, characterized by $r$, $v$, and $\ph$ (the latter  generates a Killing direction of $\Sigma_{\cal B}$  whose norm defines our radial coordinate $g_{\varphi\varphi} = r^2$).
Then for an affinely-parameterized geodesic congruences with tangent vector
$p^a = \dot{v}  \, \partial_v^a +\dot{r}  \, \partial_r^a + \dot{\ph}  \, \partial_\ph^a$,
 it is convenient to define the `energy' $E$, `angular momentum' $L$, and norm of the tangent vector $\kappa$:
\begin{equation}
\begin{split}
E &\equiv - p_a \, \partial_v^a = f \, \dot{v} - \dot{r} \\
L &\equiv p_a \, \partial_\ph^a = r^2 \, \dot{\ph} \\
\kappa &\equiv p_a \, p^a 
      = - f  \, \dot{v}^2 + 2 \, \dot{v} \, \dot{r} + r^2 \, \dot{\ph}^2
      = - \frac{ E^2 }{ f } + \frac{ \dot{r}^2 }{ f }  + \frac{L^2}{r^2}
\label{ELkappa}
\end{split}
\end{equation}	
where $\dot{} \equiv \frac{d}{d\lambda}$. Note that we are considering full congruences smeared in the directions orthogonal to $\partial_\ph$ in $\Sigma_{\cal B}$ to exploit the symmetry.

For affinely-parameterized null or spacelike or timelike geodesic, $\kappa = 0$ or 1 or -1, respectively; in particular it is a constant of motion.  Since  $\partial_\ph^a$ is a Killing field, $L$ is a conserved along the full geodesic. On the other hand, since $\partial_v^a$ is not a Killing field, $E$ is in general {\it not} conserved.  However, in the thin shell limit it is conserved for each piece of the geodesic (inside the shell and outside the shell) individually, which we will exploit.  In particular, for constant $E$, the geodesic $\{ v(\lambda) \, , \, r(\lambda) \, , \, \ph(\lambda) \}$ can be obtained by integrating \req{ELkappa}.

While the first-order equations \req{ELkappa} are  convenient to use in finding the geodesics analytically, we can't solve them globally using integrals when $E$ is not constant.  The second-order geodesic equations valid for generic $f(r,v)$ are
\begin{equation}
\begin{split}
& \ddot{v} + \frac{1}{2} \, f_{,r} \, \dot{v}^2 - r \, \dot{\ph}^2 = 0 \\
& \ddot{r} + \frac{1}{2} \,
\left(  f \, f_{,r} - f_{,v} \right) \, \dot{v}^2  -  f_{,r} \, \dot{r}\, \dot{v} - r \, f  \, \dot{\ph}^2= 0 \\
& \ddot{\ph} + \frac{2 }{ r} \, \dot{r} \, \dot{\ph} = 0
\label{VaidyaGeodEqs}
\end{split}
\end{equation}	
where we use the shorthand $f_r \equiv \frac{\partial f}{\partial r}(r,v)$ etc..  In order to solve \req{VaidyaGeodEqs} to obtain a specific geodesic, we need to supply two initial conditions for each of the three coordinates.  In terms of the initial position $\{ v_0, r_0, \ph_0 \}$ and the initial velocity, specified by $\kappa$, $L$, and initial energy $E_0$, and also a discrete parameter $\eta=\pm1$ which specifies whether the geodesic is initially ingoing or outgoing, these are
\begin{equation}
\begin{split}
v(0) = v_0 \ , \qquad
	& \dot{v}(0) = \frac{1 }{f(r_0,v_0)} \, \left[  
	E_0 +\eta \, \sqrt{ E_0^2 + f(r_0,v_0) \, \left( \kappa - \frac{L^2}{r_0^2} \right)} \, \right] \ \cr
r(0) = r_0 \ , \qquad
	& \dot{r}(0) = \eta \, \sqrt{ E_0^2 + f(r_0,v_0) \, \left( \kappa - \frac{L^2}{r_0^2} \right)}  \ \cr
\ph(0) = \ph_0 \ , \qquad
	&\dot{\ph}(0) = \frac{L}{r_0^2}
\end{split}
\label{}
\end{equation}	
Usually one can exploit symmetries to set $\ph_0=0$. For any given $f(r,v)$, we can solve these numerically to find any geodesic through the bulk. 

Though the coordinates $\{v,r,\ph\}$ are useful for finding geodesics, they are not the best for visualization since the AdS boundary is at $r=\infty$ and $v$ is a null coordinate; hence on our spacetime diagrams (such as right panel of \fig{f:CWsketch}, \fig{f:CWradprof}, as well as many of the following figures) we present in this paper, 
we plot
$\rho =\arctan r$ 
radially and  $v - \rho + \frac{\pi}{2}$ vertically, so that ingoing radial null geodesics are straight lines at 45$^\circ$.  (Note that except for pure AdS spacetimes, the  time delay which outgoing radial geodesics experience when climbing out of gravitational potential well is manifested by these being  generically steeper than  45$^\circ$ lines.)

\paragraph{Jump across thin shell:}

We now consider the thin shell limit.  Since we can solve \req{ELkappa} by integration in each part of the spacetime where $E$ is constant, all that remains is to account for the jump across the shell. As discussed in \cite{Hubeny:2006yu} (see Appendix F), the jump follows immediately from \eqref{VaidyaGeodEqs}.  With 
$f(r,v)=r^2+K+\Theta(v) \, m_0/r^{d-2}$,  
 while $f$ and $f_{,r}$ remain finite with a discrete jump,  $f_{,v} \sim \delta(v)$ diverges at $v=0$.\footnote{Without loss of generality, we henceforth set $t_s =0$.}  Thus for a geodesic crossing the shell, since the coordinates $\{v(\lambda) ,r(\lambda),\ph(\lambda) \}$ are continuous across the shell, $\{ \dot{v} , \dot{r} , \dot{\ph} \}$ must also remain finite (as is evident from \req{ELkappa}).  This means that from \req{VaidyaGeodEqs}, $\ddot{v}$ and $\ddot{\ph}$ must remain finite as well, which in turn implies that $\dot{v}$ and $\dot{\ph}$ are in fact continuous across the shell.  On the other hand, $\ddot{r}$ has a $\delta(v)$ piece from the  $f_{,v}$ term, so $\dot{r}$ must jump across the shell.  We can easily compute this jump by direct integration, 
\begin{equation}
\ddot{r} \sim \frac{1}{2} \, \mu(r) \, \dot{v}^2 \, \delta(v)
\qquad \Longrightarrow \qquad
\dot{r} = \int \ddot{r} \, d \lambda = \int \frac{\ddot{r}}{\dot{v}} \, dv = \frac{1}{2} \, \mu(r) \, \dot{v} \ ,
\label{}
\end{equation}	
with $\mu(r) = m_0/r^{d-2}$, which means that the jump in $\frac{dr}{dv}$ across the shell is 
\begin{equation}
\frac{dr_i}{dv_i} - \frac{dr_o}{dv_o} = \frac{1}{2} \, (f_i - f_o) = \frac{1}{2} \, \mu(r) \ .
\label{dvdrjump}
\end{equation}	
It is however even simpler to read off the jump in $E$ directly from the fact that 
\begin{equation}
\dot{v} =  \frac{1 }{f} \, \left[  
	E +\eta \, \sqrt{ E^2 + f \, \left( \kappa - \frac{L^2}{r^2} \right)} \, \right]
\label{vdot}
\end{equation}	
is continuous across the shell.  A bit of algebra then gives
\begin{equation}
E_o = \frac{1}{2\, f_i} \left[ (f_i + f_o) \, E_i  -\eta \,
	 (f_i - f_o) \sqrt{E_i^2 + f_i \, \left( \kappa - \frac{L^2}{r^2} \right)}\right] \ ,
\label{EoEidirect}
\end{equation}	
from which we recover
\begin{equation}
E_i - E_o = \frac{1}{2} \,  ( f_i - f_o)  \, \dot{v} \ .
\label{Ejump}
\end{equation}	
We now have all the information required to explore the properties of $\chi_{\cal A}$ in the Vaidya-\AdS{} spacetimes explicitly.

\section{Shell collapse in three dimensions}
\label{s:adsv3}

Having gleaned some general features of $\chi_{\cal A}$ in time dependent geometries, we now turn to the specific example of null shell collapse in \AdS{3}, modeled by \eqref{vaidya} with $d=2$. In this case we take $d{\bf \Sigma}_1^2 \equiv d\varphi^2$ with the spatial circle parametrized by $\varphi \simeq \varphi + 2\pi$, and 
\begin{equation}
f(r,v) = \begin{cases}
	r^2+1, &\text{ for }\; v=t < 0\\
r^2 - \rh^2 ,& \;\,\text{for}\; v=t > 0	
\end{cases}
\label{}
\end{equation}	
so that the spacetime is global \AdS{3} before the insertion of an operator deformation at $\ts=0$ and BTZ with horizon radius $\rh$ afterwards. This could be achieved for instance by homogeneously injecting energy along the spatial circle. The region ${\cal A}$ is then taken to be an arc of length $2\,\varphi_{\cal A}$, without loss of generality lying between $\pm\varphi_{\cal A}$. This region will be taken to lie entirely at constant time $t = t_{\cal A}$ on the boundary. 

The general strategy for finding the causal wedge is as described in \sec{s:prelims}. Since the boundary spacetime is  ESU$_2$ (whose metric is flat), the domain of dependence of ${\cal A}$ is given by 
$\domd = J_{\partial}^-[\ddtipf] \cap J_{\partial}^+[\ddtipp]$, and correspondingly the causal wedge for $\CA$  merely extends this construction into the bulk,
$\cwedge = J^-[\ddtipf] \cap J^+[\ddtipp]$.
Hence to find $\partial_\pm(\cwedge)$, and therefore $\Xi_\CA$, we need to find future-directed null geodesics from $\ddtipp$ and past-directed null geodesics in $\ddtipf$.  These future and past tips lie at 
\begin{equation}
\begin{array}{cccc}
\ddtipf: & \quad t_\wedge = v_\wedge = \tA+\phA \ , \quad &  \ph_\wedge = 0 \ ,  \quad &   r = \infty \ , \\
\ddtipp: & \quad t_\vee = v_\vee = \tA-\phA \ , \quad &  \ph_\vee = 0 \ ,  \quad &   r = \infty \ .
\end{array}
\label{endpts3}
\end{equation}	

The expressions for the null geodesics themselves in the \AdS{3}  part of the geometry are given by the following expressions ($\eta = \pm1$ for outgoing/ingoing respectively and $\ph_\infty = 0 $ w.l.o.g.):
\begin{align}
v(r) &= v_{\infty} - \frac{\pi}{2} \, (\eta+1) 
	+ \eta \, \arctan\left( \sqrt{(1-\ellk^2) \, r^2 - \ellk^2}\right)
	+ \arctan r
\nonumber \\
\ph(r) &=  \eta \, \left[ \arctan\left( \frac{\sqrt{(1-\ellk^2) \, r^2 - \ellk^2}}{\ellk}\right)
	 - \frac{\pi}{2} \, {\rm sign}(\ellk) \right]
\label{vphirAdS}
\end{align}	
with $k = L/E$ for simplicity (using the scaling freedom of the null geodesic affine parameter). Likewise, the BTZ null geodesics are (now with $j= L/E$):
\begin{align}
v(r) &= v_{\infty} + \frac{1}{2 \, \rh} \, \ln \left[ \frac{(r-\rh)}{(r+\rh)}  \,
	 \frac{\left(\sqrt{(1-j^2) \, r^2 + j^2 \,  \rh^2}-\eta \, \rh\right)}
	 {\left(\sqrt{(1-j^2) \, r^2 + j^2 \,  \rh^2}+ \eta \, \rh\right)} \right]
\nonumber \\
\ph(r) &=  \eta \,  \frac{1}{2 \, \rh} \,
	 \ln \left[ \frac{\sqrt{(1-j^2) \, r^2 +j^2 \,  \rh^2}-j \, \rh}
	 {\sqrt{(1-j^2) \, r^2 +j^2 \,  \rh^2}+j\, \rh} \right]
\label{vphrBTZ}
\end{align}	
To keep track of various geodesic congruences, it is useful to adopt suggestive\footnote{
Here we envision the boundary as being on the right, as in \fig{f:CWradprof}.
}  labels:
\begin{itemize}
\item $v_\og(r,\ell)$ and $\ph_\og(r,\ell)$ describe outgoing congruence terminating at $\ddtipf$ at the boundary.
\item $v_\ig(r,\ell)$ and $\ph_\ig(r,\ell)$ describe ingoing congruence starting from $\ddtipp$.
\end{itemize}
In these expressions $\ell$ stands for the (normalized) angular momentum $L/E$ along the given geodesic seqment (which for notational convenience we call $k$ in AdS and $j$ in BTZ); because of the refraction \req{EoEidirect}, the value of $\ell$ will change between $k$ and $j$ as the geodesic passes through the shell.

\subsection{Construction of $\csf{\CA}$}
\label{s:xi3}
Having the explicit expressions for the geodesics at hand, the desired surface $\csf{\CA}$ (which is a curve in $d=2$)  can be found easily.  One obvious quantity of interest is the minimal radial position attained along the curve $\Xi$; we denote this by $\rX$ in what follows. 
It is useful to demarcate our discussion into four different time intervals for the temporal location of $\CA$, corresponding to the four rows of Table \ref{t:tAbreakdown}. We consider these in turn:

\paragraph{1. Vacuum ($\tA < - \phA$):} 
Here $t_\wedge, t_\vee < 0$ implying that the entire causal wedge  and thus $\csf{\cal A}$ is in the \AdS{3} part of the spacetime. Using \eqref{vphirAdS} with the initial points \eqref{endpts3} we chart out the surface of $\partial(\cwedge)$; see right panel of \fig{f:CWsketch} for the actual shape when $\phA= \pi/3$.
The explicit expressions are unnecessary for our purposes (and can be found in \cite{Hubeny:2012wa}). To determine $\rX$, it suffices to consider purely radial geodesics $L=k=0$. Equating $v_\og(r,0)$ with $v_\ig(r,0)$ we immediately find
\begin{equation}
\rX = \cot\phA \ .
\label{rxiads3}
\end{equation}	
Furthermore, one can conveniently characterize $\Xi_\CA$ itself by
\begin{equation}
\sin \rho \, \cos \ph = \cos \phA
\label{XiinAdS}
\end{equation}	
where $\rho \equiv \arctan r$.  On our spacetime plots such as \fig{f:CWsketch}, 
$\Xi_\CA$ would be a horizontal curve at $v - \rho + \frac{\pi}{2}= \tA$.
As discussed in \cite{Hubeny:2012wa}, in pure AdS (and hence in the present ``vacuum" regime of Vaidya-AdS), the causal information surface $\Xi_\CA$ in fact coincides with the extremal surface $\extr{\CA}$; in 3 dimensions this is given by a spacelike geodesic with energy $E=0$ and angular momentum $L= \cot{\phA}$.  In \cite{Hubeny:2012ry} this surface was characterized by 
\begin{equation}
r^2(\ph) = \frac{L^2}{\cos^2 \ph - L^2 \, \sin^2 \ph} \ ,
\label{}
\end{equation}	
which, as can be easily checked, is equivalent to \req{XiinAdS}.

\paragraph{2. Shell encounter by $\bcwedgef$ only: ($-\phA < \tA < 0$):} 
In this time interval, the ingoing congruence which generates $\bcwedgep$, specified by\footnote{
We now distinguish the angular momenta characterizing the top and bottom of the causal wedge $\partial_\pm(\cwedge)$ by subscript $k_\pm$ for AdS  and $j_\pm$ in BTZ.
}  $\{v_\ig(r,k_-) , \ph_\ig(r,k_-)\}$, still lies entirely in the \AdS{3} geometry as explained in \sec{s:pgenexp}, cf.\ Table \ref{t:tAbreakdown}.  On the other hand, since $v_\wedge >0$, the outgoing congruence generating $\bcwedgef$ has segments in both the AdS and the BTZ part of the spacetime. Let us denote the segments in the two regions then as 
$\{v_\og(r,k_+) , \ph_\og(r,k_+)\}$ and $\{v_\og(r,j_+) , \ph_\og(r,j_+)\}$
respectively, accounting now for the fact that the energies in the two spacetimes will differ (while $L$ along an individual geodesic remains constant). 

Starting with the outgoing congruence which terminates at $\ddtipf$,
for each outgoing geodesic, labeled by $j_+$,  we need to find the spacetime point  $p_s = \{v_s=0 \, , \, r_s \, , \, \ph_s\}$ where it hits the shell, as well as how does it refract there, specified by the relation between $j_+$ and $k_+$. Using the fact that the segment 
$\{v_\og(r,j_+) , \phi_\og(r,j_+)\}$
 connects $p_s$ to $q^\wedge$ we find that 
\begin{align}
r_s = \rh \,  \left( \coth (\rh \, v_\wedge ) + \frac{1}{\sqrt{1-j_+^2}} \, 
{\rm csch} (\rh \, v_\wedge ) \right) \,, \qquad 
e^{\rh \, \ph_s} = \frac{e^{\rh\, v_\wedge} \, \sqrt{1-j_+} +  \sqrt{1+j_+} }
{e^{\rh\, v_\wedge}  \, \sqrt{1+j_+} +  \sqrt{1-j_+} } \,.
\label{rphs1}
\end{align}	
With the knowledge of $r_s$, we can then solve the refraction condition \eqref{EoEidirect} with $j_+ = L/E_o$ and $k_+ = L/E_i$ to find that 
\begin{equation}
k_+ = \frac{2 \, j_+ \, r_s \, (r_s^2 - \rh^2)}{r_s \, (2 \, r_s^2 + 1 - \rh^2 ) + (\rh^2 + 1) \, \sqrt{(1-j_+^2) \, r_s^2 + j_+^2 \,  \rh^2}} \ ,
\label{jpkprel}
\end{equation}	
where  $r_s$ itself depends on  $j_+$ as given by \req{rphs1}.

The main distinguishing feature of the time interval under focus is that $k_+(j_+)$
from \eqref{jpkprel} spans the entire range $\pm1$. This in turn implies that 
we can view $k_+$ as the data characterizing the full angular span of $\bcwedgef$, and confirms that $\csf{\CA}$ lies entirely in the \AdS{3} part of the spacetime. 

Having described how the outgoing congruence refracts through the shell, it only remains to find where it intersects with the ingoing congruence emanating from $\ddtipp$.  For each pair of intersecting geodesics, we denote their intersection by $p_\Xicurve = \{v_\Xicurve \, , \, r_\Xicurve \, , \, \ph_\Xicurve \}$, which we can determine by solving
\begin{equation}
 v_\og(r_\Xicurve,k_+) = v_\ig(r_\Xicurve,k_-)
 =v_\Xicurve 
\ , \qquad
 \ph_\og(r_\Xicurve,k_+) = \ph_\ig(r_\Xicurve,k_-)
 =\ph_\Xicurve
\ .
\label{Ximatchingvph}
\end{equation}	
While the expressions themselves are easy to write down and solve explicitly as we describe in \App{s:3dXi}, it is convenient to solve  \eqref{Ximatchingvph}  numerically to find 
$\Xi_\CA$.  
Note that \req{Ximatchingvph} gives a one-parameter family of solutions for $p_\Xicurve$ (with corresponding angular momenta $k_\pm$), which determines $\Xi_{\CA}$.  We can naturally take $\Xi_{\CA}$  to be parameterized by $j_+ \in (-1,1)$, or more conveniently by $r_\Xicurve \in (\rX,\infty)$ (for each half of $\Xi$). 
Since $k_+ =0 $ when $j_+ =0$, we can easily find the minimal radial position $\rX$ attained by $\Xi_{\CA}$ analytically:
\begin{equation}
\rX = \tan\left(\frac{t_{\CA}- \varphi_{\CA}}{2} + \arctan\left[ \rh \, \coth \left(\rh \, \frac{t_{\CA}+ \varphi_{\CA}}{2}\right) \right]\right) \ .
\label{rxiint}
\end{equation}	
Note that in the relevant regime, $\rX$ is a monotonically increasing function of $\tA$ (for fixed $\phA$ and $\rh$).

\begin{figure}
\begin{center}
\includegraphics[width=6in]{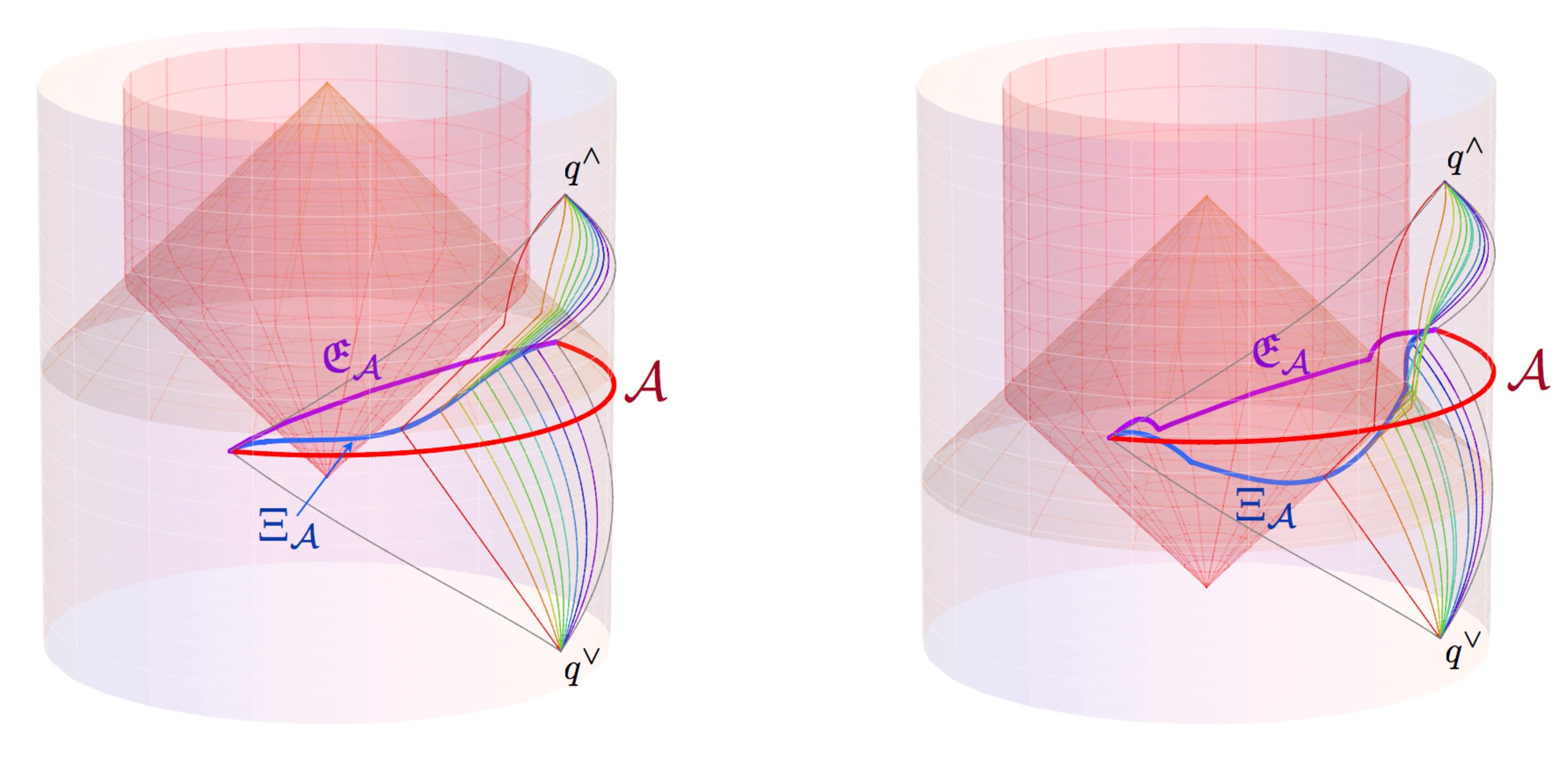}
\caption{
A plot of the causal  information surface  $\Xi_\CA$ (thick blue curve) along with representative generators of $\partial_\pm(\cwedge)$ (thin null curves, color-coded by $r_\Xicurve$), in Regime 2 (left) and 3 (right), as discussed in text.  For orientation we also show the boundary, imploding shell, corresponding event horizon whose final size is $\rh=2$, the region $\CA$ (thick red curve) whose size is $\phA=\frac{2\pi}{5}$ and time $\tA=-0.1$ (left) and $\tA=0.6$ (right), the corresponding domain of dependence $\domd$ (thin grey curves) with its future and past tips $\ddtipf$, $\ddtipp$ as marked, as well as the extremal surface $\extr{\CA}$ (thick purple curve) for comparison.
}
\label{f:CWAB}
\end{center}
\end{figure}

In the left panel of \fig{f:CWAB} we plot $\Xi_\CA$ (thick blue curve) for a region $\CA$ (thick red curve), along with representative generators of $\partial_\pm(\cwedge)$ (thin null curves, color-coded by $r_\Xicurve$),  for $\phA=2\pi/5$ and the final black hole size $\rh=2$. 
(Hence the radial null geodesics drawn as thin red curves are precisely analogous to the curves demarcating the causal wedge profile in \fig{f:CWradprof}.)
For comparison, we also show the extremal surface $\extr{\CA}$ (thick purple curve).
We see that unlike the previous case, in this regime $\Xi_\CA$ is no longer plotted as purely horizontal curve, but rather bends outward and downward - i.e.\ to the past of the constant $t = v - \rho + \frac{\pi}{2}$ surface.   On the other hand, the extremal surface $\extr{\CA}$ remains undeformed since, being anchored at $\tA < t_s=0$, it cannot yet `know' about the shell.
%
\begin{figure}
\begin{center}
\includegraphics[width=6.4in]{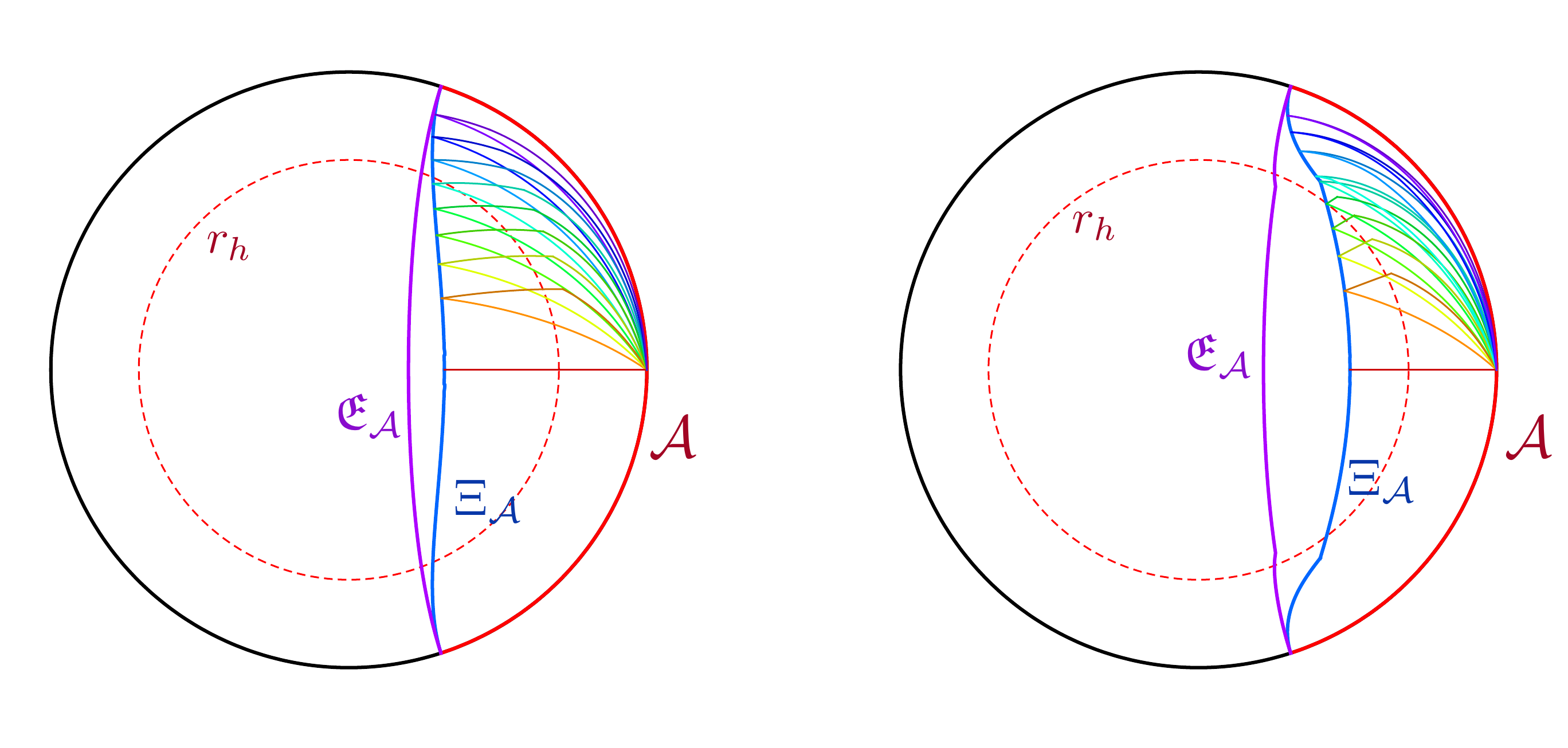}
\caption{
Top view of the same plot as in \fig{f:CWAB} (with the same color-coding scheme), i.e.\ all curves are projected onto the Poincare disk.  For orientation, we also indicate the final black hole size $\rh$ (dashed red curve). 
}
\label{f:PDAB}
\end{center}
\end{figure}
While the downward bend of $\Xi_\CA$ is easy to see in \fig{f:CWAB}, the outward deformation is more apparent from when viewed from a different angle.  To that end, in \fig{f:PDAB} we present the same constructs as in \fig{f:CWAB}, but viewed from top, i.e.\ projected onto a constant time slice.  This projection is known as Poincare disk, where $\rho$ is the radial coordinate and $\ph$ the angular one.  Here it is evident that $\Xi_\CA$ lies closer to the boundary than $\extr{\CA}$.  For orientation we also show the final black hole size $\rh$, even though the generators are not directly dependent on it.  On the other hand, we can see that the generators of $\bcwedgef$ are refracted by the shell (which we don't show since its projection covers the full Poincare disk and each generator intersects it different time and radial position).

\paragraph{3. Shell encounter by $\Xi_\CA$: ($0 < t_{\CA} < \phA$):}
 We now come to the most complicated regime of interest (cf.\ 3rd line of Table \ref{t:tAbreakdown}). For the time interval under consideration, while we still have $v_\vee<0$ and $v_\wedge> 0$, there is a qualitative change in the behavior. This is because $\csf{\cal A}$ itself will intersect the shell, which one can argue for as follows. Along the ingoing congruence $\{v_\ig(r,k_-)  , \phi_\ig(r,k_-)\}$, the radial null geodesic ($k_-=0$) stays at $v= v_\vee$ and thus is parallel to the shell's trajectory and has to remain in AdS. On the other hand, the geodesics with $k_-  \approx \pm1$  lie close to the boundary and these must intersect the congruence from $q^\wedge$ at $v =t_{\CA} >0 $ on the boundary. The only way for this to happen is for the ingoing congruence itself to cross over through the shell and sample regions with both signs of $v$.  
 
Hence in this regime, both the congruences generating  $\bcwedgef$ and  $\bcwedgep$ refract through the shell.  The transition (determined by where $\Xi_\CA$ itself intersects the shell) is given by some critical angular momenta $j^*_+$ and $k^*_-$
demarcating the transfer of refraction from the top boundary of the causal wedge to its bottom boundary.  We have to determine these to find $\csf{\cal A}$. 

The analysis however is straightforward; start with the radial geodesics for which only the refraction of the $j_+ = k_+ = 0$ geodesic matters. This is of course similar to what we encountered in Regime 2 and it hence follows that the minimal radial position attained along $\Xi_\CA$ continues to be given by \eqref{rxiint}. We then increase $j_+$ and follow $\Xi$ along its path through the \AdS{} region as before. At the same time we monitor the ingoing geodesics along  $\bcwedgep$ and ask when they hit the shell. This happens for 
\begin{align}
r_s =\cot v_{\vee}  - \frac{1}{\sqrt{1-k_-^2}} \, 
\csc v_{\vee} \,, \qquad 
 \ph_s =\frac{\pi}{2} \, {\rm sign}(k_-) +  \arctan \left( \frac{ \cos v_\vee - \sqrt{1-k_-^2}}{k_- \, \sin v_\vee } \right) \,.
 \label{rphs2}
\end{align}	
The critical angular momenta $j_+^\ast$ and $k_-^\ast$ at which $\Xi_\CA$ crosses the shell is then obtained by equating $r_s$ and $\phs$ in \eqref{rphs2} with the corresponding result in the black hole part \eqref{rphs1}.
Denoting the spacetime point where these critical geodesics with  $j_+^\ast$ and $k_-^\ast$ intersect (which is simultaneously the point where $\Xi_\CA$ intersects the shell) by $p_X = \{v_X \, , \, r_X \, , \, \ph_X \}$, 
we have $v_X=0$, $r_X \equiv r_s(j_+^\ast) = r_s (k_-^\ast)$ and $\ph_X \equiv \ph_s(j_+^\ast) = \ph_s(k_-^\ast)$. 

The strategy for finding $\Xi_\CA$ then is similar to what was employed in \req{Ximatchingvph}. For $| j_+| < |j_+^\ast|$ or equivalently for $\rX \leq r \leq r_X$, the previous analysis carries over unchanged. For larger values of angular momenta ($r > r_X$), we must first account for the refraction of the  ingoing congruence from $q^\vee$ through the shell, by employing the relation between $k_-$ and $j_-$, analogous to \eqref{jpkprel} and obtained from the same refraction condition \eqref{EoEidirect}, now with $\eta = -1$, $j_- = L/E_o$ and $k_- = L/E_i$:

\begin{equation}
j_- = \frac{2 \, k_- \, r_s \, (r_s^2 +1)}{r_s \, (2 \, r_s^2 + 1 - \rh^2 ) + (\rh^2 + 1) \, \sqrt{(1-k_-^2) \, r_s^2 - k_-^2}} \ .
\label{jpkprelpast}
\end{equation}	
 The analog of \req{Ximatchingvph}  which we need is simply obtained by replacing $k_- \to j_-$ and $k_+ \to j_+$  since the intersection happens in the BTZ part of the spacetime now. 

In the right panel of \fig{f:CWAB} we plot $\Xi_\CA$ along with representative generators of $\partial_\pm(\cwedge)$ for this regime, as well as the extremal surface $\extr{\CA}$ for comparison.
We can see that $\Xi_\CA$ now deforms to an even larger extent than in Regime 2 (cf.\ the left panel), being pushed further outward and downward, as well as kinked by the shell.
The behavior of the extremal surface  $\extr{\CA}$ is likewise more complicated   than in the previous two cases (whether or not $\extr{\CA}$ crosses the shell depends on the interplay of $\tA$ and $\phA$; in the present case it does), and its detailed structure will be presented elsewhere \cite{Hubeny:2013uq}.  However, we can make the general statement that $\extr{\CA}$  does not coincide with $\Xi_\CA$, and reaches deeper into the bulk, as characterized by the minimum  radius attained along $\extr{\CA}$, $\rE<\rX$.  
We will revisit this point in the Discussion.

\paragraph{4. Thermal ($\tA > \phA$):}
Finally, let us consider the regime $v_\wedge, v_\vee >0$, so the entire causal wedge is in the BTZ part of the geometry.   As demonstrated in \cite{Hubeny:2012wa}, the causal information surface $\Xi_\CA$ now again coincides with the extremal surface $\extr{\CA}$; both are deformed outward and downward by the presence of the black hole, such that $\rX=\rE>\rh$.
  By similar arguments as for regime 1, we now find the minimal radius reached to be\footnote{ Note that we have denote  the minimal radius attained by $\csf{\cal A}$ and $\extr{\cal A}$ in BTZ spacetimes as  $\rxi$ for future convenience.}
 \begin{equation}
\rX = \rh \, \coth\left(\rh \, \varphi_{\cal A}\right)  \equiv \rxi \,.
 \label{rxibtz}
 \end{equation}	
The static case expressions \req{rxiads3} and \req{rxibtz} are in fact the special limits of \req{rxiint} as $\tA \to \pm \phA$, respectively.  As remarked above, $\rX$ increases monotonically with $\tA$, so in particular the thermal result \req{rxibtz} is larger than the vacuum result \req{rxiads3}.

\begin{figure}
\begin{center}
\includegraphics[width=2.5in]{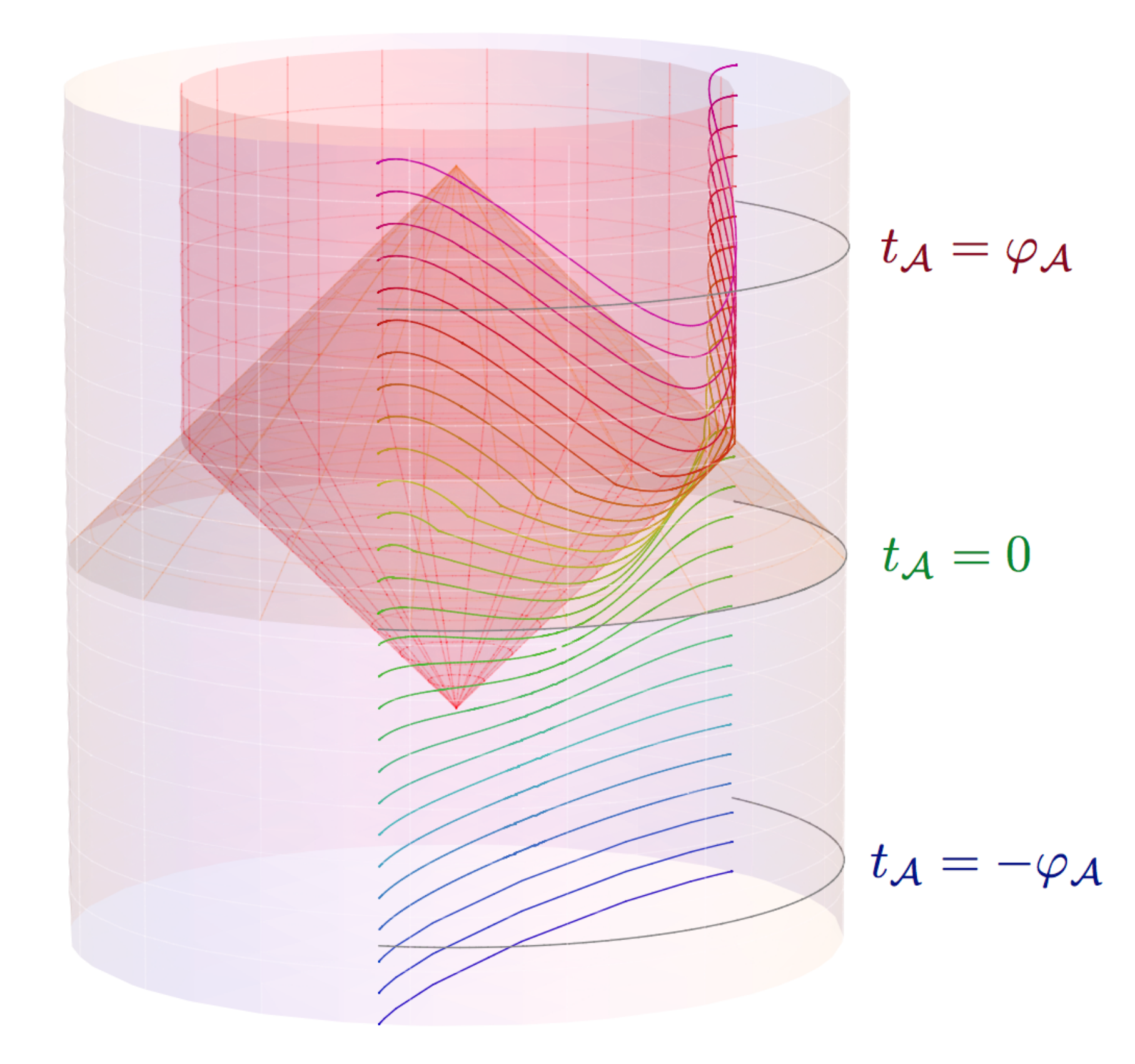}
\hspace{1cm}
\includegraphics[width=2.5in]{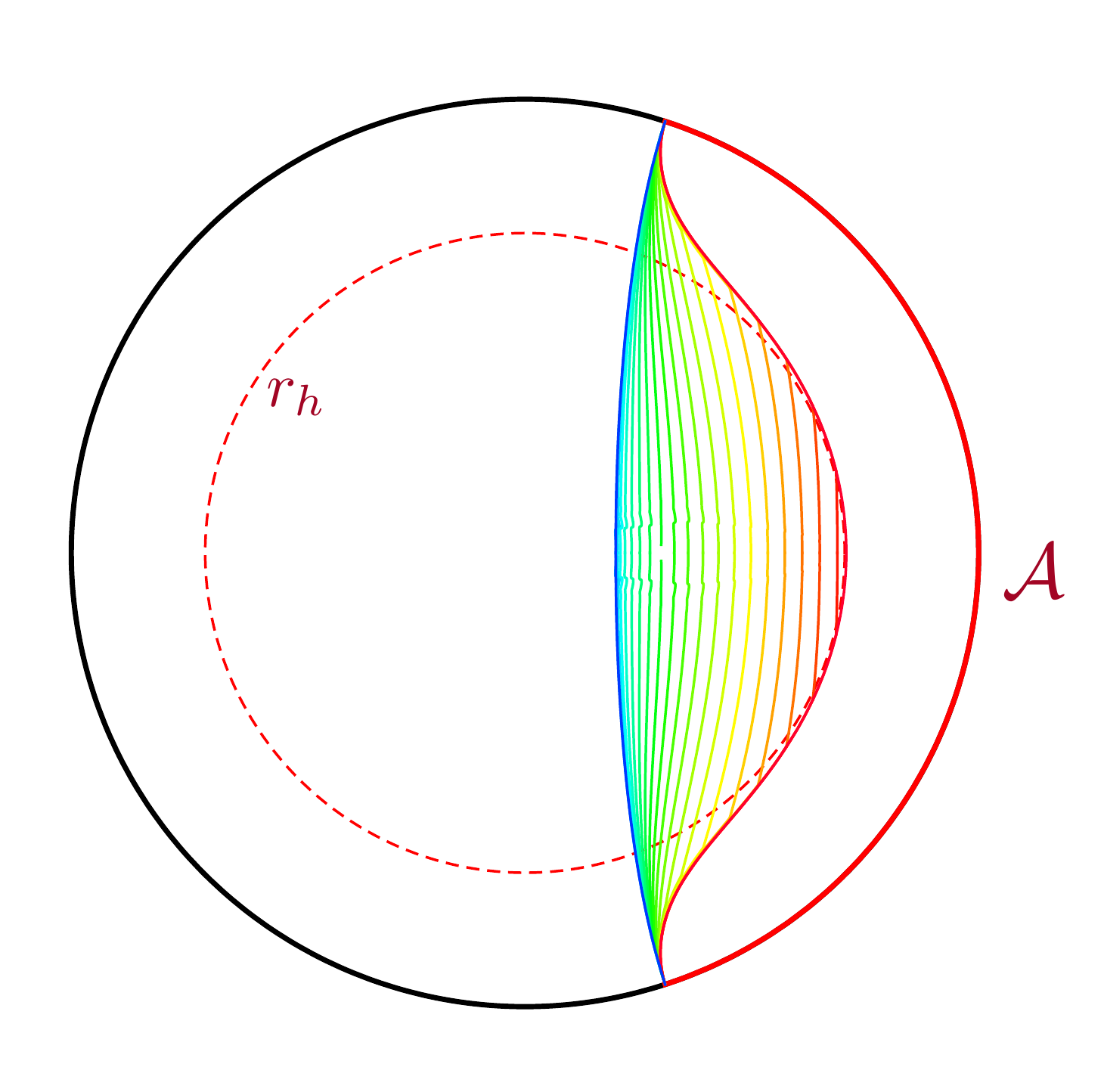}
\caption{
(Left): The curves  $\Xi_\CA$ (color-coded by $\tA$) for a range of $\tA$ sampling across the 4 regimes (separated by the three transitions at $\tA=-\phA , 0, \phA$ as labeled; the thin gray curves represent $\CA$ at those times) in increments of $0.1 \phA$, for $\phA=2\pi/5$ and  $\rh=2$.
(Right):  Same curves  $\Xi_\CA$ projected onto the Poincare disk, analogous to \fig{f:PDAB}.
}
\label{f:Xievol}
\end{center}
\end{figure}

Now that we have covered all 4 qualitatively distinct regimes, we summarize our results.
In the left panel of \fig{f:Xievol} we plot the actual surfaces $\Xi_\CA$ (now color-coded by $\tA$) as $\tA$ varies across the 4 regimes, again for a fixed value of $\phA=2\pi/5$ and the final black hole size $\rh=2$ (so the thick blue curves $\Xi_{\CA}$ in \fig{f:CWAB} are specific examples of these).  We present the same curves $\Xi_\CA$ both on a spacetime plot (left) as well as its projection onto the Poincare disk (right).  We can clearly see how the surfaces deform outward and downward so as to remain outside the event horizon.  The 4 regimes are demarcated by the regions $\CA$ for $\tA=-\phA , 0, \phA$ as labeled in the left panel, and we can see that in regimes 1 and 4 the shape of $\Xi_\CA$ remains the same, while in regimes 2 and 3 the shape of $\Xi_\CA$ changes with $\tA$ as expected.  The qualitative difference between the latter two regimes can be seen if we shift all $\Xi$'s such that they are anchored at the same position on the boundary.  Then one can confirm that in regime 2, all 
$\Xi$'s lie on the same null surface, while in regime 3 they don't.

%

\begin{figure}
\begin{center}
\includegraphics[width=4.in]{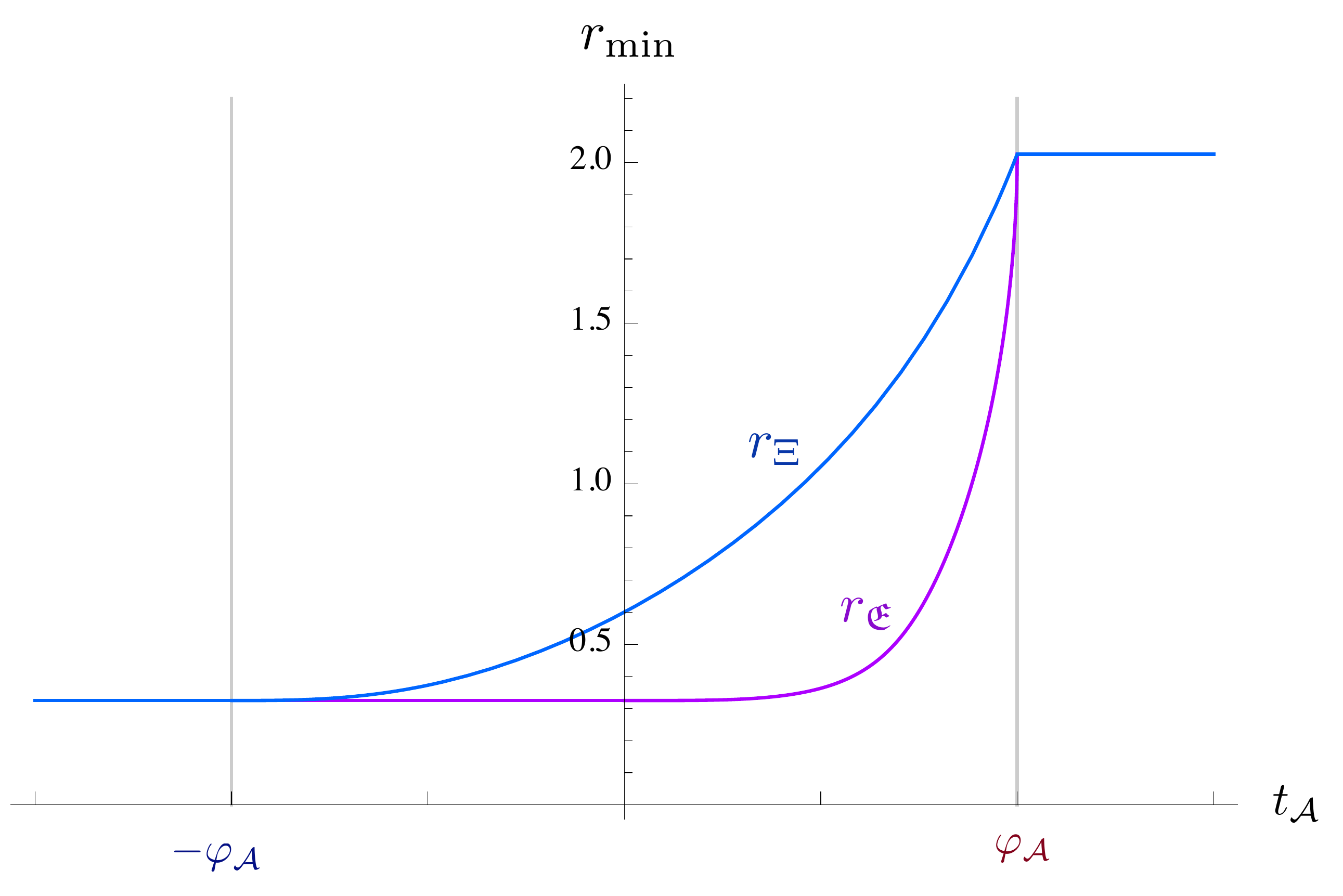}
\caption{
Comparison of minimum radii $\rX$ (blue curve) and $\rE$ (purple curve) attained by the causal information surface $\Xi_\CA$ and the extremal surface $\extr{\CA}$, respectively, as a function of $\tA$,  for the same parameters as in \fig{f:Xievol}, namely $\phA=2\pi/5$ and  $\rh=2$.  
The regimes 1,2,3,4 are again demarcated by $\tA= -\phA,0,\phA$.
}
\label{f:rminXEcomp}
\end{center}
\end{figure}
To characterize the change in $\Xi_\CA$ under variations of $\tA$, it is better to concentrate on one salient feature of $\Xi_\CA$ rather than its entire shape.  One such handy quantity is the bulk depth to which $\Xi_\CA$ penetrates.
In \fig{f:rminXEcomp} we plot the  minimum radius $\rX$ (blue curve) and $\rE$ (purple curve) attained by the causal information surface $\Xi_\CA$ and the extremal surface $\extr{\CA}$, respectively, as  $\tA$ varies across the 4 different regimes discussed above, again for $\phA=2\pi/5$ and final black hole size $\rh=2$. We clearly see that the expectations explained in \sec{s:pgenexp} pan out:  $\rX$ coincides with $\rE$ in regimes 1 (AdS) and 4 (BTZ) and differs in regimes 2 and 3 (shell encounter); in particular $\rX>\rE$ (i.e.\ doesn't penetrate as deep into the bulk) in the latter cases.  Moreover, in regime 2 ($\tA<t_s=0$), while $\rE$ remains at its AdS value by causality, $\rX$ already starts to vary, illustrating the quasi-teleological behavior of $\cwedge$.

\subsection{Determining $\chi_{\CA}$ and comparison with $S_\CA$}
\label{s:chi3}

Now that we have analysed how the causal wedge $\cwedge$ and the corresponding causal information surface $\Xi_{\CA}$ `evolves' during a thin shell collapse, let us turn to its proper area, the causal holographic information $\chi_\CA$.  In particular, we would like to compare the regulated value of $\chi_\CA$ with the regulated entanglement entropy $S_\CA$.  One might naively expect that $\chi_\CA$ and $S_\CA$ would evolve with $\tA$ in a manner which is qualitatively analogous to that of $\rX$ and $\rE$ plotted in \fig{f:rminXEcomp}.  Indeed, we find that in  regimes 1 and 4 (AdS and BTZ), $\chi_\CA$ and $S_\CA$ must coincide, since the actual surfaces whose area these quantities measure also coincide.

In particular, as can be verified by explicit computation, in the AdS (`vacuum') case,\footnote{The expressions are written in terms of the field theory central charge $c_\text{eff}$ which is related to the gravitational Newton's constant via the standard Brown-Henneaux result $c_\text{eff} = \frac{3 \, R_{AdS}}{2\, G_N^{(3)}}$. \label{fn:centralcharge}}
\begin{equation}
{\rm Regime \ 1:} \qquad \qquad
\chi_{\CA}= S_{\CA} = \frac{c_\text{eff}}{3}\; \log[2 \, r_\infty\,  \sin \phA ] \equiv {\mathscr S}_\text{AdS}(\phA)
\label{chivac}
\end{equation}	
while in the BTZ (`thermal') case, 
 \begin{equation}
{\rm Regime \ 4:} \qquad \qquad
\chi_{\CA}= S_{\CA} = \frac{c_\text{eff}}{3} \; \log\left[\frac{2 \,r_\infty}{ \rh}\, \sinh\left(\rh \, \phA \right) \right] \equiv {\mathscr S}_\text{BTZ}(\phA,\rh)
 \label{chitherm}
 \end{equation}	
where $r_\infty$ is the radial UV cut-off to regulate the standard divergence encountered in the expressions. We have also introduced new definitions for the values of the $\chi$ and $S$ in the AdS and BTZ geometries respectively for future convenience.

In the intermediate regime (Regimes 2 and 3) where the causal wedge encounters the shell and $\Xi_\CA$ no longer coincides with $\extr{\CA}$, we can compute $\chi_\CA$ numerically.
(If fact, in the present case we can also use a trick, explained in \sec{s:trick}, to obtain $\chi_\CA$ almost analytically.)
Using the more obvious numerical method, we integrate the length element induced from \eqref{vaidya} onto the curve $\Xi$.
This boils down to integrating the proper length in \AdS{} for $\rX \leq r \leq r_X$ (which for Regime 2 is $r_X=\infty$ so this gives the full answer) and using the BTZ metric for $r\geq r_X$. We regulate the result  by integrating the length element up to $ r = r_\infty$; since at the end of the day  we are going to use background subtraction, all we need to do is to ensure that  we pick the same UV regulator for the AdS spacetime.

\begin{figure}
\begin{center}
\includegraphics[width=6.5in]{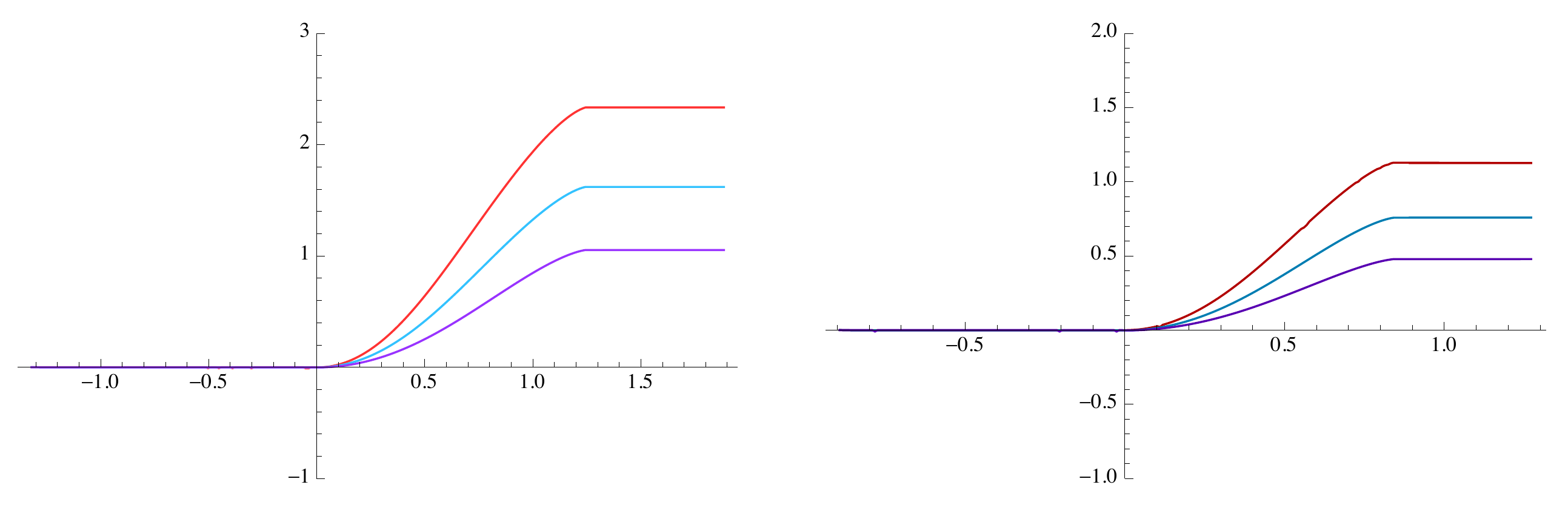}
\begin{picture}(0,0)
\setlength{\unitlength}{1cm}
\put(-0.3,1.6) {{\scriptsize $t_{\cal A}$}}
\put(-9,1.2) {{\scriptsize $t_{\cal A}$}}
\put(-5,5.3) {{\scriptsize $\delta\chi_{\cal A}$}}
\put(-5,5.3) {{\scriptsize $\delta\chi_{\cal A}$}}
\put(-13.6,5.4) {{\scriptsize $\delta\chi_{\cal A}$}}
\put (-10.3,4.45) {{\scriptsize $\color{red}{\rh = 2.0}$}}
\put (-10.3,3.62) {{\scriptsize $\color{hue55}{\rh = 1.5}$}}
\put (-10.3,2.93) {{\scriptsize $\color{hue75}{\rh = 1.0}$}}
\put (-1.8,3.84) {{\scriptsize $\color{red}{\rh = 2.0}$}}
\put (-1.8,3.25) {{\scriptsize $\color{hue55}{\rh = 1.5}$}}
\put (-1.8,2.8) {{\scriptsize $\color{hue75}{\rh = 1.0}$}}
\put (-14,-0.2) {${\phA = 1.256}$}
\put (-5.5,-0.2) {${\phA = 0.85}$}
\end{picture}
\caption{The variation of $\chi_\CA$ with time $\tA$. We plot the absolute value of $\chi_{\cal A}$ 
evaluated with a radial cut-off $r_\infty = 100$ in the Vaidya-\AdS{3} spacetime. The shell implodes from the boundary at $t_s = 0$.
The plots are shown for two different region sizes indicated above and for different final black hole size for each choice of $\phA$. }
\label{f:chioftA}
\end{center}
\end{figure}

In \fig{f:chioftA} we plot the behavior of $\chi_\CA$ for two different values of $\phA$. 
While we see that $\chi_\CA$ indeed  monotonically interpolates between the AdS value and the BTZ value, we encounter a surprise:  {\it $\chi_\CA$ behaves causally:} it does not start to grow till the shell encounter!  In other words, $\chi_{\cal A} = S_{\cal A}$ even in Regime 2, {\it despite the fact that the surfaces $\Xi_\CA$ and $\extr{\CA}$ differ}.

The reason that $\chi_\CA$ remains at the AdS value in Regime 2, and only starts to vary in Regime 3, is the following.  As explained in \sec{s:pgenexp} (cf.\ Table \ref{t:tAbreakdown}), $\bcwedgep$ lies entirely in AdS, so $\Xi_{\CA}$ lies on the same null surface $\partial J^+[\ddtipp]$ as the extremal surface $\extr{\CA}$ (the latter lying on $\partial J^+[\ddtipp]$ by virtue of coinciding with the causal holographic surface in pure AdS).  In particular, past-directed outgoing null geodesics emanating in a normal direction to $\extr{\CA}$ thus generate $\partial J^+[\ddtipp]$, with $\Xi_\CA$ lying along these generators between $\extr{\CA}$ and $\ddtipp$.

Since $\partial J^+[\ddtipp]$ is a boundary of a causal set, its  generators must be null geodesics which reach the boundary at $\ddtipp$ without encountering caustics along the way.  By Raychaudhuri's equation, this in turn implies that these generators cannot contract towards the boundary, i.e.\ that their expansion along past-directed (outgoing) direction must be non-negative, but cannot increase.
On the other hand, as shown in \cite{Hubeny:2007xt}, the extremal surface is precisely the one with null normal congruences (both ingoing and outgoing ones, or both future and past-directed ones) having zero expansion.  Since the generators of $\partial J^+[\ddtipp]$ start out at $\extr{\CA}$ with zero expansion towards the boundary, they have to maintain zero expansion all along the entire $\partial J^+[\ddtipp]$, which can be also checked by explicit calculation (as we do in \sec{s:adsvd}).

Having established zero expansion along null generators of $\partial J^+[\ddtipp]$, the final ingredient in our argument is translating this into comparison of areas of $\extr{\CA}$ and $\Xi_{\CA}$.  Since the expansion $\Theta $ is the differential increase in area along the `wavefront' of these generators, $\Theta = \frac{d}{d\lambda} \log A(\lambda)$,
if $\Theta(\lambda)=0$, then the `wavefront' area $A(\lambda)$ stays constant.
Furthermore, since we can think of $\extr{\CA}$ as lying at $\lambda =\lambda_{\extr{}}$ and $\Xi_{\CA}$ as lying at  $\lambda= \lambda_\Xi > \lambda_{\extr{}}$ (using, if necessary, the freedom of overall rescaling of affine parameter along a null geodesic, and noting that at the boundary, finite $\lambda$ flow degenerates to a point, so that all constant $\lambda$ wavefronts remain pinned to $\partial \CA$), the increase of area between $\extr{\CA}$ and $\Xi_{\CA}$ must vanish, i.e.\
$\chi_{\cal A}= S_{\cal A}$.

Note that the above argument crucially relied on the fact that $\extr{\CA} \in \partial J_+[\ddtipp]$.  This situation is general for $d=2$ where our region $\CA$ is just an interval, since then $\domd$ is specified by the two points $\ddtipp$ and $\ddtipf$ for any $\CA$.  On the other hand, as pointed out in \sec{s:prelims}, this
 clearly does not hold in higher dimensions for generic shapes of $\CA$.  
It is only for special (round) regions $\CA$ that $\extr{\CA}$ coincides with $\Xi_\CA$ in AdS and hence can lie on $\partial J_+[\ddtipp]$.
For generic (non-round) $\CA$, $\extr{A}$ does not lie on $\partial J_+[\ddtipp]$, so we do not have a handy curve on $\partial J_+[\ddtipp]$ on which we are guaranteed to have zero expansion.
On the other hand, this lack of proof does not necessarily imply inequality between $S_{\CA}$ and $\chi_\CA$ a-priori.
To see whether $\chi_\CA$ does behave teleologically as expected, or whether it still maintains causality for a more subtle reason, in \sec{s:adsvd} we examine these quantities explicitly in higher-dimensional thin-shell Vaidya-AdS for both round and non-round regions.

\subsubsection{Trick to evaluate $\chi_{\cal A}(\tA)$}
\label{s:trick}

Above we have argued that in the thin shell Vaidya-AdS$_3$ spacetime, $\chi_\CA$ behaves causally, in the sense that it stays at the AdS value for all $\tA \le 0$, i.e.\ up till the appearance of the shell.  However, it is also clear that between $\tA = 0 $ and $\tA=\phA$ (when $\chi_\CA$ saturates to the BTZ value ${\mathscr S}_\text{BTZ}$), there is a non-trivial variation in $\chi(\tA)$ which we evaluated numerically; cf.\ \fig{f:chioftA}. The numerical computation follows the logic outlined earlier to find $\csf{\cal A}$ and then evaluating its area (further details are presented in \App{s:3dXi}).

In the present special case of thin shell Vaidya-\AdS{3} we however can exploit a trick to give a simple compact expression for $\chi_{\cal A}$ which only uses the critical angular momenta  $j_+^\ast$ and $k_-^\ast$ discussed above \req{rphs2} and the forms of null and (zero-energy) spacelike geodesics in pure AdS and BTZ. While the expressions for $j_+^\ast$ and $k_-^\ast$ require solving transcendental equations, we know analytically the expressions for null and spacelike geodesics in \AdS{3} and BTZ spacetimes, which suffices to bring the expression for $\chi_{\cal A}$ into a convenient compact form.

The basic idea is simply an extension of the one used to argue that $\chi(\tA<0) = \chi(\tA<-\phA)$, now applied to light cones in both AdS and BTZ.
Fix $\tA \in (0 , \phA)$, and consider the curve $\Xi_{\cal A}\equiv \Xi$ (we drop the subscript ${\cal A}$ for the time being). This is composed of a central piece which resides in AdS and the edge pieces which reside in BTZ and these intersect at $p_X = \{ v=0, r_X, \, \pm \ph_X \}$.  In fact, since everything is reflection-symmetric around $\ph=0$, it will be convenient to deal with only one side (say for positive $\ph$); we'll denote the respective parts of one half of the curve by 
$\Xi_\text{AdS}$ and  $\Xi_\text{BTZ}$ respectively, and  correspondingly their proper lengths by $ {\cal L}_\text{AdS}$ and $ {\cal L}_\text{BTZ}$, respectively.
The total length of $\Xi$ then determines $\chi_{\cal A} \propto 2 \left( {\cal L}_\text{AdS} + {\cal L}_\text{BTZ}\right)$.  

Since the segments $\Xi_\text{AdS}$ and $\Xi_\text{BTZ}$ lie in different geometries, it is useful to 
tackle them separately. A-priori to compute the two contributions to $\chi$ we would be satisfied with any mechanism for computing the respective lengths without actually knowing the form of the curves themselves. While this is usually tricky, in the present case we can map the computation of the lengths ${\cal L}_\text{AdS}$ and $ {\cal L}_\text{BTZ}$, to computations to lengths of two other known curves in the AdS and BTZ spacetime.

Imagine cutting (half of) $\Xi$ at the intersection with the shell into its two segments at $p_X$. Since $\Xi$ lies on the boundary of the causal wedge, it follows that $\Xi_\text{AdS}$ lies on $\partial J^+[\ddtipp]$; similarly $\Xi_\text{BTZ}$ lies on $\partial J^-[\ddtipf]$. We are going to slide the two segments along these light cones to a point where we encounter some known curves whose length is easy to compute. 

First we however have to understand why we are free to slide the curve along the light cones. Consider the AdS part of $\Xi$: by construction, $\Xi_\text{AdS}$ lies in the AdS part of the geometry, on the light cone 
$\bcwedgep \in \partial J^+[\ddtipp]$, whose null generators have zero expansion (as argued above).  
This means that the length of $\Xi_A$ is the same as the length of any other curve on $\partial J^+[\ddtipp]$ which traverses the same set of generators, namely  those null generators of $\partial J^+[\ddtipp]$ with sub-critical angular momentum $k \le k_-^\ast$ (we take by convention $k_-^\ast >0$ on $\Xi_\text{AdS}$ without loss of generality). So as long as we slide $\Xi_\text{AdS}$ up by the same affine time along the generators of $\partial J^+[\ddtipp]$ with the ends on the generator $k_-^\ast$ we won't change its length. 

Among curves that lie on this light cone in AdS, a particularly convenient one is the zero-energy spacelike geodesic in pure AdS, $\extr{\CA}$. We know it coincides with $\Xi$ in AdS and therefore lies on the future light cone from $\ddtipp$, and moreover its length is given simply by its affine parameter which is easy to evaluate. 
Such spacelike geodesic $\extr{\CA}$ would lie at constant $t$ in AdS, and in particular encounters the critical null generator (i.e.\ one from $\ddtipp$ with angular momentum $k_-^\ast$) at $t= \tA$. So ${\cal L}_\text{AdS}$ is given by the affine parameter $\lambda(r)$ of a spacelike $E=0$ geodesic in AdS anchored on $\partial {\cal A}$ (i.e.\ with angular momentum $L=r_{\extr{}}=\cot \phA$), evaluated at the value of $r$ at which this geodesic intersects the critical null generator with $k=k_-^\ast$.

Using explicit expressions for the null geodesics in AdS \eqref{vphirAdS}, we learn that the element of the ingoing congruence with angular momentum $k_-^\ast$ starting from $\ddtipp$ makes it to $t=\tA$ at a radial position $r_\ast^2 = \frac{r_{\extr{}}^2+ (k_-^\ast)^2}{1- (k_-^\ast)^2}$. Then it is a simple matter to integrate spacelike geodesic $r(\lambda)$ to infer $\lambda(r_*)$. We use $\frac{dr}{d\lambda} = \sqrt{(r^2+1)\, (r^2-r_{\extr{}}^2)}$ for a $E=0$ spacelike geodesic with $L$ replaced by the minimal radial position attained. The integral is simple to evaluate and we obtain
\begin{equation}
{\cal L}_\text{AdS} = \frac{1}{2} \, \log \frac{1+k_-^\ast}{1-k_-^\ast} \ .
\label{halfchiA}
\end{equation}	
Note that \req{halfchiA} is independent of $\phA$, which it has to be by scaling invariance of AdS.  Also note that for $\tA\le0$, we have $k_-^\ast=1$, so when the entire $\Xi$ lies in AdS, we recover the usual divergence in its length.

Let us now turn to the outer piece of $\Xi$, namely $\Xi_\text{BTZ}$.  This part of $\Xi$ lies entirely in the BTZ part of the geometry, and in particular on the past light cone 
$\bcwedgef \in \partial J^-[\ddtipf]$. We again slide this down to a convenient position staying on this light cone; the main difference is that we are interested only in the segment of the light cone generated by null geodesics with $-1<j_+ < j_+^\ast$ with $j_+^\ast$ indicating being the anchor point of our slide (having chosen positive $k_-$ we now need to choose negative $j_+$).

 Since in pure BTZ, an extremal surface $\extr{\CA}$ also coincides with the causal information surface $\Xi_\CA$, the generators of $\partial J^-[\ddtipf]$ must have zero expansion everywhere in BTZ, by the same type of argument as for the AdS light cones: $\extr{}$ forces the generators to start with zero expansion, and Raychaudhuri equation ensures that the subsequent expansion does not grow and does not become negative -- i.e.\ it has to stay zero.
So it then follows that the length ${\cal L}_\text{BTZ}$ of $\Xi_\text{BTZ}$ is the same as the length of any other curve on $\partial J^-[\ddtipf]$ which traverses the same generators, in this case characterized by super-critical angular momentum, $j_+ < j_+^\ast $.   

The calculation then proceeds exactly as above; we can pick the spacelike $E=0$ geodesic in pure BTZ geometry ending at $t =\tA$, and find where it intersects with the null generator of the past light cone from $\ddtipf$ with angular momentum $j_+^\ast$. Using \eqref{vphrBTZ} for the explicit form of the null geodesics in BTZ we  find that $r_*^2 = (r_\xi^2 - \rh^2\, (j_+^\ast)^2)/(1-(j_+^\ast)^2)$.\footnote{Note that since we are moving the segments of the curves $\Xi_\text{AdS}$ and $\Xi_\text{BTZ}$ the radial positions $r_\ast$ in AdS and BTZ are unrelated.}  Integrating the expression for the spacelike geodesic with $E=0$ and $L = r_\xi$ (again set by the minimal radius attained) which takes the simple form $\frac{dr}{d\lambda} = \sqrt{(r^2 -\rh^2)\, ( r^2 -\rxi^2 )}$ in the BTZ spacetime,  between $r_*$ and the radial cut-off $r_\infty$, we learn that 
\begin{equation}
{\cal L}_\text{BTZ}
	= \log\left(\frac{2\,r_\infty}{\rh} \, \sinh(\rh\,\phA\right) - \frac{1}{2} \, \log \frac{1-j_+^\ast}{1+ j_+^\ast} \ .
\label{halfchiB}
\end{equation}	

From these two simple expressions it follows that in the regime $0 < t_{\cal A} < \phA $ we have a compact expression for $\chi_{\cal A}$
\begin{equation}
\chi_{\cal A}= {\mathscr S}_\text{BTZ}  + \frac{c_\text{eff}}{6} \log \left(\frac{1+k_-^\ast}{1-k_-^\ast} \; \frac{1 + j_+^\ast}{1-j_+^\ast} \right)
\label{chiintA}
\end{equation}	
where we have written the expression in terms of the BTZ value of $\chi$ cf., \eqref{chitherm}. Note that  $k_-^\ast >0$ and $j_+^\ast <0$  additive logarithmic piece can a-priori be positive or negative. 
However, since the presence of the black hole effectively repels geodesics, $|j_+^\ast |>|k_-^\ast|$, which forces the second term in \req{chiintA} to be negative.
Moreover, explicit numerical solutions for the critical angular momenta confirm that is always negative and  $\chi_{\cal A} < {\mathscr S}_\text{BTZ}$ which is consistent with the numerical results of \fig{f:chioftA}. We would like to emphasize that this is a-priori rather remarkable since the surface $\csf{\cal A}$ lies nowhere near any extremal surface in the bulk, as is evident from \fig{f:CWAB}. Despite the apparent non-locality in the definition of the causal holographic information, the final result is manifestly local. We will return to this point in \sec{s:discuss}.

\subsubsection{The behaviour of $\chi_{\cal A} - S_{\cal A}$}
\label{s:cmins}

Having understood how to compute $\chi_{\cal A}$, let us finally consider the difference between $\chi_\CA$ and $S_\CA$ in regime 3, which is the only domain in $t$ where it is different from zero.

\begin{figure}
\begin{center}
\includegraphics[width=4.2in]{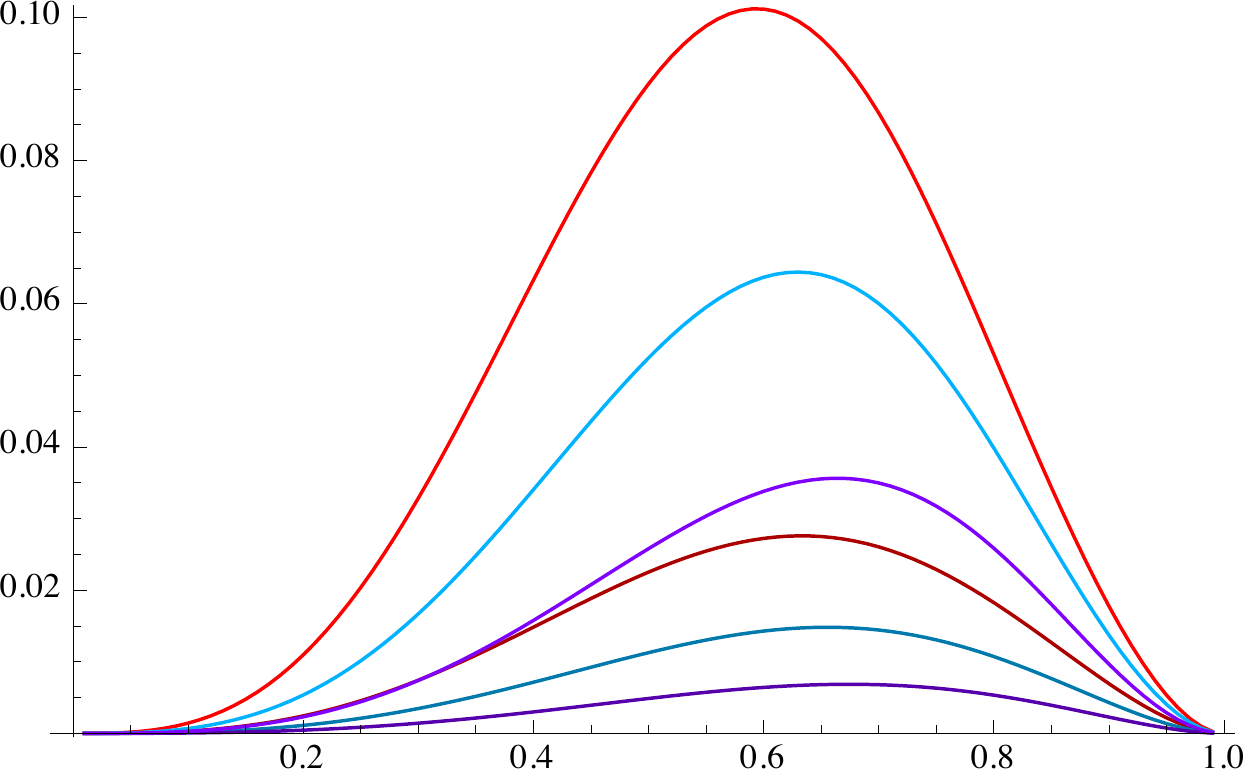}
\begin{picture}(0,0)
\setlength{\unitlength}{1cm}
\put(0,0) {{ $t_{\cal A}/\phA$}}
\put(-11,7) {{ $\chi_{\cal A}- S_{\cal A}$}}
\put(-1.2,6.5){$\downarrow$}
\put (-0.8,6.5) {{\scriptsize $\color{red}{\rh = 2.0,\; \phA =1.257}$}}
\put (-0.8,5.5) {{\scriptsize $\color{hue55}{\rh = 1.5,\; \phA =1.257}$}}
\put (-0.8,4.5) {{\scriptsize $\color{hue75}{\rh = 1.0,\; \phA =1.257}$}}
\put (-0.8,3.5) {{\scriptsize $\color{red}{\rh = 2.0,\; \phA =0.85}$}}
\put (-0.8,2.5) {{\scriptsize $\color{hue55}{\rh = 1.5,\; \phA =0.85}$}}
\put (-0.8,1.5) {{\scriptsize $\color{hue75}{\rh = 1.0,\; \phA =0.85}$}}
\end{picture}
\caption{
The variation of $\chi_\CA-S_\CA$ with time $\tA$ for the Vaidya-\AdS{3} spacetime. The plots are presented for different values of region size $\phA$ and final black hole radius $\rh$ for comparison. These have been obtained by using \eqref{chiintA} and \eqref{hee theta} which are evaulated numerically. We note that the result is in good agreement with data obtained by numerically solving for $\csf{\cal A}$ as described in \App{s:3dXi} and thence computing $\chi_{\cal A}$ and similary for $S_{\cal A}$
}
\label{f:chimS}
\end{center}
\end{figure}

First of all we recall that $\chi_{\cal A}$ and $S_{\cal A}$ have a leading area law divergence in the UV which is replaced by the logarithmic behaviour in $d=2$. Unlike the higher dimensional examples, here neither has any further divergences, so it makes sense to consider the difference $\chi_{\cal A} - S_{\cal A}$ in the present case. One  naively expects \cite{Hubeny:2012wa} that in this regime  $\chi_\CA>S_\CA$, since the surface $\csf{\cal A}$ lies closer to the boundary and hence ought to have greater (unregulated) length.\footnote{The divergent logarithmic contribution comes from the fact that the curves approach  the boundary normally.} However, it is clear that this cannot be the entire story since we have argued that $\chi_{\cal A} = S_{\cal A}$ in regime 4. It therefore must follow that the difference $\chi_{\cal A} - S_{\cal A}$ is non-monotonic.  Indeed, our explicit computation bears this expectation out. In \fig{f:chimS} we plot  variation of $\chi_\CA-S_\CA$ with time $\tA$.
We see that this vanishes at both endpoints of this regime, $\tA=0$ and $\tA=\phA$, and is positive in between.  Moreover, the slope vanishes at both ends (though the numerics are not well under control there). To get a handle on the behaviour of $\chi_{\cal A}-S_{\cal A}$ near $t_{\cal A}=0$ and $t_{\cal A} =\phA$ we turn to an examination of the two quantities in these regimes in a perturbation expansion in time.

\paragraph{The behaviour for $t_{\cal A} \simeq 0$:}
Firstly, consider the behaviour near $t_{\cal A} = 0$. As we explain in \App{s:3dS} it is quite straightforward to work out the rate of growth of $S_{\cal A}$ from the vacuum value. We find:
\begin{equation}
 S_{\CA} (\tA,\phA)
\,=\, {\mathscr S}_\text{AdS}
+ \frac{c_\text{eff}}{6} \left[ \frac{\rh^2+1}{2}\,\tA^2
-\frac{\rh^2+1}{48} \left(6 \, \csc^2\phA  +\rh^2-3\right)\tA^4
+ \cdots \right] \ ,
\label{Snear0}
\end{equation}
indicating a quadratic growth in the holographic entanglement entropy about its vacuum value.\footnote{As far as we are aware this is the first analytic result in the literature regarding the rate of growth of $S_{\cal A}$ at early times for finite region size.  The linear behaviour in the intermediate times has been noted before since this matches the CFT computation quite nicely. We also note that earlier \cite{Aparicio:2011zy} derived an universal expression for the early-time growth focussing on arbitrarily large regions in the context of field theories in ${\mathbb R}^{1,1}$. More specifically our results are valid for $\tA \ll \{ \rh , \phA\}$, with no hierarchy implied between the thermal scale set by $\rh$ and the region size $\phA$, whereas consideration of arbitrarily large regions requires $\tA \ll \rh \ll \phA$. The latter is only sensitive to the IR part of the entanglement entropy and does not for example see the saturation to the late time thermal value as we describe next. We thank Esperanza Lopez for discussions on this issue. For completeness we  present in \App{s:3dS} the general behaviour of the growth of $S_{\cal A}(t)$ at early times starting from various initial configurations (global or Poincar\'e vacuum and thermal state)  cf., \eqref{hee tA=0 btz} and \eqref{hee tA=phA btz}. \label{fn:earlyt}
} 

The behaviour of $\chi_{\cal A}$ can be computed directly using \eqref{chiintA} which is a clean local formula. Were it not for this it would be quite hard to estimate the change in $\chi_{\cal A}$ about its vacuum value. We solve \eqref{rphs1} and \eqref{rphs2} for $j_+^\ast$ and $k_-^\ast$ for small $t_{\cal A}$, which can be done analytically; plugging the result into \eqref{chiintA} we have (with $r_{\extr{}} = \cot \phA $ and $\rxi$ defined in \eqref{rxibtz}):
\begin{equation}
\chi_{\cal A}(t_{\cal A}, \phA) \,=\,
{\mathscr S}_\text{AdS} + \frac{c_\text{eff}}{6} \left[\frac{\rh^2+1}{2}\,\tA^2
+ \frac{1}{4} (r_{\extr{}} - \rxi) (\rxi^2 - r_{\extr{}}^2 -2 -2\,\rh^2 ) \,\tA^3 + \cdots \right] \ .
\label{chinear0}
\end{equation}	
From \eqref{Snear0} and \eqref{chinear0} we conclude that the leading and first subleading terms in the growth of $\chi_{\cal A}$ and $S_{\cal A}$ cancel each other off leaving behind a cubic growth:
\begin{equation}
\chi_{\cal A}- S_{\cal A}=  \frac{c_\text{eff}}{24} \; (\cot\phA - \rh\, \coth\rh \phA) (\rh^2\, \text{csch}^2\,\rh \phA - \csc^2\phA -\rh^2-1 ) \,\tA^3 + \cdots
\label{}
\end{equation}	
The coefficient of the cubic is positive definite guaranteeing that $\chi_{\cal A} > S_{\cal A}$ in the neighbourhood of the origin.

\paragraph{The behaviour for $t_{\cal A} \sim \phA$:} At the other end where $t_{\cal A}$ approaches $\phA$, the quantities $\chi_{\cal A}$ and $S_{\cal A}$ tend to their BTZ values $ {\mathscr S}_\text{BTZ}$ respectively. We can use a perturbation expansion in $\varepsilon\equiv \phA - t_{\cal A}$ to figure out the rate of approach. For $S_{\cal A}$ this is described in  \App{s:3dS}; the upshot of the computation is that $S_{\cal A}$ approaches the thermal value as a power law  with leading exponent $\frac{3}{2}$. Specifically,
\begin{equation}
S_{\CA} (\tA,\phA)
\,=\,
{\mathscr S}_\text{BTZ} - \frac{c_\text{eff}}{6} 
\; \frac{\rh^2+1}{\sqrt{2\,\rxi}} 
\left[ 
\frac{4}{3} \,\varepsilon^\frac{3}{2}+ \frac{\sqrt{2}}{3}\, \frac{\rh^2+1}{\rxi^\frac{3}{2}} \; \varepsilon^2
+ \cdots \right] \ .
\label{snearpha}
\end{equation}

Likewise we can use the result \eqref{chiintA} to figure out the behaviour of $\chi_{\cal A}$ in this regime. The strategy involves solving \eqref{rphs1} and \eqref{rphs2} for $j_+^\ast$ and $k_-^\ast$ perturbatively in $\phA - t_{\cal A}$ and then plugging the result back into the expression for $\chi$. A straightforward algebraic exercise  leads to 
\begin{equation}
\chi_{\cal A} = {\mathscr S}_\text{BTZ} -  \frac{c_\text{eff}}{6} \; 
\frac{\rh^2+1}{\sqrt{2\,\rxi}} 
\left[
\frac{4}{3} \,\varepsilon^\frac{3}{2}- \frac{5\,\rxi^2 - 4\,\rh^2 -1}{5\, \rxi}\; \varepsilon^\frac{5}{2} 
+ \cdots \right]\ .
\label{chinearpha}
\end{equation}	
From \eqref{chinearpha} and \eqref{snearpha} it follows that 
\begin{equation}
\chi_{\cal A}- S_{\cal A}= \frac{c_\text{eff}}{6}\;  
\frac{(\rh^2+1)^2}{3\,\rh^2 \coth^2(\rh \phA)} \, 
\big(\phA-\tA\big)^2 + \cdots
\label{chiSnearpha}
\end{equation}	
implying that the curves in \fig{f:chimS} approach the axis quadratically. 

We should note that in the vicinity of $\tA = 0$ and $\tA = \phA$ the difference between $\chi_{\cal A}$ and $S_{\cal A}$ is smaller than would be anticipated. In the former regime whilst both deviate from their vacuum value quadratically, the leading deviations cancel and the cubic term in $\chi_{\cal A}$ dominates. On the other hand for $\tA \to \phA^-$ it is the quadratic term in $S_{\cal A}$ which gives the rate of approach to the thermal answer with the leading $\frac{3}{2}$ power canceling out.  We would like to suggest that the smallness of the difference between $\chi_{\cal A}$ and $S_{\cal A}$ has to do with the specific nature of entanglement in $1+1$ dimensional CFTs, a point we will return to in \sec{s:discuss}.

Let us also note that from the numerical analysis we see that the difference $\chi_{\cal A}- S_{\cal A}$ has a characteristic peak, which lies  in the vicinity of $t_{\cal A}^* \approx \frac{2}{3} \, \phA$.  The location of the peak is mildly dependent on both the black hole size and the size of the interval.

\section{Shell collapse in higher dimensions}
\label{s:adsvd}

Let us now turn to the higher dimensional examples.  Here we have a richer set of choices for the shape of the region ${\cal A}$. We will however restrict attention to two simple examples (disks and strips) to illustrate the basic features of the causal wedges and $\chi_{\cal A}$. The choice of regions is dictated both by  tractability and to motivate the general lessons about the causal construction one can infer from them.

\paragraph{(i). Spherical entangling region:} We choose either a ball shaped round region $ {\cal A} \subset {\mathbb R}^{d-1}$ in Poincar\'e AdS,  or slices of the boundary sphere at constant latitude ${\cal A} \subset {\bf S}^{d-1}$ in global AdS, depending on whether we want to consider field theories on Minkowski space or on the Einstein Static Universe. 

For such spherical entangling regions it was argued in \cite{Hubeny:2012wa} that the causal information surface $\csf{\cal A}$ coincides with the extremal (in fact minimal) surface $\extr{\cal A}$. So in the vacuum we expect that $\chi_{\cal A} = S_{\cal A}$. However, this ceases to be true once we excite the state. In the process of the collapse, which we continue to model by the Vaidya-\AdS{d+1} geometry, we expect to see both $\chi_{\cal A}$ and $S_{\cal A}$ increase monotonically from their vacuum values. For $S_{\cal A}$, which evolves causally with $\delta S_{\cal A} =0$ for  $t_{\cal A} \leq 0$ this was seen originally in \cite{Hubeny:2007xt} and has been throughly explored in the recent investigations of holographic quench scenarios mentioned in the Introduction. Explicit computations confirm a similar result to hold for $\chi_{\cal A}$. In particular, in the regime $t_{\cal A} \leq 0$ the difference $\delta \chi_{\cal A}=0$ implying the $\chi_{\cal A}$ also behaves causally for the same reason as for the $d=3$ case described in the previous section. 

One can in fact prove this analytically; for completeness and to bolster our arguments in \sec{s:chi3}, we take a brief detour to show why the area of $\Xi_\CA$ remains unchanged for all $\tA \le 0$.  
As explained above, in this regime, the surface $\Xi_\CA$ lies entirely in AdS, and moreover lies on the same light cone $\partial J^+[\ddtipp]$ as $\extr{\cal A}$.  
Without loss of generality, consider \AdS{d+1} in  static coordinates 
\begin{equation}
ds^2 = \frac{-dt^2 + d\rbdy^2 + \rbdy^2 \, d\Omega_{d-2}^2 + dz^2 }{z^2}
\label{}
\end{equation}	
where $\rbdy$ is the boundary radial coordinate, and $z$ is the bulk Fefferman-Graham radial coordinate.  Let the $d-1$-dimensional region $\CA$ on the boundary $z=0$ be at $t=a$, $\rbdy = a$, so that the light cone in question (corresponding to $\partial J^+[\ddtipp]$) is simply the one from origin, described by
\begin{equation}
-t^2 + \rbdy^2  + z^2 = 0 \ .
\label{lccond}
\end{equation}	
As a co-dimension 1 surface, the light cone is parameterized by any two of these three coordinates and all the angles in $\Omega_{d-2}$, which just come along for the ride.  Now, since both the metric as well as $\CA$ is spherically symmetric in the $\Omega_{d-2}$, so will be any putative surface $\Xi_\CA$; such spacelike $d-1$ dimensional surface will then be parameterized by e.g.\ $z$ (or $\rbdy$ or $t$) and the $d-2$ angles, but it won't depend on the angles.  That means that $\Xi_\CA$ is specified by a single function, $\rbdy(z)$.  It can be drawn as a curve in $\{t,\rbdy,z\}$ space, with $t(z)$ determined from \req{lccond}.
We will now show that the area of {\it any}  surface $\Psi$ specified by arbitrary function $\rbdy(z)$  is in fact independent of $\rbdy(z)$, and indeed even independent of $\rbdy(0) = a$.
The area of $\Psi$ given by
\begin{equation}
A_{\Psi} =  V_\Omega \int_0^\zmax
dz \, \frac{\rbdy(z)^{d-2}}{z^{d-1}} \, \sqrt{-t'(z)^2 + \rbdy'(z)^2 + 1}
= 
 V_\Omega \int_0^\zmax
dz \, \frac{\rbdy^{d-2}}{z^{d-1}} \, 
\frac{(\rbdy - z \, \rbdy')}{\sqrt{\rbdy^2 + z^2}} \ ,
\label{APsirz}
\end{equation}	
where $V_\Omega= 2 \pi^{\frac{d-1}{2}}/\Gamma(\frac{d-1}{2})$ is the area of unit $S^{d-2}$, $\zmax$ is the maximal reach of $\Psi$
and the last term was obtained by using \req{lccond} to simplify $t'(z)$.
We can now use the change of variables $u= \frac{\rbdy(z)}{z}$, to rewrite \req{APsirz} as
\begin{equation}
A_{\Psi} =  V_\Omega \int_0^\infty  \frac{u^{d-2}}{ \sqrt{u^2 + 1}} \, du \ .
\label{APsiu}
\end{equation}	
Note that not only the integrand, but also the limits of integration, are independent of the shape of $\Psi$: the lower limit follows from $\rbdy(\zmax) = 0$ and the upper limit from $z=0$ while $\rbdy(0)=a>0$.
The expression \req{APsiu} is useful to obtain the actual area in $d+1$ dimensions.

Note that the total area $A_{\Psi}$ is of course infinite, so we regulate the expression using a finite $z$ cutoff.  Then slight care is needed to correctly compare the regulated areas of different surfaces $\Psi$.  A consistent cutoff must be one which is mapped between different surfaces $\Psi$ by the null generators of the light cone, and is implemented by keeping $u_\infty \equiv \rbdy(\zcutoff)/\zcutoff$ fixed.  For example, if we fix the cutoff $\varepsilon$ for the original surface given by $\rbdy_{\extr{}}(z) = \sqrt{a^2 - z^2}$, then along another surface $\Psi$ specified by $\rbdy_{\Psi}(z) \approx a + \rbdy_{\Psi}''(0) \, z^2 + \ldots$, the new cutoff is modified at quadratic order,
\begin{equation}
\zcutoff = \varepsilon \, \left( 1 
  + \varepsilon^2 \,  \frac{1+ a\, \rbdy_{\Psi}''(0)}{2 \, a^2}  + \ldots \right) \ .
\label{}
\end{equation}	

However for $t_{\cal A}> 0$ we expect to find $\chi_{\cal A} \neq S_{\cal A}$ since the corresponding values differ in the thermal state \cite{Hubeny:2012wa} even for these spherical entangling surfaces. Each individually evolves monotonically to their final thermal values as a function of $t_{\cal A}$.

\paragraph{(ii). Strip-like region:} Our second example is a strip like region ${\cal A}\subset {\mathbb R}^{d-1}$ in Poincar\'e AdS; we take ${\cal A}$ to be a segment of the real line in one of the directions (call it $x$) and translationally invariant in all other spatial directions. The problem of computing the causal wedge in this case still continues to be an effective three dimensional problem. In this case it is known that $\chi_{\cal A} \neq S_{\cal A}$ even in the vacuum \AdS{d+1} geometry \cite{Hubeny:2007xt, Hubeny:2012wa}, so this makes for a good example to illustrate the general features we argued for in \sec{s:pgenexp}. We anticipate as described there that $\chi_{\cal A}$ will evolve teleologically  and numerical checks show that it indeed  does so. Note that while it is still true that in the analog of regime 2 ($t_{\cal A} \leq 0$) the surface $\csf{\cal A}$ continues to lie on $\partial J^+[\ddtipp]$, the fact that $\extr{\cal A}$ lies outside $\cwedge$ in \AdS{d+1} implies that the generators of $\partial J^+[\ddtipp]$ have positive expansion towards the boundary. With the introduction of the shell, $\csf{\cal A}$ bends down and toward the boundary, thus moving in the direction of the expanding generators and thereby ends up having greater area consistent with the general expectations outlined in \sec{s:pgenexp}. 
 
\section{Discussion}
\label{s:discuss}

We have explored properties of bulk causal constructs of \cite{Hubeny:2012wa} which are naturally associated with a specified spatial region $\CA$ in the boundary field theory.  In particular, we studied how the causal wedge $\cwedge$, causal information surface $\Xi_\CA$, and the causal holographic information $\chi_\CA$ behave in time-dependent bulk geometries, in order to glean further hints for what these constructs might correspond to  in the field theory.

While we do not yet have the answer to this important question, and therefore there might be no apparent motivation for field theorists to study these constructs, it is useful to draw a lesson from the story of entanglement entropy: the work on understanding properties of co-dimension two extremal surfaces could likewise have been largely ignored were it not for the connection with an important field theory quantity, the entanglement entropy; yet most of our insight into the behavior of entanglement entropy and related quantities derives from understanding the behavior of the bulk surfaces.  Certain important properties, such as strong subadditivity, which are difficult to prove directly on the field theory side, become manifest in the bulk description.  But suppose that we did not have this connection between entanglement entropy and bulk extremal surfaces yet.  Nevertheless, results about extremal surfaces would secretly contain important insights, waiting to be realized, about the field theory.  While extremal surfaces $\extr{\CA}$ enjoy the status of having an associated field theory quantity $S_\CA$ already identified, our causal constructs $\cwedge$, $\Xi_\CA$, and $\chi_\CA$ fall into the latter category:  We do not yet know what field theory quantity they correspond to.  We study their properties to gather hints about what such a dual quantity might be, but we do not offer a definitive answer.  Nevertheless, the results we uncover may bear more fruit later when the dual of $\chi$ and $\Xi$ are finally identified.

The path towards elucidating the nature of $\chi$ and $\Xi$ which we chose to follow in the present work focuses on a specific class of spacetimes, namely the thin-shell Vaidya-AdS geometries, which describe maximally rapid collapse from pure AdS to a black hole.  This choice not only made our calculations tractable, but offered explicit results in case of physical interest which is in some sense furthest removed from equilibrium.  Before reviewing these results, let us remark on one potential drawback to this approach.  In particular, the large amount of symmetry which rendered the calculation tractable simultaneously renders such cases somewhat special, so that greater caution is needed in drawing general lessons.  We have however exercised this caution and explicitly identified where and why the calculation simplifies.
Moreover, these cases also enjoy an important physical significance, as already noted in \cite{Hubeny:2012wa}.

In the 3-dimensional case, where the bulk geometry is pure \AdS{3} before/inside the shell and BTZ after/outside, the causal information surface $\Xi_\CA$ coincides with the extremal surface $\extr{\CA}$ only in Regimes 1 and 4 (identified in Table  \ref{t:tAbreakdown}) when $\tA \le -\phA$ or $\tA \ge \phA$, respectively, which implies that $\chi_\CA = S_\CA$ in these regimes.  \fig{f:CWAB} explicitly demonstrates that $\Xi_\CA$ and $\extr{\CA}$ differ in Regimes 2 and 3, i.e.\ for $-\phA < \tA < \phA$.   On general grounds, we might then have expected that $\chi_\CA > S_\CA$ in  these regimes.  Nevertheless, we have seen that in fact $\chi_\CA = S_\CA$ in Regime 2 as well, namely when $\tA\le0$.  This is because the corresponding causal information surface $\Xi_\CA$ lies on the same light cone as $\extr{\CA}$ whose null generators have zero expansion.
The important implication of this result is that while we might expect on causal grounds that $\chi_\CA$ behaves quasi-teleologically in $\tA$, in the present case it is completely causal: by measuring $\chi_\CA$, one cannot determine the presence of the shell until the shell has been injected on the boundary.

On the other hand, in Regime 3, where $\tA>0$,  $\Xi_\CA$ is necessarily kinked by the shell.  (The  extremal surface $\extr{\CA}$ is likewise kinked by the shell, but it no longer lies on the same light cone as $\Xi_\CA$.)  Here we indeed confirm that  $\chi_\CA > S_\CA$ (cf.\ \fig{f:chimS}), and in the process discover another surprise:  It is easier to find $\chi$ than to find $S$!  Ordinarily one would have expected that due to the temporally non-local nature of the causal wedge, finding $\chi$ is more complicated, since it requires us to know the spacetime sufficiently far into the future and past of $\tA$, whereas once $\extr{\CA}$ is determined, changing the metric to its future or past does not affect it.  (In fact, direct numerical computation of $\chi$ is indeed more involved than that of $S$.)  However, in this case, $\Xi_\CA$ lies piecewise either on (future) light cone in AdS or (past)  light cone in BTZ, and we can therefore evaluate its length by computing the length of different curves on the same light cones, connected by flows along non-expanding generators.  In particular, the computation of $\chi$ then reduces to finding affine parameter along spacelike $E=0$ geodesics in pure AdS or BTZ, plus finding the intersection of  $\Xi$ with the shell, which yields the simple result  \req{chiintA}.  On the other hand,
 computation of $S$  requires finding a geodesic which refracts through the shell, and is moreover posed as a boundary-value problem.

The analytic simplification in evaluating $\chi$ allows us to find its scaling behaviour in the regions  near its initial (vacuum) and final (thermal) values.
We find that both $\chi$ and $S$ start growing quadratically at small $\tA$, with cubic correction for $\chi$ and quartic correction for $S$.  On the other hand, near the saturation point $\tA = \phA$, both  $\chi$ and $S$ exhibit a faster $3/2$ power-law behaviour,  with greater subleading correction to $S$ than to $\chi$. It would be interesting to understand the significance of these exponents from a field theory perspective (especially for $S$).
 In our 3-dimensional setting, we can in fact compare $S$ and $\chi$ directly, since both have the same divergence structure.  Evaluating $\chi - S$ (presented in \fig{f:chimS}), we find that not only the divergent parts, but in fact both the leading finite piece and the first subleading pieces cancel, so that $\chi - S$ grows only as $\tA^3$ when $\tA\to 0^+$ and as $(\phA-\tA)^2$ when $\tA \to \phA^-$.
It is quite remarkable that despite the geometric differences in the construction of $\Xi_\CA$ and $\extr{\CA}$, their lengths $\chi_\CA$ and $S_\CA$ agree to such high order.  We believe that this is related to the point discussed below of why these cases are so special.

The fact that the deviation from the vacuum value near $\tA=0$ is slower than the deviation from the thermal value near $\tA=\phA$ indicates where the effect of the shell is greatest.\footnote{
Although the strength of the shell (e.g.\ as measured by the amount of refraction of geodesics which cross it) blueshifts into the bulk, this might be offset by the portion of the curve which traverses the other side of the shell.
}
 Near $\tA=0$, both $\Xi_\CA$ and $\extr{\CA}$ cross the shell near the boundary, where the shell is weak.  As $\tA$ increases, these surfaces cross the shell deeper, where it gains more strength.  Near $\tA = \phA$, only the tips of $\Xi_\CA$ and $\extr{\CA}$ (i.e.\ their parts at small $\ph < \ph_s \ll \pi/2$) feel the shell.  Although this constitutes a tiny region of the full surface which is appreciably affected by the shell, the intersection $r_s$ is radially deeper, and the effect of the shell correspondingly stronger.  The latter effect is the more important one, causing $\chi$ and $S$ to deviate from their static values more quickly near the thermal end than near the vacuum end.

Having enjoyed the simplifications specified above in three dimensional setting of \sec{s:adsv3}, ultimately afforded by the fact that $\Xi_\CA$ and $\extr{\CA}$ coincide both in \AdS{3} as well as in BTZ, we briefly considered higher dimensional situations in \sec{s:adsvd}. 
There one of the simplifications disappears but we can still consider special cases where another simplification prevails.
In particular,  for spherical regions $\CA$, the bulk causal wedge $\cwedge = J^-[\ddtipf] \cap J^+[\ddtipp]$ is generated simply, and $\Xi_\CA$ still coincides with $\extr{\CA}$ in pure \AdS{d+1} for any $d$.  
This implies that in Regime 2, $\Xi_\CA$ is still on the same light cone in AdS as $\extr{\CA}$ (along which the null generators always have zero expansion as shown in \cite{Hubeny:2007xt}), so that $\chi_{\cal A}=S_\CA$ for $\tA \le 0$, just  as in the 3-dimensional case.
However, it is no longer true that the extremal surface  $\extr{\CA}$ coincides with the causal information surface $\Xi_\CA$ in the higher-dimensional black hole, \SAdS$_{d+1}$ for $d>2$.  This means not only that $\chi_\CA \ne S_\CA$ in Regime 4, but also that we can no longer find $\chi_\CA$ as simply in Regime 3.  In particular, while it is still true that $\chi$ is piecewise either on a (future) light cone from the boundary in AdS or a (past) light cone from the boundary in \SAdS, we can no longer evaluate the area of the latter by  the same trick of simply comparing with the corresponding part of the extremal surface.

To take another step towards the generic situation, in last part of \sec{s:adsvd} we drop the other simplification as well by considering non-spherical entangling regions $\partial \CA$, while nevertheless retaining tractability of the computation.  Specifically, in the case of the strip in planar Vaidya-AdS$_{d+1}$ for $d>2$, we find that $\Xi$ and $\extr{}$ do not coincide even for pure \AdS{d+1}, as already noted in \cite{Hubeny:2012wa}.  In this case we confirm that $\chi_\CA$ differs from its AdS value already in Regime 2 -- in other words, here $\chi_\CA$ {\it does behave quasi-teleologically}.  As pointed out in \sec{s:pgenexp}, this is a very mild form of teleology: $\chi_\CA$ knows about the shell only short time before the shell (on the timescale of light-crossing  transversely across the strip).
Nevertheless, this is an important data point to keep in mind when searching for plausible field theory duals to this construct.
 
As pointed out in \cite{Hubeny:2012wa}, one lesson of recent findings is that $\chi_A$ and $S_A$ coincide whenever the degrees of freedom in $\cal A$ are maximally entangled with those in its complement ${\cal A}^c$, namely when the region ${\cal A}$ is a spherical ball in planar or global AdS in all dimensions (or an arc of the boundary circle in the BTZ spacetime). The fact that the coincidence extends slightly beyond these stationary cases is suggestive. We believe that as a result of this maximal entanglement in the \AdS{} geometry for the said regions, it is impossible for $\chi_{\cal A}$ to grow from its vacuum value until the disturbance has come to play on the boundary. It is for this reason in the field theory that despite the deformation of the causal wedge in a quasi-teleological fashion, $\chi_{\cal A}$ nevertheless evolves causally. 

So far, we have restricted our attention to a specific class of configurations, namely the thin-shell Vaidya-AdS spacetimes, serving a convenient toy model of quantum quench in the field theory.
A complementary approach to elucidating the nature of $\chi$ and $\Xi$ is to maximally relax the assumptions about the bulk spacetime, and consider general global properties that these constructs must satisfy.  Since this approach is rather more formal, we have chosen to present the results separately in a companion paper \cite{Hubeny:2013fk}.  As a preview, here we simply note some of the key features.

It is sometimes useful to consider disconnected regions, ${\cal A} = \CA_1 \cup \CA_2$ with $\CA_1 \cap \CA_2 = \emptyset$.  If $\CA_1$ and $\CA_2$ are taken to lie at the same time in the boundary field theory, then it follows trivially that the causal wedges for the two parts are disjoint, and the causal holographic information is simply the sum of the two parts.  More generally (for intersecting regions) it is easy to establish subadditivity, though as demonstrated by explicit examples in \cite{Hubeny:2012wa}, strong subadditivity is {\it not} universally satisfied.  
In higher dimensions we can also consider a single but non-simply-connected region $\CA$.  In such a case, $\cwedge$ is likewise non-simply-connected, and $\Xi_\CA$ consists of multiple components. More curiously, even for simply-connected regions $\CA$, $\cwedge$ itself can have non-trivial topology, and $\Xi_\CA$ may consist of arbitrarily many disconnected components. We explain and explicitly demonstrate this in \cite{Hubeny:2013fk}. 

While the above remarks might lead the reader to expect that ``anything goes" and that causal wedges have properties which are hard to characterize globally, there are some features which hold in full generality.  One such important feature, already alluded to above, is that causal wedges can never penetrate the event horizon of a black hole.  This follows simply from causality: 
causal curves from within the black hole can never reach the boundary and therefore the interior of a  black hole cannot be contained in the causal wedge of any boundary region.  Nevertheless,  $\cwedge$ can reach close (or up to) the horizon for suitable regions $\CA$.  

In this context, there is an important difference between global and planar asymptotically AdS geometries:  For the field theory living on Minkowski background, there is no upper bound on the size of $\CA$, and $\Xi_\CA$ can penetrate arbitrarily close to the horizon for arbitrarily large regions $\CA$.  In fact, in this regime it is easy to see that the finite piece of $\chi_\CA$ scales extensively with the volume of $\CA$.  
On the other hand, for the field theory living on Einstein Static Universe, the region $\CA$ can at best wrap the sphere.  That means that the extent of the domain of dependence is either bounded by $\phA<\pi$ or fills up the entire boundary spacetime.  In the former  case, $\Xi_\CA$ only reaches a finite distance from the black hole.
On the other hand,
in the latter case, (for both ESU and Minkowski boundary geometries) when $\CA$ covers the entire Cauchy slice of the spacetime, the boundary of the causal wedge by definition coincides with the event horizon.  In this case $\chi_\CA$ is precisely the black hole entropy.

Finally, let us contrast this feature of the causal information surface $\Xi_\CA$ with the extremal surface $\extr{\CA}$.
While the black hole deforms $\extr{\CA}$ outward as compared to the AdS case (with same $\phA$), and in static bulk black hole spacetime extremal surfaces anchored on the boundary must lie outside the horizon
(both of these features were recently demonstrated by e.g.\ \cite{Hubeny:2012ry}), in a time-dependent situation the extremal surface {\it can} actually penetrate the event horizon. This was argued already in \cite{Hubeny:2012ry,Hubeny:2002dg}, and seen explicitly in \cite{AbajoArrastia:2010yt} for the planar Vaidya-AdS case, but is also manifest in the right panel of \fig{f:CWAB}.  
These issues are examined further in the forthcoming work  \cite{Hubeny:2013uq}.
Hence while causal holographic information is not cognisant of the causally disconnected region inside a bulk black hole, the entanglement entropy does sample at least a bit of the spacetime inside.

\acknowledgments 
It is a pleasure to thank Matt Headrick and Hong Liu for useful discussions about causal constructs in AdS. ET would like to thank Marco Beria for useful discussions. MR and VH would like to thank (i) SISSA for hospitality during the initial stages of this project, and (ii) the organizers of the MCTP workshop ``Entanglement, RG flows and holography'' for a stimulating meeting. MR would also like to thank organizers of the ``Indian Strings Meeting 2012'' at Puri, India and ``Quantum Aspects of Black Holes'' worskop at Sogang/CQUeST, Seoul, Korea for their hospitality during the course of the project. ET would like to thank Scuola Normale Superiore and the Department of Physics of the University of Pisa for the warm hospitality during parts of this work. VH and MR are supported in part by  the STFC Consolidated Grant ST/J000426/1.

\appendix

\section{Computational details for $\Xi_\mathcal{A}$ in Vaidya-\AdS{3}}
\label{s:3dXi}

In this appendix we present some details about the computation of quantities which are relevant for the explicit determination of $\Xi_\CA$ in thin shell geometries.

Let us begin by collecting expressions of the null geodesics occurring in the determination of $\Xi_\CA$ for $-\phA < \tA < \phA$. In order to shorten the formulae, we find it convenient to introduce $V^\alpha(r) \equiv r^2-(L/E_\alpha)^2 f_\alpha(r)$; explicitly 
\begin{equation}
V^i(r) \equiv (1-k^2)r^2-k^2 \ ,
\hspace{2cm}
V^o(r) \equiv (1-j^2)r^2+j^2 \rh^2 \ .
\label{Vio def}
\end{equation}
In the following, these quantities will occur with a further subindex which is either $\og$ or $\ig$, indicating whether the corresponding geodesic is respectively outgoing or ingoing. 

\begin{itemize}
	
\item  An outgoing geodesic lying entirely outside the shell and arriving at $\ddtipf$ reads
\begin{eqnarray}
\label{v out jp}
v_\og(r,j_+)  &=&
\frac{1}{2\rh}\,\log\left(
\frac{\big(\sqrt{V_\og^o(r)}- \rh\big)}{\big(\sqrt{V_\og^o(r)} + \rh\big)}\,
 \frac{(r-\rh)}{(r+\rh)} 
\right) +v_\wedge 
\hspace{.7cm}
\\
\label{phi out jp}
\rule{0pt}{.8cm}
\varphi_\og(r,j_+)  &=&
\frac{1}{2\rh}\,\log\left(
\frac{\sqrt{V_\og^o(r)}- j_{+} \rh}{
\sqrt{V_\og^o(r)}+ j_{+} \rh} \right)
\end{eqnarray}
\item Likewise, an outgoing geodesic lying inside the shell and connecting $\Xi_\CA$ to a point of the shell with coordinates $p_s=(v_s=0, \rs, \phs)$ is given by
\begin{eqnarray}
\label{v out kp A}
v_\og(r,k_+)  &=&
\arctan \sqrt{V^i_\og(r)}  +\arctan (r) 
- \arctan \sqrt{V^i_\og(\rs)}  - \arctan (\rs) 
\\
\rule{0pt}{.7cm}
\label{phi out kp A}
\varphi_\og(r,k_+)  &=&
\arctan \Big(\tfrac{1}{k_{+}}\sqrt{V^i_\og(r)}\,\Big) 
- \arctan \Big(\tfrac{1}{k_{+}}\sqrt{V^i_\og(\rs)}\,\Big) +  \phs
\end{eqnarray}
The choice of the integration constants guarantee that $v_\og(\rs,k_+) =0$ and $\varphi_\og(r,k_+) =\varphi_{s}$. \\
\item An ingoing geodesic outside the shell which connects a point of the shell to a point of $\Xi_\CA$ reads
\begin{eqnarray}
\label{v in jm B}
v_\ig(r,j_{-})  &=&
\frac{1}{2\rh}\,\log\left(
\frac{\sqrt{V^o_\ig(r)} + \rh}{\sqrt{V_\ig^o(r)} - \rh}\; \frac{r-\rh}{r+\rh} 
\right)
-
\frac{1}{2\rh}\,\log\left(
\frac{\sqrt{V^o_\ig(\rs)} + \rh}{\sqrt{V_\ig^o(\rs)} - \rh}\; \frac{\rs-\rh}{\rs+\rh} 
\right)
\hspace{.7cm}
\\
\rule{0pt}{.7cm}
\label{phi in jm B}
\varphi_\ig(r,j_{-}) &=&
\frac{1}{2\rh}\,\log\left(
\frac{\sqrt{V_\ig^o(r)} + j_{-} \rh}{\sqrt{V_\ig^o(r)} - j_{-} \rh} \;
\frac{\sqrt{V_\ig^o(\rs)} - j_{-} \rh}{\sqrt{V_\ig^o(\rs)} + j_{-} \rh} 
\right)
+  \varphi_{s}
\end{eqnarray}
The integration constants are obtained by imposing $v_\ig(\rs ,j_{-})=0$ and $\varphi_\ig(\rs ,j_{-}) =\varphi_{s}$.\\
\item Finally, an ingoing geodesic which starts at $q^\vee$ is given by
\begin{eqnarray}
\label{v in km}
v_\ig(r,k_-)  &=&
v_{\vee}-\,\arctan \Big(\sqrt{V^i_\ig(r)}\,\Big) +\arctan (r) 
\\
\label{phi in km}
\rule{0pt}{.6cm}
\varphi_\ig(r,k_-)   &=&
-\arctan \Big(\tfrac{1}{k_{-}}\sqrt{V^i_\ig(r)}\,\Big) +\frac{\pi}{2}\, \textrm{sign}(k_{-})
\end{eqnarray}
\end{itemize}

\subsection{Solving the refraction conditions}

Having explicit solutions for the geodesics we turn to the refraction conditions. We discuss an alternative way to find the solutions of the refraction conditions across the shell which is slightly different with respect to the one described in \sec{s:geodvaidya} (and serves to provide a complementary viewpoint). The equations to solve are
\begin{equation}
\label{ref law global}
\frac{1}{v_i'(\rs)} - \frac{1}{v_o'(\rs)} \,=\,\frac{1+r_+^2}{2} \ ,
\hspace{2cm}
\frac{v_i'(\rs)}{\varphi_i'(\rs)} \,=\,  \frac{v_o'(\rs)}{\varphi_o'(\rs)} \ .
\end{equation}
Given a geodesic crossing the shell at the point $p_s$ from one side, these equations tell us which is the corresponding geodesic on the other side of the shell. We find it convenient first to find the solution from the second equation of  (\ref{ref law global}) and, subsequently, employ the first one as a consistency check. \\
In order to deal with the second equation of (\ref{ref law global}) notice that
\begin{equation}
\label{vprime over phiprime}
\frac{v_\alpha'(\rs)}{\varphi_\alpha'(\rs)} 
=
\frac{\rs\big(\rs+\eta \sqrt{V^\alpha(r)}\,\big)}{(L/E_\alpha) f_\alpha(\rs)}
\equiv
\rs \, C_\alpha(\rs, L/E_\alpha) \ .
\end{equation}
The second equation of (\ref{ref law global}) can be written as $C_{i}(\rs, k)=C_{o}(\rs, j)$.
From this equation we can extract either $k(\rs, j)$ or $j(\rs, k)$ finding
\begin{eqnarray}
\label{ref k(j)}
k &=& \frac{2\rs C_o(\rs,j)}{C^2_o(\rs,j) f_i(\rs)+1}
\;=\;
\frac{2 j \,\rs  f_o(\rs)}{\rs \big[f_i(\rs)+f_o(\rs)\big]+\eta \big[f_i(\rs)-f_o(\rs)\big] \sqrt{V^o(\rs)} }
\\
\rule{0pt}{.7cm}
\label{ref j(k)}
j &=& \frac{2\rs C_i(\rs,k)}{C^2_i(\rs,k) f_o(\rs)+1}
\;=\;
\frac{2 k\, \rs  f_i(\rs)}{\rs \big[f_o(\rs)+f_i(\rs)\big]+\eta \big[f_o(\rs)-f_i(\rs)\big] \sqrt{V^i(\rs)} }
\end{eqnarray}
The second expression in (\ref{ref k(j)}) and (\ref{ref j(k)}) is obtained multiplying the first one respectively by $1=\tfrac{\rs-\eta\sqrt{V^o(\rs)}}{\rs-\eta\sqrt{V^o(\rs)}}$ and $1=\tfrac{\rs-\eta\sqrt{V^i(\rs)}}{\rs-\eta\sqrt{V^i(\rs)}}$.
Notice that (\ref{ref k(j)}) and (\ref{ref j(k)}) can be interchanged by exchanging the inside and outside quantities, as expected.

\subsection{Regime 2: $-\phA < \tA <0$}

In this regime the whole refraction curve belongs to $\partial_+(\cwedge)$ and therefore $\Xi_\CA$ lies entirely  inside the shell.

\subsubsection{Refraction curve on $\partial_+(\cwedge)$}

The outgoing geodesic arriving at $\ddtipf$ crosses the shell at the point $p_s$, whose radial coordinate $\rs$ is defined by $v_\og(\rs,j_+) =0$. From (\ref{v out jp}), this equation reads
\begin{equation}
\label{eq rs plus}
\frac{\big(\sqrt{V_\og^o(\rs)}- \rh\big)(\rs-\rh)}{\big(\sqrt{V_\og^o(\rs)} + \rh\big)(\rs+\rh)}\,
=  e^{-2 \rh v_\wedge} \ .
\end{equation}
For the radial case $j_+ =0$, the l.h.s. of (\ref{eq rs plus}) simplifies to a square, giving $\rs = \rh \coth (\rh v_\wedge/2)$. 
In the generic case, writing (\ref{eq rs plus}) in a form where $\sqrt{V_\og^o(\rs)}$ is isolated on one side of the equation and then squaring it, we obtain an algebraic equation of fourth order in $\rs$ which admits $\rs=\pm1$ as solutions. Thus, we are left with the following second order equation
\begin{equation}
\label{eq rs plus 2th}
\rs^2-2\rh \coth(\rh v_\wedge)\, \rs
+ \frac{1-j_{+}^2\coth^2(\rh v_\wedge)}{1-j_{+}^2}\,\rh^2 
= 0 \ ,
\end{equation}
whose largest root is the first formula in (\ref{rphs1}).
The angular coordinate $\phs = \varphi_\og(\rs,j_+) $ can be found from (\ref{phi out jp}). By observing that $V_\og^o(\rs)$ reduces to a perfect square, one obtains the second formula of (\ref{rphs1}).\\
The outgoing geodesic of $\partial_+(\cwedge)$ which refracts at the point $p_s$ of the shell is made by the part inside and the part outside the shell, which are characterized by $k_+$ and $j_+$ respectively. These coefficients are related one to each other by (\ref{ref k(j)}) and (\ref{ref j(k)}) with $\eta=+1$ (see e.g. (\ref{jpkprel})).

\subsubsection{$\Xi_\mathcal{A}$ inside the shell}
\label{sec app Xi A}

In the regime we are considering  $\Xi_\mathcal{A}$ is entirely inside the shell. The curve is the solution of \eqref{Ximatchingvph}.  In the first equation, the term $\arctan(r_\textrm{x}^i)$ cancels because it occurs both in (\ref{v out kp A}) and (\ref{v in km}).
In order to deal with (\ref{Ximatchingvph}), first we bring the terms dependent on $r_\textrm{x}^i$ on one side of the equations, then we take the $\tan$ of them.
Employing the addition formula $\tan(a+b) =\tfrac{\tan a+\tan b}{1-\tan a \,\tan b}$ and the property $\tan(x\pm\pi/2) = -\cot(x)$, we find
\begin{eqnarray}
\label{Tv in def}
& &\hspace{-1.6cm}
\frac{\sqrt{V^i_\og(r_\textrm{x}^i)} +  \sqrt{V^i_\ig(r_\textrm{x}^i)}}{
1- \sqrt{ V^i_\og(r_\textrm{x}^i) V^i_\ig(r_\textrm{x}^i)}}
\,=\,
\tan\left(v_{\vee} + \arctan \sqrt{V^i_\og(\rs)} + \arctan (r_{s}) \right)
\,\equiv\, T^i_v
\\
\rule{0pt}{.9cm}
\label{Tphi in def}
& &\hspace{-1.6cm}
\frac{\sqrt{V^i_\og(r_\textrm{x}^i)}/k_+ +  \sqrt{V^i_\ig(r_\textrm{x}^i)}/k_-}{
1- \sqrt{ V^i_\og(r_\textrm{x}^i) V^i_\ig(r_\textrm{x}^i)}/(k_+ k_-)}
\,=\,
-\,\cot\left( \arctan \Big(\sqrt{V^i_\og(r_s)}/k_{+}\Big) - \varphi_{s} \right)
\,\equiv\, T^i_\varphi 
\end{eqnarray}
In the special case of the radial geodesics $k_{+}=k_{-}=0$, the equation (\ref{Tv in def}) simplifies to 
\begin{equation}
\frac{2r_\Xi}{1-r_\Xi^2} \,=\, \tan\big(v_{\vee} +2 \arctan (r_{s}) \big) \ .
\end{equation}
Then, using the duplication formula for $\tan$ in this equation, we find the solution (\ref{rxiint}).\\ 
For non radial geodesics, we can obtain simpler expressions for (\ref{Tv in def}) and (\ref{Tphi in def})
multiplying their l.h.s.'s by $1=\tfrac{\sqrt{V^i_\og(r_\textrm{x}^i)} - \sqrt{V^i_\ig(r_\textrm{x}^i)}}{\sqrt{V^i_\og(r_\textrm{x}^i)} -  \sqrt{V^i_\ig(r_\textrm{x}^i)}}$ and $1=\tfrac{\sqrt{V^i_\og(r_\textrm{x}^i)}/k_{+} - \sqrt{V^i_\ig(r_\textrm{x}^i)}/k_{-}}{\sqrt{V^i_\og(r_\textrm{x}^i)}/k_{+} - \sqrt{V^i_\ig(r_\textrm{x}^i)}/k_{-}}$, respectively. Besides the radial case, this step is not allowed for $k_{+}=k_{-}$, but this is never realized in this regime.
This trick leads to the important simplification of a factor $(r_\textrm{x}^i)^2 +1$, allowing us to write (\ref{Tv in def}) and (\ref{Tphi in def}) as a linear system in terms of $\sqrt{V^i_\og(r_\textrm{x}^i)}$ and $\sqrt{V^i_\ig(r_\textrm{x}^i)}$, whose solution reads
\begin{eqnarray}
\label{sqrt V outgoing A}
 \sqrt{V^i_\og(r_\textrm{x}^i)}  & = &
 \frac{(k_{+} + k_{-})( k_{-}  T^i_\varphi - T^i_v )}{(1-k^2_{-}) \, T^i_v T^i_\varphi }
\\
\rule{0pt}{.7cm}
\label{sqrt V ingoing A}
 \sqrt{V^i_\ig(r_\textrm{x}^i)}  & = &
  \frac{(k_{+} + k_{-})(k_{+} T^i_\varphi - T^i_v)}{(1-k^2_{+}) \, T^i_v T^i_\varphi}
\end{eqnarray}
Taking the square of these two equations, we can find $(r_\textrm{x}^i)^2$ and a consistency condition
\begin{eqnarray}
\label{rx A analytic}
&& \hspace{-1.7cm} 
r_\textrm{x}^i \,=\,\frac{1}{\sqrt{1-k^2_{+}}}
\left[ k^2_{+}+\left(
 \frac{(k_{+} + k_{-})(T^i_v - k_{-} T^i_\varphi)}{(1-k^2_{-}) \, T^i_v T^i_\varphi }
\right)^2\right]^{\frac{1}{2}}
\\
\label{consistency A analytic}
\rule{0pt}{.8cm}
&& \hspace{-1.7cm}
k^2_{+}+\left(
 \frac{(k_{+} + k_{-})(T^i_v - k_{-}  T^i_\varphi)}{(1-k^2_{-}) \, T_v T_\varphi }
\right)^2
\,=\,
\frac{1-k^2_{+}}{1-k^2_{-}}
\left[ k^2_{-}+\left(
  \frac{(k_{+} + k_{-})(T^i_v - k_{+}  T^i_\varphi)}{(1-k^2_{+}) \, T^i_v T^i_\varphi}
\right)^2\right]
\end{eqnarray}
In order to understand these relations, we recall that $(\rs,\phs)$ and $k_+$ depend on $(j_+, v_\wedge)$ through (\ref{rphs1}) and (\ref{jpkprel}).
Then, from (\ref{Tv in def}) and (\ref{Tphi in def}) we write $T^i_v=T^i_v(j_+, v_\wedge,v_\vee)$ and $T^i_\varphi=T^i_\varphi(j_+, v_\wedge)$. 
Plugging these dependences into (\ref{consistency A analytic}), and inverting (numerically) this relation we obtain $k_{-}=k_{-}(j_{+},v_\wedge,v_\vee)$, telling us which geodesic of $\partial_-(\cwedge)$ intersects the geodesic of $\partial_+(\cwedge)$ characterized by $j_{+}$.
Substituting this result into (\ref{rx A analytic}) we finally find  $r_\textrm{x}^i=r_\textrm{x}^i(j_{+},v_\wedge,v_\vee)$. 
The angular coordinate $\varphi_\textrm{x}^i = \varphi_{\og}(r_\textrm{x}^i,k_+) = \varphi_\textrm{x}^i (j_{+},v_\wedge,v_\vee)$ is then obtained through (\ref{phi out kp A}).\\
As a check of (\ref{rx A analytic}) and (\ref{consistency A analytic}), notice that in this regime of $\tA$ and $\forall j_{+}$ we have $v^i_\textrm{x} < 0$, namely that $\Xi_\CA$ is entirely inside the shell.


\subsection{Regime 3: $0 < \tA  < \phA$}

In this regime, the central part of the refraction curve belongs to $\partial_+(\cwedge)$. The corresponding part of $\Xi_\CA$ is inside the shell and it can be found as explained in  \sec{sec app Xi A}. Thus, the results found above must be applied only for $|j_+| < |j_+^\ast| <1$ or, equivalently for $|k_-| < k_-^\ast <1$, where $j_+^\ast$ (or $k_-^\ast$) characterizes the critical geodesics.\\
For $|j_+^\ast| < |j_+| <1$ (or $k_-^\ast < |k_-| <1$ equivalently), the refraction curve belongs to $\partial_-(\cwedge)$ and therefore the corresponding parts of $\Xi_\CA$ (they are symmetric w.r.t. to the radial geodesics) are outside the shell. For these geodesics, the results showed in this subsection must be used in order to find $\Xi_\CA$ outside the shell.

\subsubsection{Refraction curve on $\partial_-(\cwedge)$ and critical geodesics}

The ingoing geodesic starting from $\ddtipp$ at the boundary crosses the shell at $p_s$ first. The radial coordinate $\rs$ of this point is such that $v_\ig(\rs, k_{-})=0$, i.e. from (\ref{v in km})
\begin{equation}
\label{eq rs B}
\arctan \Big(\sqrt{V^i_\ig(\rs)}\,\Big) - \arctan (\rs) = v_\vee \ .
\end{equation}
First, we take the $\tan$ of this equation employing also the subtraction formula for the $\tan$. Then, multiplying the l.h.s. of the resulting equation by $1=\tfrac{\sqrt{V^i(\rs)} +\rs}{\sqrt{V^i(\rs)} +\rs}$, we find  expression where $\sqrt{V^i(\rs)} $ occurs only in the denominator. Isolating $\sqrt{V^i(\rs)} $ and then taking the square, we obtain the following second order equation for $\rs$
\begin{equation}
\label{eq rs B 2th}
\rs^2 - 2 \cot(v_\vee) \,r_s -\frac{1+k_{-}^2 \cot^2(v_\vee)}{1-k_{-}^2} = 0 \ ,
\end{equation}
whose largest solution is given in the first equation of (\ref{rphs2}). The angular coordinate $\phs = \varphi_\ig(\rs ,k_-) $ given in the second equation of (\ref{rphs2}) can be computed through (\ref{phi in km}), by further noticing that $V^i_\ig(\rs)$ becomes a perfect square.\\
The ingoing geodesic starting from $\ddtipp$, characterized by $k_{-}$, becomes an ingoing geodesic characterized by $j_{-}$ and propagating inside the shell. The function $j_{-}(k_{-})$ is given in (\ref{jpkprelpast}) and it is obtained from (\ref{ref j(k)}) with $\eta=-1$.\\
In this regime of $\tA$, two symmetric critical geodesics on $\partial_-(\cwedge)$ occur. They meet both the shell and $\partial_+(\cwedge)$ at the same point. This implies that the corresponding critical value $k_{-}^\ast$ satisfies (\ref{eq rs B 2th}) with $r_\textrm{x}^i$ given in (\ref{rx A analytic}) instead of $\rs$, namely
\begin{equation}
\label{eq km critical}
(r_\textrm{x}^i)^2 - 2 \cot(v_\vee) \,r_\textrm{x}^i -\frac{1+(k_{-}^\ast)^2 \cot^2(v_\vee)}{1-(k_{-}^\ast)^2} = 0
\qquad \Longrightarrow \qquad 
j^\ast_{+} \ .
\end{equation}
As discussed in the end of \sec{sec app Xi A}, from (\ref{rx A analytic}) and (\ref{consistency A analytic}) we find $k_{-}^\ast=k_{-}^\ast(j^\ast_{+},\tA,\phA)$ and  $r_\textrm{x}^i=r_\textrm{x}^i(j^\ast_{+},\tA,\phA)$. Substituting these results into (\ref{eq km critical}), it becomes an equation for $j^\ast_{+}$ that we can (numerically) invert, getting $j^\ast_{+}=j^\ast_{+}(\tA,\phA)$, where $0 < \tA  < \phA$.

\subsubsection{$\Xi_\mathcal{A}$ outside the shell}

The curve $\Xi_\mathcal{A}$ outside the shell is given by $(r_\textrm{x}^o,\varphi_\textrm{x}^o)$ satisfying
\begin{equation}
\label{eqs Xi out}
v_\ig(r_\textrm{x}^o,j_{-}) \,=\,v_\og(r_\textrm{x}^o,j_{+}) \,\equiv\, v_\textrm{x}^o \ ,
\hspace{1.5cm} 
\varphi_\ig(r_\textrm{x}^o,j_{-}) \,=\,\varphi_\og(r_\textrm{x}^o,j_{+}) \,\equiv\, \varphi_\textrm{x}^o
\end{equation}
where the geodesics of $\partial_+(\cwedge)$ are given by (\ref{v out jp}) and (\ref{phi out jp}), while the ones belonging to $\partial_-(\cwedge)$ are described by (\ref{v in jm B}) and (\ref{phi in jm B}). 
Notice that in the first equation of (\ref{eqs Xi out}) the term $\tfrac{1}{2\rh} \log \tfrac{r_\textrm{x}^o-\rh}{r_\textrm{x}^o+\rh}$ simplifies. Writing (\ref{eqs Xi out}) as equations involving the arguments of the $\log$'s, they become
\begin{eqnarray}
\label{eq v Xi out 1}
& & \hspace{-1cm}
\frac{\big(\sqrt{V^o_\ig(r_\textrm{x}^o)} + \rh\big)
\big(\sqrt{V^o_\ig(r_{s})} - \rh\big)(r_{s}+\rh)}{
\big(\sqrt{V^o_\ig(r_\textrm{x}^o)} - \rh\big)
\big(\sqrt{V^o_\ig(r_{s})} + \rh\big)(r_{s} - \rh)
}
\,=\,
e^{2\rh v_\wedge} \, 
\frac{\sqrt{V^o_\og(r_\textrm{x}^o)} - \rh}{\sqrt{V^o_\og(r_\textrm{x}^o)} + \rh}
\\
\label{eq phi Xi out 1}
\rule{0pt}{.9cm}
& & \hspace{-1cm}
\frac{\big(\sqrt{V^o_\ig(r_\textrm{x}^o)} + j_{-} \rh\big)\big( \sqrt{V^o_\ig(r_{s})} - j_{-} \rh \big) }{
\big( \sqrt{V^o_\ig(r_\textrm{x}^o)} - j_{-} \rh\big) \big( \sqrt{V^o_\ig(r_{s})} + j_{-} \rh \big) }
\;
e^{2\rh \varphi_{s}}
\,=\,
\frac{\sqrt{V^o_\og(r_\textrm{x}^o)}- j_{+} \rh}{\sqrt{V^o_\og(r_\textrm{x}^o)}+ j_{+} \rh}
\end{eqnarray}
Multiplying the l.h.s.\ of  (\ref{eq v Xi out 1})  by $1=\tfrac{\sqrt{V^o_\ig(r_\textrm{x}^o)} + \rh}{\sqrt{V^o_\ig(r_\textrm{x}^o)} + \rh}$ and its r.h.s.\ by $1=\tfrac{\sqrt{V^o_\og(r_\textrm{x}^o)} - \rh}{\sqrt{V^o_\og(r_\textrm{x}^o)} - \rh}$, the equation simplifies. 
A similar simplification occurs in (\ref{eq phi Xi out 1}) when we multiply its l.h.s. by $1=\tfrac{\sqrt{V^o_\ig(r_\textrm{x}^o)}- j_{-} \rh}{\sqrt{V^o_\ig(r_\textrm{x}^o)}- j_{-} \rh}$ and its r.h.s.\ by $1=\tfrac{\sqrt{V^o_\og(r_\textrm{x}^o)}- j_{+} \rh}{\sqrt{V^o_\og(r_\textrm{x}^o)}- j_{+} \rh}$. In the resulting equations, the dependence of $r_\textrm{x}^o$ can be isolated on one side, which becomes the square of a simple rational function in terms of $\sqrt{V^o_\ig(r_\textrm{x}^o)}$. 
Taking the square root of the two equations in such form, they can be written respectively as follows
\begin{eqnarray}
\label{Tvout def}
& & \hspace{-1cm}
\frac{\sqrt{V^o_\ig(r_\textrm{x}^o)}+\rh}{\sqrt{V^o_\og(r_\textrm{x}^o)}-\rh}
\,=\; 
e^{\rh v_\wedge}
\sqrt{\frac{(1-j_{-}^2)(r_{s} - \rh)\big(\sqrt{V^o_\ig(r_{s})} + \rh\big)}{(1-j_{+}^2)(r_{s} + \rh)\big(\sqrt{V^o_\ig(r_{s})} - \rh\big)}}
\,\equiv\, T^o_v
\\
\rule{0pt}{.8cm}
\label{Tphiout def}
& & \hspace{-1cm}
\frac{\sqrt{V^o_\ig(r_\textrm{x}^o)} + j_{-}\rh}{\sqrt{V^o_\og(r_\textrm{x}^o)}-j_{+}\rh}
\,=\,
e^{-\rh \varphi_{s}}
\sqrt{\frac{(1-j_{-}^2)\big(\sqrt{V^o_\ig(r_{s})} + j_{-} \rh\big)}{
(1-j_{+}^2)\big(\sqrt{V^o_\ig(r_{s})} - j_{-}\rh\big)}}
\,\equiv\, T^o_\varphi
\end{eqnarray}
It is now clear that these equations can be written as a linear system in $\sqrt{V^o_\og(r_\textrm{x}^o)}$ and $\sqrt{V^o_\ig(r_\textrm{x}^o)}$, which can be easily inverted, giving
\begin{eqnarray}
\label{sqrtVout plus}
& & 
\sqrt{V^o_\og(r_\textrm{x}^o)}  \,=\,
\rh \, \frac{1- j_{-} +T^o_v- j_{+}\, T^o_\varphi}{T^o_v - T^o_\varphi}
\\
\rule{0pt}{.7cm}
\label{sqrtVout minus}
& &
\sqrt{V^o_\ig(r_\textrm{x}^o)}  \,=\,
\rh \, \frac{T^o_\varphi - j_{-}\, T^o_v +(1-j_{+})T^o_v T^o_\varphi}{T^o_v - T^o_\varphi}
\end{eqnarray}
Squaring both these equations, we obtain $(r_\textrm{x}^o)^2$ and a consistency condition
\begin{eqnarray}
\label{rx B analytic}
& & \hspace{-1.5cm}
r_\textrm{x}^o\,=\,
\frac{\rh}{\sqrt{1-j_{+}^2}}  \left[
\left(
\frac{1 - j_{-} + T^o_v - j_{+}\, T^o_\varphi }{T^o_v - T^o_\varphi} 
\right)^2 - j_{+}^2
\,\right]^{\frac{1}{2}}
\\
\rule{0pt}{.8cm}
\label{consistency B analytic}
& & \hspace{-1.5cm}
\left(
\frac{1 - j_{-} + T^o_v - j_{+}\, T^o_\varphi}{T^o_v - T^o_\varphi}
\right)^2 - j_{+}^2 
\,=\,
\frac{1 - j_{+}^2}{1 - j_{-}^2}
\left[ \left( 
 \frac{T^o_\varphi - j_{-}\, T^o_v +(1-j_{+})T^o_v T^o_\varphi}{T^o_v - T^o_\varphi}
 \right)^2 - j_{-}^2\,\right]
\end{eqnarray}
These equations allow to find the part of $\Xi_\CA$ outside the shell, as well as (\ref{rx B analytic}) and (\ref{consistency B analytic}) lead to the part of $\Xi_\CA$ inside the shell. Indeed, first we observe that
$(\rs,\phs)$ and $j_-$ depend on $(k_-, v_\vee)$ through (\ref{rphs2}) and (\ref{jpkprelpast}).
Then, the definitions  (\ref{Tvout def}) and (\ref{Tphiout def})  tell us $T^o_v=T^o_v(k_-, v_\vee, v_\wedge)$ and $T^o_\varphi=T^o_\varphi(k_-, v_\vee)$. 
By inserting all these functions into the consistency condition (\ref{consistency B analytic}), it becomes an equation which gives us (through numerical inversion) $j_+ = j_+(k_-, v_\vee, v_\wedge)$, namely the geodesic of $\partial_+(\cwedge)$ intersecting the geodesic of $\partial_-(\cwedge)$ characterized by $k_{-}$.
Given this result, (\ref{rx B analytic}) allows to obtain $r_\textrm{x}^o=r_\textrm{x}^o(k_{-},v_\vee, v_\wedge)$. The angular coordinate $\varphi_\textrm{x}^o = \varphi_{\ig}(r_\textrm{x}^o,j_-) = \varphi_\textrm{x}^o (k_{-},v_\vee, v_\wedge)$ is then obtained through (\ref{phi in jm B}).\\
We can check (\ref{rx B analytic}) and (\ref{consistency B analytic}) by verifying that in this regime of $\tA$ and for $|k_-| > k_-^\ast$ we have $v^o_\textrm{x} > 0$, namely that $\Xi_\CA$ is outside the shell.

\section{Computational details for $S_\CA$ }
\label{s:3dS}

In this appendix we give some details for the analytical computation of $S_\CA$ in three dimensions. We generalize the discussion of \cite{Balasubramanian:2011ur}, allowing for the geometry inside the shell to be either Poincar\'e AdS or global AdS or another BTZ spacetime (corresponding to heating up a preexisting thermal state). While many of the results follow from our discussion in the main text and \App{s:3dXi} modified appropriately to spacelike geodesics, it is useful to work these out explicitly to obtain compact expressions for the entanglement entropy.

\subsection{Spacelike geodesics and refraction conditions}

While for the determination of $\Xi_\CA$ we needed the null geodesics, for the holographic entanglement entropy $S_\CA$ we have to compute the spacelike geodesics anchored on the boundary at the endpoints of the interval $\CA$.

We can be sufficiently general and consider a metric like (\ref{vaidya}) with $d=2$ and $f(r,v) = f_i(r) + \Theta(v) [f_o(r) - f_i(r) ]$, where we always consider a BTZ geometry outside the shell (i.e. $f_o(r)=r^2-r_{h,o}^2$), while inside the shell we choose either Poincar\'e AdS ($f_i(r)=r^2$) or global AdS ($f_i(r)=r^2+1$) or
another BTZ ($f_i(r)=r^2-r_{h,i}^2$) with $r_{h,i} < r_{h,o}$ in order to satisfy the null energy condition.

As discussed in \sec{s:geodvaidya}, the spacelike geodesics ($\kappa =1$ in (\ref{ELkappa})) are characterized by the pair $(E,L)$. The ones we are interested in are made by three branches: two disconnected and symmetric ones outside the shell and one inside the shell. The coordinates of the meeting points are $(v_s=0, r_s, \pm \phs)$. The symmetry of the problem allows us to consider only a half of the geodesic, the outgoing one, going from $(v_{{\mathfrak E}}, \rE,0)$ to the point $(\tA, \infty ,\phA)$ of the boundary.
Being $\dot{\varphi}$ continuos across the shell, $L_i =L_o$, while the jump of $E$ is given by (\ref{Ejump}), where $\kappa=1$ occurs through $\dot{v}$ at the shell (see (\ref{vdot})).
Since we are considering equal-time endpoints (for endpoints at different times on the boundary see \cite{Aparicio:2011zy}), inside the shell we have $(E_i, L_i)=(0,\rE)$. Then, from (\ref{vdot}) we find that 
$\dot{v}$ at the shell for spacelike and outgoing geodesics ($\eta=1$) reads $\dot{v}_s = [(r_s^2 - \rE^2) / (r_s^2 f_i(r_s))]^{1/2}$ and, through (\ref{Ejump}), this leads to $E_o$. Thus, the pair $(E_o,L_o)$ reads
\begin{equation}
\label{EL out spacelike geods}
E_o = 
\frac{[f_o(r_s) - f_i(r_s)] \sqrt{r_s^2 - \rE^2}}{2\, r_s \sqrt{f_i(r_s)}}  \ ,
\hspace{2cm}
L_o =  L_i = \rE \ .
\end{equation}

The equations to solve in order to find the spacelike geodesics are
\begin{equation}
\label{eqs spacelike geods}
t' = \frac{\eta E_\alpha}{f_\alpha \sqrt{E_\alpha^2 + f_\alpha (1-L_\alpha/r^2)}} \ ,
\hspace{2cm}
\varphi' = \frac{\eta L_\alpha}{r^2 \sqrt{E_\alpha^2 + f_\alpha (1-L_\alpha/r^2)}} \ ,
\end{equation}
where $\alpha$ is either $i$ or $o$, and $f_i(r)$ is one of the three choices described right above.\\
Outside the shell we have a BTZ geometry and the solutions of (\ref{eqs spacelike geods}) are
\begin{eqnarray}
t_o(r)  &=&
\frac{1}{2r_h} \,\log\left(
\frac{r^2-(r_h + E_o)r_h + \eta \sqrt{D_o(r)}}{r^2-(r_h - E_o)r_h + \eta \sqrt{D_o(r)}}
\right) +C_t
\;\equiv\; \tilde{t}(r) + C_t 
\\
\rule{0pt}{.8cm}
\label{varphi(r) btz}
\varphi_o(r)  &=&
\frac{1}{2r_h} \,\log\left(
\frac{r^2-r_h L_o + \eta \sqrt{D_o(r)}}{r^2+ r_h L_o + \eta \sqrt{D_o(r)}}
\right) +C_\varphi
\;\equiv\; \tilde{\varphi}(r) + C_\varphi
\end{eqnarray}
where we introduced $D_o(r) \equiv  
E_o^2 r^2+(r^2-r_h^2)(r^2-L_o^2)$.
It is important to remark that the integration constants  $C_t$ and $C_\varphi$ provide the boundary data, namely
\begin{equation}
C_t = t_\mathcal{A} \ ,
\hspace{3.5cm} 
C_\varphi = \varphi_\mathcal{A} \ .
\end{equation}
These parameters are also related to $(r_s, \varphi_s)$ of the point at the shell by the conditions $v_o(r_s) = 0$ and $\varphi_i(r_s) = \varphi_o(r_s)\equiv \phs$, which give respectively 
\begin{equation}
\label{tA and phiA}
t_\mathcal{A} = - \,\tilde{t}(r_s) -\frac{1}{2r_h} \log \left(\frac{r_s-r_h}{r_s+r_h}\right) \ ,
\hspace{3cm}
\varphi_\mathcal{A} = \varphi_s - \tilde{\varphi}(r_s) \ .
\end{equation}
Let us focus on the first equation of (\ref{tA and phiA}) and try to extract $\rs$. First, notice that,
from (\ref{EL out spacelike geods}), we find that $D_o(r_s) = (r_s^2-\rE^2)[f_o(r_s)+f_i(r_s)]^2/[4 f_i(r_s)]$. This result gives
\begin{equation}
\label{ttilde(rshell)}
\tilde{t}(r_s)  = \frac{1}{2r_h} \,\log\left(
\frac{2(r_s^2-r_h^2)r_s +\sqrt{\tfrac{r_s^2-\rE^2}{f_i(r_s)}}
\,\big[ \,r_s \big| f_o(r_s)+f_i(r_s) \big| - r_h\big(f_o(r_s) - f_i(r_s)\big)\big]}{
2(r_s^2-r_h^2)r_s +\sqrt{\tfrac{r_s^2-\rE^2}{f_i(r_s)}}
\,\big[ \,r_s \big| f_o(r_s)+f_i(r_s) \big| - r_h\big(f_o(r_s) + f_i(r_s)\big)\big]}
\right) \ .
\end{equation}
Following \cite{Balasubramanian:2011ur}, we find it convenient to introduce a new parameter $\theta$ which mixes $\rs$ and $\rE$. Its definition depends on the geometry inside the shell
\begin{equation}
\label{cos theta def}
\cos \theta \equiv
\sqrt{\frac{r_s^2-\rE^2}{f_i(r_s)}}
=
\left\{\begin{array}{lcl}
\sqrt{1-\rE^2/r^2_s} & & \textrm{Poincar\'e AdS} \\
\sqrt{1-(\rE^2 +1)/(r^2_s+1)} & & \textrm{global AdS} \\
\sqrt{1-(\rE^2 - r_{h,i}^2)/(r^2_s- r_{h,i}^2)} & & \textrm{BTZ} 
\end{array}
\right.
\end{equation}
Notice that the cases of Poincar\'e AdS and global AdS are obtained from the expression for BTZ by substituting respectively $r_{h,i}^2=-1$ and $r_{h,i}^2=0$. This feature persists also in the expressions hereafter, unless otherwise specified. 
From (\ref{cos theta def}), notice that $\rE^2$ can be written in terms of $r_s^2$ and $\theta$ as  $\rE^2 = r_s^2-(r_s^2 - r_{h,i}^2) \cos^2\theta =
r_s^2 \sin^2\theta + r_{h,i}^2 \cos^2\theta$.\\
Plugging the definition (\ref{cos theta def}) into (\ref{ttilde(rshell)}) with the assumption $f_o(r_s)+f_i(r_s)  > 0$, the first equation of (\ref{tA and phiA}) leads us to write $\rs$ as a function of $t_\mathcal{A}$ and $\theta$. 
After some algebra, one finds the following algebraic equation of the second order in $r_s$
\begin{equation}
\label{eq II order rs}
\left(\frac{r_s}{r_{h,o}}\right)^2 - \coth(r_{h,o} t_\mathcal{A})\, \frac{r_s}{r_{h,o}}\
+\frac{\cos\theta}{1+\cos\theta}
\left(1-\frac{r^2_{h,i}}{r^2_{h,o}}\right)
=0
\end{equation}
whose largest root reads
\begin{equation}
\label{rs p}
r_s =
\frac{r_{h,o}}{2}
\left(
\coth(r_{h,o} t_\mathcal{A}) +
\sqrt{\coth^2(r_{h,o} t_\mathcal{A}) -\frac{2\cos \theta(1-r_{h,i}^2/r_{h,o}^2)}{1+\cos\theta}}
\,\right) \ .
\end{equation}
Notice that $r_s$ is contained also in the r.h.s. of (\ref{rs p}) through $\cos\theta$, and this means that we have not fully inverted the first equation of (\ref{tA and phiA}). Nevertheless, the formula (\ref{rs p}) gives $r_s$ in terms of $t_\mathcal{A}$ and $\theta$.\\
From the second equation in (\ref{tA and phiA}) we can write $\varphi_\mathcal{A}$ in a more compact form.
From (\ref{varphi(r) btz}) with $\eta=+1$ and $D_o(\rs)$ given above, we obtain 
\begin{equation}
\label{tilde phi rs}
\tilde{\varphi}(r_s) = \frac{1}{2r_{h,o}}\,\log\left(
\frac{2(1+\cos\theta)r_s^2 -2r_{h,o}\sqrt{r_s^2\sin^2\theta + r_{h,i}^2\cos^2\theta}-(r_{h,o}^2+r_{h,i}^2)\cos\theta}{2(1+\cos\theta)r_s^2 + 2r_{h,o}\sqrt{r_s^2\sin^2\theta + r_{h,i}^2\cos^2\theta}-(r_{h,o}^2+r_{h,i}^2)\cos\theta}
\right) \ ,
\end{equation}
where the term with the square root is simply $\rE$ extracted from (\ref{cos theta def}).\\
In order to find $\varphi_s = \varphi_i(r_s)$, we need the solution $\varphi_i(r)$ of the second equation in (\ref{eqs spacelike geods}) for the different choices of $f_i$. They read
\begin{equation}
\varphi_i(r) = \left\{\,
\begin{array}{lcl}
\displaystyle 
\frac{\sqrt{r^2-\rE^2}}{\rE\, r}
& &\textrm{Poincar\'e AdS}
\\
\rule{0pt}{.9cm}
\displaystyle 
\frac{1}{2} \,\arcsin\left(\frac{(1-\rE^2)r^2 -2\rE^2}{(1+\rE^2)r^2}\right) +\frac{\pi}{4}
 & &\textrm{global AdS}
\\
\rule{0pt}{1.2cm}
\displaystyle 
\frac{1}{2r_{h,i}}\,
\log\left(
\frac{r^2-r_{h,i} \, \rE +\sqrt{(r^2-r^2_{h,i})(r^2 -\rE^2)}}{r^2+r_{h,i} \,\rE +\sqrt{(r^2-r^2_{h,i})(r^2 -\rE^2)}}\;\frac{\rE +r_{h,i}}{\rE - r_{h,i}}
\right)
& \hspace{.0cm} &
\textrm{BTZ}
\end{array}
\right.
\label{phrspgeods}
\end{equation}
where the integration constants have been fixed by imposing that $\varphi_i(\rE) =0$.
Now we can write $\varphi_i(r_s) =\varphi_s$ in terms of $r_s$ and $\theta$ as follows
\begin{equation}
 \label{phis theta}
\varphi_s = \left\{\,
\begin{array}{lcl}
\displaystyle 
\frac{\cot\theta}{r_s}
& &\textrm{Poincar\'e AdS}
\\
\rule{0pt}{.9cm}
\displaystyle 
\arctan\left(\frac{\cos\theta}{\sqrt{\rs^2 \sin^2\theta -\cos^2\theta}} \right) 
& &\textrm{global AdS}
\\
\rule{0pt}{1.2cm}
\displaystyle 
\frac{1}{2r_{h,i}}\,
\log\left(
\frac{\sqrt{r_s^2 \sin^2\theta+r^2_{h,i} \cos^2\theta} +r_{h,i} \cos\theta}{\sqrt{r_s^2 \sin^2\theta+r^2_{h,i} \cos^2\theta}- r_{h,i} \cos\theta}
\right)
& &\textrm{BTZ}
\end{array}
\right.
\end{equation}
For $\theta \in (0,\pi/2)$, the first expression is the limit $r_{h,i} \rightarrow 0$ of the third one, as expected. As for the second one, it can obtained by setting $r_{h,i}=i$ (the imaginary unit) in the third one and then using that $\tfrac{1}{2i} \log(z/\bar{z}) = \textrm{arg}(z)$ for any complex number $z$.
\\
At this point we remark that
\begin{equation}
\label{phiA dep}
\varphi_\mathcal{A} = \varphi_\mathcal{A} (r_s, \theta) 
= \varphi_\mathcal{A} (t_\mathcal{A}, \theta) \ .
\end{equation}
The first equality comes from the second formula in (\ref{tA and phiA}), (\ref{tilde phi rs}) and the expression of $\varphi_s$ in (\ref{phis theta}) corresponding to the geometry inside the shell, while the last step is found by further substituting $r_s = r_s(t_\mathcal{A}, \theta) $ given in (\ref{rs p}). Thus, by (numerically) inverting (\ref{phiA dep}), one can find $\theta=\theta(t_\mathcal{A},\varphi_\mathcal{A})$.

\paragraph{Behaviour near $\tA =0$ and $\tA =\phA$:}
It is instructive to perform such inversion through  series expansions around $t_\CA=0$ and $t_\CA=\phA$. In both these cases we use that 
\begin{equation}
\label{tan phA}
\tan \varphi_\mathcal{A} = 
\frac{\tan \phs(t_\mathcal{A}, \theta)  - \tan \tilde{\varphi}(t_\mathcal{A}, \theta) }{1+ \tan \phs(t_\mathcal{A}, \theta)  \, \tan \tilde{\varphi}(t_\mathcal{A}, \theta) }
\end{equation}
where $\phs(t_\mathcal{A}, \theta) $ is the expression in (\ref{phis theta}) corresponding to BTZ and $\tilde{\varphi}(t_\mathcal{A}, \theta)  \equiv \tilde{\varphi}(\rs) $ is obtained by plugging (\ref{rs p}) into (\ref{tilde phi rs}). We stress that all the following results provide the corresponding quantities also for Poincar\'e AdS and global AdS inside the shell. They are easily obtained as the special cases of $r_{h,i} = i$ (the imaginary unit) and $r_{h,i} = 0$ respectively.

About $t_\CA=0$, using that $\theta(0,\varphi_\mathcal{A})=0$, we can write
\begin{equation}
\label{tA=0 expansion phA}
\theta(t_\mathcal{A},\varphi_\mathcal{A}) \equiv \sum_{k=1}^{+\infty} \Theta_k(\phA)\,\tA^k
\qquad \Longrightarrow \qquad
\tan \varphi_\mathcal{A} 
\equiv
\sum_{n=0}^{+\infty} \Phi_n(\phA)\,\tA^n
\end{equation}
where for $\tan \varphi_\mathcal{A} $ (\ref{tan phA}) is applied and therefore the coefficients $\Phi_n$'s depend on $\phA$ through the $\Theta_k$'s. 
Since the l.h.s. of the second equation in (\ref{tA=0 expansion phA}) is independent of $\tA$, it can be solved order by order, namely $\tan \varphi_\mathcal{A} = \Phi_0(\phA)$
and $0= \Phi_n(\phA)$ for $n\geqslant 1$. From these equations we find respectively
\begin{equation}
\label{Thetas soln}
\Theta_1 = \frac{r_{h,i}}{\sinh (r_{h,i}\,\phA)}\;,
\hspace{.8cm}\Theta_2= 0 \;,
\hspace{.8cm}\Theta_3= \frac{\Theta_1(2\Theta_1 -r_{h,o}^2+3 r^2_{h,i})}{12} \;,
\hspace{.8cm}\Theta_4= 0 \;,
\hspace{.8cm}\dots
\end{equation}
which give the expansion of $\theta(t_\mathcal{A},\phA) $ around $\tA=0$. 

As for the expansion of $\theta(\tA ,\phA) $ around $t_\CA=\phA$, notice that $\theta(\phA,\phA)=\pi/2$. Now we need to introduce half-integer powers, namely (recall that $0<\tA<\phA$)
\begin{equation}
\label{tA=phA expansion phA}
\theta(t_\mathcal{A},\varphi_\mathcal{A}) \equiv 
\frac{\pi}{2} - \sum_{k=1}^{+\infty} \tilde{\Theta}_{k/2}(\phA)\,(\phA-\tA)^{k/2}
\qquad \Longrightarrow \qquad
\tan \varphi_\mathcal{A} 
\equiv
\sum_{p=0}^{+\infty} \tilde{\Phi}_{p/2}(\phA)\,(\phA-\tA)^{p/2} \ .
\end{equation}
As above, we can solve the second equation order by order. The first order gives no information since from the expansion of the r.h.s. of (\ref{tan phA}) with the first formula of (\ref{tA=phA expansion phA}) plugged in we find $\tilde{\Phi}_{0}(\phA) = \tan \varphi_\mathcal{A} $. For $p>0$ we impose $\tilde{\Phi}_{p/2}(\phA)=0$, finding that
\begin{align}
\label{Thetas tilde soln}
&\tilde{\Theta}_{1/2} = \sqrt{2r_{h,o} \coth(r_{h,o} \, \phA)}\,,
\hspace{1.5cm}
\tilde{\Theta}_{1} = -\frac{r_{h,o}^2-r_{h,i}^2}{3\,r_{h,o}}\, \tanh(r_{h,o}\, \phA)\,, 
\hspace{1.5cm}
\nonumber \\
&\tilde{\Theta}_{3/2} = 
\frac{-2 \,(5\,r_{h,o}^4 - 4\,r_{h,i}^2\, r_{h,o}^2 + 2\,r_{h,i}^4) + 3\,r_{h,o}^4\,\text{csch}^2(r_{h,o}\,\phA) + 4\, (r_{h,o}^2-r_{h,i}^2)^2\, \text{sech}^2(r_{h,o}\,\phA)}{18\,r_{h,o}^2\, \sqrt{2\,r_{h,o} \coth(r_{h,o}\,\phA)}}
\end{align}
and so on, which provide the expansion of $\theta(t_\mathcal{A},\phA) $ around $\tA=\phA$.

In order to find e.g. $\rE=\rE(\tA,\phA)$, we first write $\rE^2 = \rs^2-(\rs^2-r_{h,i}^2) \cos\theta $ from (\ref{cos theta def}), then we plug in $\rs(\tA,\theta)$ given in (\ref{rs p}) and finally we employ $\theta =\theta(t_\mathcal{A},\varphi_\mathcal{A})$ obtained by inverting (\ref{phiA dep}), as explained above. A plot of $\rE=\rE(\tA,\phA)$ as a function of $\tA$ for a fixed $\phA$  is shown in   \fig{f:rminXEcomp}.
Through the expansions of $\theta(t_\mathcal{A},\phA) $ found above, we can write the first terms of the expansion of $\rE=\rE(\tA,\phA)$ when $\tA \rightarrow 0$ and $\tA \rightarrow \phA$ for a BTZ spacetime inside the shell. They read respectively
\begin{eqnarray}
\rE(\tA,\phA) &=& r_{h,i} \coth(r_{h,i} \,\phA) 
+\frac{(r_{h,o}^2-r_{h,i}^2) r_{h,i}^3 \coth(r_{h,i} \,\phA)}{8 \,\sinh^2(r_{h,i} \,\phA)}\;\tA^4 +\dots
\\
\rule{0pt}{.8cm}
\rE(\tA,\phA) &=&r_{h,o} \coth(r_{h,o}\,\phA) 
-\frac{(r_{h,o}^2 - r_{h,i}^2)\,\sqrt{\phA-\tA}}{\sqrt{2r_{h,o} \coth(r_{h,o}\, \phA) }}
-\frac{(r_{h,o}^2- r_{h,i}^2)^2 (\phA-\tA)}{3\,r_{h,o}^2  \coth^2(r_{h,o}\, \phA) }
+\dots\ .
\nonumber\\
& & 
\end{eqnarray}
We can also find $r_s$ in terms of $(v_\mathfrak{E}, \rE)$ at the turning point.
Indeed, inside the shell we have
\begin{equation}
v_i(r) = t_i(r) +p_i(r) +B_i
\end{equation}
where $B_i$ is a constant and  $p_i(r)$ is the solution of $p'=1/f_i(r)$ given by $p_i(r)=\tfrac{1}{2r_{h,i}}
\log\frac{r-r_{h,i}}{r+r_{h,i}}$ for BTZ, $p_i(r) = \arctan r$ for global AdS and $p_i(r) = -1/r$ for Poincar\'e AdS.
Since $E_i=0$, we have that $t_i(r) =0$ for every $r$ inside. In particular, considering the point $r=\rE$, we obtain $B_i = v_\mathfrak{E} -p_i(\rE)$. Then, by imposing $v_i(r_s) =0$, we find that $p_i(r_s)  = p_i(\rE) - v_\mathfrak{E}$
and therefore we can write $r_s$ in terms of $(v_\mathfrak{E}, \rE)$.

\subsection{Holographic entanglement entropy}

According to the prescription of \cite{Ryu:2006bv,Ryu:2006ef,Hubeny:2007xt}, the holographic entanglement entropy $S_\CA$ is given by the length of the spacelike geodesic studied in the previous subsection. The functional giving the length of a piece of curve which is either entirely inside ($\alpha=i$) or entirely outside the shell ($\alpha=o$)  reads
\begin{equation}
\label{proper length}
\mathcal{L}_\alpha = \int_{r_1}^{r_2} \sqrt{- f_\alpha (t')^2 +\frac{1}{f_\alpha} +r^2 (\varphi')^2}\, dr \ .
\end{equation}
Integrating (\ref{proper length}) with (\ref{eqs spacelike geods}) plugged in, a primitive is
\begin{equation}
\label{primitive btz}
\mathcal{L}_{\alpha}(r)
\equiv
\frac{1}{2} \,\log 
\Big(
E^2-L^2+2r^2-r_h^2+2\sqrt{E^2 r^2+(r^2-r_h^2)(r^2-L^2)}\,
\Big) \ .
\end{equation}
We recall that if $\alpha=i$, then  $\rh^2=r_{h,i}^2$ in this expression and, in particular, $r_{h,i}^2=-1$ for global AdS and $r_{h,i}^2=0$ for Poincar\'e AdS inside the shell. In the latter case the argument of the logarithm becomes a perfect square. \\
The holographic entanglement entropy is given by
\begin{equation}
\label{hee}
4\,G_N^{(3)} \, S_\mathcal{A} = 
2\big[\mathcal{L}_{i}(r_s)-\mathcal{L}_{i}(\rE)\big]
+
2\big[\mathcal{L}_{\textrm{\tiny BTZ}}(r_\infty) -\mathcal{L}_{\textrm{\tiny BTZ}}(r_s)\big]
\end{equation}
where $G_N^{(3)}$ is the three dimensional Newton constant and $r_\infty\rightarrow \infty$ is the UV cutoff in the boundary theory (cf., Footnote \ref{fn:centralcharge} for conversion between $G_N^{(3)}$ and the boundary central charge $c_\text{eff}$).

Outside the shell, by using (\ref{primitive btz}) and that $\mathcal{L}_{\textrm{\tiny BTZ}}(r_\infty) = \log(2\, r_\infty) + O(r_\infty^{-2})$ asymptotically, we find
\begin{eqnarray}
\label{hee out part}
2\big[\mathcal{L}_{\textrm{\tiny BTZ}}(r_\infty)-\mathcal{L}_{\textrm{\tiny BTZ}}(r_s)\big]
&=&
2\log(2\, r_\infty) -
\log 
\Big(
E_o^2-\rE^2+2r_s^2-r_{h,o}^2+2\sqrt{D_o(r_s)}\,
\Big)\\
& & \hspace{-5cm}=\;
2\log(2\, r_\infty) 
- \log \Bigg(
\frac{[f_o(r_s) - f_i(r_s)]^2 (r_s^2 - \rE^2)}{4r_s^2 f_i(r_s)}
- \rE^2+2r^2_s - r_{h,o}^2 
+ \sqrt{\frac{r_s^2 - \rE^2}{f_i(r_s)}} \,\big| f_o(r_s) + f_i(r_s)\big|
 \Bigg)
 \nonumber
\end{eqnarray}
where $D_o(r)$ has been defined after (\ref{varphi(r) btz}). In the second step, (\ref{EL out spacelike geods}) and the expression of $D_o(\rs)$ given after (\ref{tA and phiA}) have been used.
Inside the shell, since $E_i = 0$, the argument of the logarithm in (\ref{primitive btz}) becomes a perfect square and therefore we find
\begin{equation}
\label{hee inside part}
2\big[\mathcal{L}_{i}(r_s)-\mathcal{L}_{i}(\rE)\big]
=
2 \log \Bigg( \frac{\sqrt{f_i(r_s)} + \sqrt{r_s^2 - \rE^2}}{\sqrt{\rE^2 -r_{h,i}^2}} \,\Bigg)\,.
\end{equation}
Combining together the two contributions (\ref{hee out part}) and (\ref{hee inside part}), and discarding the divergent term $2\log(2\, r_\infty) $, we find for the regularized holographic entanglement entropy
\begin{eqnarray}
\label{hee reg}
4 G_N^{(3)}  S_{\mathcal{A},\textrm{reg}} 
&=&
2 \log \Bigg( \frac{\sqrt{r_s^2 - r_{h,i}^2} + \sqrt{r_s^2 - \rE^2}}{\sqrt{\rE^2 -r_{h,i}^2}} \,\Bigg)
\\
& & \hspace{-.5cm}
- \,\log \Bigg(
\frac{\big(r_{h,o}^2-r_{h,i}^2\big)^2 (r_s^2 - \rE^2)}{4r_s^2 (r_s^2 - r_{h,i}^2)}
- \rE^2+2r^2_s - r_{h,o}^2 
+ \sqrt{\frac{r_s^2 - \rE^2}{r_s^2 - r_{h,i}^2}} \,\big( 2r_s^2 - r_{h,o}^2-r_{h,i}^2\big)
 \Bigg) .
 \nonumber
\end{eqnarray}
It is interesting to rewrite this formula by employing (\ref{cos theta def}). First, notice that
\begin{equation}
\label{hee theta in}
 \frac{\sqrt{f_i(r_s)} + \sqrt{r_s^2 - \rE^2}}{\sqrt{\rE^2 -r_{h,i}^2}} 
 =
 \frac{1+\cos\theta}{\sin\theta} \ .
\end{equation}
After some algebra the argument of the logarithm in the second line of (\ref{hee reg}) can be written as 
\begin{equation}
\label{hee theta out}
r_{h,o}^2(1+\cos\theta)^2
\left[\left(
\frac{\cos \theta (1-r_{h,i}^2/r_{h,o}^2)}{2(1+\cos\theta)r_s/r_{h,o}}
+\frac{r_s}{r_{h,o}}\right)^2
-1\,\right]
=
\frac{r_{h,o}^2(1+\cos\theta)^2}{ \sinh^2(r_{h,o} t_\mathcal{A})}
\end{equation}
where the second step is obtained by isolating $\coth(r_{h,o} t_\mathcal{A})$ in (\ref{eq II order rs}) first and then by using the identity $\coth^2 x -1 =1/\sinh^2 x$.
Thus, putting together (\ref{hee theta in}) and (\ref{hee theta out}), one can write the regularized holographic entanglement entropy (\ref{hee reg}) in a compact form. Adding also the divergent term coming from (\ref{hee out part}), the  holographic entanglement entropy reads
\begin{equation}
\label{hee theta}
 S_{\CA} (\tA,\phA)
\,=\,\frac{c_\text{eff}}{3}\,\log\left(\frac{2\, r_\infty\, \sinh( r_{h,o} t_\mathcal{A})}{r_{h,o} \sin\theta (t_\mathcal{A},\varphi_\mathcal{A}) }\right)
\end{equation}
where we remarked that $\theta= \theta(\tA,\phA)$, as discussed after (\ref{phiA dep}).
For Poincar\'e AdS inside the shell, we recover the result of \cite{Balasubramanian:2010ce,Balasubramanian:2011ur}.
Notice that the expression (\ref{hee theta}) is formally the same for all the choices of the geometry inside the shell that we considered. The definition (\ref{cos theta def}) distinguishes among them.

\paragraph{Behaviour near $\tA =0$ and $\tA =\phA$:}
It is  useful to write the first terms of the expansions of $S_{\CA} (\tA,\phA)$ around $\tA=0$ and $\tA=\phA$.\\
As for the expansion around  $\tA=0$, first we plug into (\ref{hee theta}) the expansion of $\theta(t_\mathcal{A},\varphi_\mathcal{A})$ given in (\ref{tA=0 expansion phA}) and then employ the explicit expressions for $\Theta_k(\phA)$, obtained by solving the second equation in (\ref{tA=0 expansion phA}) order by order in $\tA$ (see (\ref{Thetas soln})). Thus, we find that
\begin{eqnarray}
\label{hee tA=0 btz}
\frac{6}{c_\text{eff}}\, S_{\CA} (\tA,\phA)
&=&
2\, \log \left(\frac{2\, r_\infty}{r_{h,i}}\,\sinh(r_{h,i} \,\phA) \right)
+\frac{r_{h,o}^2-r^2_{h,i}}{2}\,\tA^2 
\\
\rule{0pt}{.8cm}
& &
-\, \frac{r_{h,o}^2-r^2_{h,i}}{48} \left(
\frac{6 \,r_{h,i}^2}{\sinh^2(r_{h,i}\,\phA)} 
+r_{h,o}^2 + 3  r_{h,i}^2
\right)\tA^4
+ O(\tA^6)\ .
\nonumber
\end{eqnarray}
When the spacetime inside the shell is global AdS (i.e. when $r_{h,i}$ is the imaginary unit), the expansion (\ref{hee tA=0 btz}) becomes (\ref{Snear0}).
Instead, the limit $r_{h,i} \rightarrow 0$ of (\ref{hee tA=0 btz}) provides the corresponding result for Poincar\'e AdS inside the shell, namely
\begin{equation}
\label{hee tA=0 poincare}
\frac{6}{c_\text{eff}}\, S_{\CA} (\tA,\phA)
\,=\,
2\, \log \left(2\, r_\infty\, \phA\right)
+\frac{\rh^2}{2}\,\tA^2
-\frac{\rh^2}{48} \left(\frac{6}{\phA^2} +\rh^2\right)\tA^4
+ O(\tA^6) \ .
\end{equation}
Discarding the constant term in $\tA$ and taking the limit of large $\phA$ of the remaining part, we get $\rh^2\tA^2/2-\rh^4\tA^4/48+O(\tA^6) $. These are the first terms of the expansion for small $\tA$ of the function $4\log(\cosh(\rh \tA/2))$ found in \cite{Aparicio:2011zy} (see also footnote \ref{fn:earlyt}).\\
A similar procedure provides the expansion of $S_{\CA} (\tA,\phA)$ for $\tA \rightarrow \phA^-$: plugging the expansion of $\theta(t_\mathcal{A},\varphi_\mathcal{A})$ defined in (\ref{tA=phA expansion phA}) into (\ref{hee theta}) and then using (\ref{Thetas tilde soln}), we find
\begin{align}
\label{hee tA=phA btz}
\frac{6}{c_\text{eff}}\, S_{\CA} (\tA,\phA)
\,&=\,
2\, \log \left(\frac{2\,r_\infty}{r_{h,o}}\,\sinh(r_{h,o}\,\phA)\right)
-\frac{2\sqrt{2}\,(r_{h,o}^2-r^2_{h,i})}{3 \sqrt{r_{h,o} \coth(r_{h,o}\, \phA)}} \,\big(\phA-\tA\big)^{3/2}
\nonumber \\
&-\frac{(r_{h,o}^2-r^2_{h,i})^2\, \tanh^2(r_{h,o}\,\phA)}{3\,r_{h,o}^2}\,\big(\phA-\tA\big)^{2} 
+ \dots
\end{align}
where the dots represent subleading contributions. Specializing this result to the case of global AdS inside the shell, we find  (\ref{snearpha}). Instead, setting $r_{h,i}=0$ we obtain the corresponding expansion for Poincar\'e AdS inside the shell.


\providecommand{\href}[2]{#2}\begingroup\raggedright\endgroup

\end{document}